\documentclass[a4paper,11pt]{article}
\pdfoutput=1 % if your are submitting a pdflatex (i.e., if you have
\usepackage[table]{xcolor}
\usepackage{lmodern}
\usepackage{jcappub} % for details on the use of the package, please
\usepackage[toc,page]{appendix}
\usepackage{subcaption}
\usepackage{mathtools}
\usepackage{orcidlink}
\usepackage{multirow}
\usepackage[T1]{fontenc} % if needed
\usepackage{float}
\usepackage{bm}
\usepackage{feynmp-auto}

\newcommand{\semibold}[1]{{\fontseries{b}\selectfont{#1}}}

\newcommand{\para}[1]{\par\vspace{2mm}\noindent\semibold{{#1.}---}\ignorespaces} 

\newcommand{\eqgraph}[3]{%
  \begin{gathered}
  \raisebox{0pt}[\dimexpr\height+#1][\dimexpr\depth+#2]{\ignorespaces#3\unskip}%
  \end{gathered}
}

\DeclareMathOperator{\sinc}{sinc}

\DeclareMathOperator{\Or}{O}

\DeclareMathOperator{\NewRe}{Re}

\renewcommand{\Re}{\NewRe}

\newcommand{\comma}{,}

\definecolor{verde}{rgb}{0,0.5,0}

\definecolor{SussexCobaltBlue}{HTML}{1d4289}
\definecolor{SussexDeepAquamarine}{HTML}{007a78}
\definecolor{SussexPowderBlue}{HTML}{7da1c4}
\definecolor{SussexCornYellow}{HTML}{f2c75c}
\definecolor{SussexChinaRose}{HTML}{be84a3}
\definecolor{SussexBurntOrange}{HTML}{dc582a}

\hypersetup{colorlinks=true,citecolor=SussexDeepAquamarine,linkcolor=SussexCobaltBlue,urlcolor=SussexBurntOrange}

\newcommand{\etal}{\emph{et al.}}

\renewcommand{\d}{\mathrm{d}}
\newcommand{\e}[1]{\mathrm{e}^{{#1}}}
\newcommand{\im}{\mathrm{i}}
\newcommand{\Mp}{M_{\mathrm{P}}}
\newcommand{\vect}[1]{\bm{\mathrm{{#1}}}}

\newcommand{\GeV}{\text{GeV}}

\newcommand{\dimlessP}{\mathcal{P}}
\newcommand{\fNL}{f_{\mathrm{NL}}}
\newcommand{\tauNL}{\tau_{\mathrm{NL}}}
\newcommand{\gNL}{g_{\mathrm{NL}}}

\newcommand{\fNLeq}{f_{\mathrm{NL}}^{\text{eq}}}
\newcommand{\fNLloc}{f_{\mathrm{NL}}^{\text{loc}}}
\newcommand{\gNLloc}{g_{\mathrm{NL}}^{\text{loc}}}

\newcommand{\zetaGaussian}{\zeta_g}

\newcommand{\bigBox}{L}
\newcommand{\kBig}{k_{\bigBox}}
\newcommand{\zetaBig}{\zeta_{\bigBox}}
\newcommand{\zetaBigSmooth}{\overline{\zetaBig}}
\newcommand{\sigmaBig}{\sigma_{\bigBox}}
\newcommand{\sigmaBigSmooth}{\bar{\sigma}_{\bigBox}}

\newcommand{\smallBox}{\ell}
\newcommand{\kSmall}{k_{\smallBox}}
\newcommand{\zetaSmall}{\zeta_{\smallBox}}
\newcommand{\sigmaSmall}{\sigma_{\smallBox}}
\newcommand{\SSmall}{S_{\smallBox}}

\newcommand{\smoothScale}[2]{\{\hspace{-0.8ex}\{{{#2}}\}\hspace{-0.8ex}\}_{{#1}}}
\newcommand{\smoothBig}[1]{\smoothScale{\bigBox}{{#1}}}
\newcommand{\smoothSmall}[1]{\smoothScale{\smallBox}{{#1}}}

\newcommand{\IndicatorFunction}{\bm{1}}

\newcommand{\Action}{S}
\newcommand{\ActionCt}{\Action_{\text{ct}}}

\newcommand{\deltact}{\delta_{\text{ct}}}
\newcommand{\tildeDeltact}{\tilde{\delta}_{\text{ct}}}

\newcommand{\Operator}[1]{\mathcal{{#1}}}
\newcommand{\OpO}{\Operator{O}}
\newcommand{\OpQ}{\Operator{Q}}
\newcommand{\OpS}{\Operator{S}}

\title{Loop corrections in the separate universe picture}

\author[1,2]{Laura Iacconi,}
\author[1]{David Mulryne,}
\author[3]{David Seery\,$^{\text{\orcidlink{0000-0003-3421-6080}}}$}

\affiliation[1]{Astronomy Unit, Queen Mary University of London, \\
Mile End Road, London, E1 4NS, UK}
\affiliation[2]{Institute of Cosmology and Gravitation, University of Portsmouth, \\
Burnaby Road, Portsmouth, PO1 3FX, UK}
\affiliation[3]{Astronomy Centre, University of Sussex, \\
Falmer, Brighton, BN1 9QH, UK}

\emailAdd{l.iacconi@qmul.ac.uk}
\emailAdd{d.mulryne@qmul.ac.uk}
\emailAdd{D.Seery@sussex.ac.uk}

\abstract{
    In inflationary models that
    produce a spike of power
    on short scales,
    back-reaction of small-scale
    substructure onto large-scale modes
    is enhanced.
    Loop corrections that quantify this back-reaction have
    been evaluated by a number of authors.
    We argue that the separate universe framework provides a
    highly convenient tool
    for such computations.
    Each loop of interest is characterized by large hierarchies
    in wavenumber and horizon exit time.
    The separate universe framework
    highlights important
    factorizations involving these 
    hierarchies. 
    We interpret each loop correction in terms of a simple,
    classical, back-reaction model,
    and clarify the meaning of the different volume scalings
    that have been reported in the literature.
    We argue that significant back-reaction
    requires \emph{both}
    short-scale nonlinearities
    and long--short couplings that modulate
    the short-scale power spectrum.
    In the absence of long--short
    couplings, only incoherent ``shot noise''-like
    effects are present, which
    are volume-suppressed.
    Dropping the shot noise,
    back-reaction from a particular scale
    is controlled by a product
    of $\fNL$-like parameters:
    an equilateral configuration measuring the
    nonlinearity of the short-scale modes,
    and a squeezed configuration measuring the
    long--short coupling.
    These
    may carry important
    scale dependence
    controlling
    the behaviour of the loop in the decoupling
    limit where the hierarchy of scales becomes large.
    In single-field models the long--short coupling
    may be controlled by this hierarchy,
    in which case the net back-reaction would
    be safely suppressed.
    We illustrate our framework using
    explicit computations in a 3-phase ultra-slow-roll scenario.
    Our analysis differs from earlier treatments of this model,
    which did not consistently include the effect of small-scale modes.
    Finally, we discuss different choices for
    the smoothing scale used in the separate
    universe framework and argue the
    effect can be absorbed into a
    renormalization of local operators.
    This complicates interpretation of the loop, because
    the analytic part of each loop integral
    is
    degenerate with unknown, ultraviolet-sensitive contributions.
}

\begin{document}
\maketitle
\newpage

\begin{fmffile}{diagrams}
\section{Introduction}
\label{sec:intro}
\noindent
It is a longstanding proposal
that the primordial power spectrum might grow significantly
on small scales relative to its value on CMB or galaxy scales.
Several independent lines of evidence have recently repopularized
this idea.
First, despite advances in detector technology, the lack
of clear evidence
for a WIMP-like dark matter particle has encouraged
consideration of alternative scenarios.
Several possibilities exist, but one option is that
(at least some of)
the dark matter
is locked into small-scale collapsed objects,
perhaps primordial black holes
formed from direct collapse of high peaks in the
inflationary density
perturbation~\cite{Carr:2020xqk,Green:2020jor,Villanueva-Domingo:2021spv}.
To produce an appreciable population of these objects would require
enhancement of the power spectrum by roughly
seven orders of magnitude at the relevant wavenumber $k$.

Second, the LIGO/Virgo/KAGRA collaboration
continues to detect significant numbers of black-hole merger events
via their gravitational radiation.
The inferred
mass and spin distribution of the progenitor objects is not easy
to reconcile with an interpretation in which all such objects form at
the endpoint of stellar evolutionary
pathways~\cite{Bird:2016dcv}.
In particular, the presence of
significant numbers
of low-spin progenitors may suggest the existence
of a population of primordial black holes.
Modelling the evolution of the spin distribution
for such a population is extremely challenging,
due to (among other issues)
uncertainties in the accretion physics.
These details are currently a topic of
active debate.
While
there are reasons to believe that the
entire population of progenitor objects
is unlikely to be primordial in origin~\cite{Hall:2020daa}, 
it remains possible
that there is a primordial component~\cite{Franciolini:2022tfm}.
This would again require an enhancement of power.

\para{Back-reaction and mode-coupling}
In either of these scenarios,
and similar ones,
we must take seriously the possibility
that back-reaction
from a spike of power at some
high wavenumber $k$
might spoil the successful prediction of
a roughly Gaussian, scale-invariant
perturbation at low wavenumbers $p \ll k$
that seed galaxy formation or the cosmic microwave background anisotropy~\cite{Planck:2018jri}.
This is because,
whatever its origin,
a realistic model for the primordial density field
must predict \emph{some} coupling between Fourier modes.
This coupling enables short-wavelength power to spill over
into the distribution
of long-wavelength field values.

There are standard tools to measure this
leakage of power.
Recently, 
%Inomata {\etal}~\cite{Inomata:2022yte} and
Kristiano \& Yokoyama~\cite{Kristiano:2022maq}
used the methods of nonequilibrium field theory
to compute a ``one-loop'' correction to the usual
tree-level power spectrum.
The terminology of the loop expansion is
most familiar from perturbative field theory
based on the equilibrium vacuum state,
in which the loops are dominated by vacuum
fluctuations and can be regarded as
capturing the effect of quantum corrections.
At least in their ultraviolet part, such loops
measure the way that a bath of short-wavelength
fluctuations
influences the behaviour of
long-wavelength modes.
This applies whether the modes in the bath
evolve according to classical or quantum
laws.
The conclusion is that loop-level terms
such as those computed by
%Inomata {\etal} and 
Kristiano \& Yokoyama
can be regarded as
a measure of back-reaction.

In an inflationary model there are
two types of contribution to the short-wavelength
bath.
Superhorizon modes (but still shortward
of the modes of interest)
arising from
the inflationary density perturbation
sometimes have long-lived coherent behaviour
because the field operators and their canonical
momenta approximately commute.%
    \footnote{This property can be disrupted by ultra-slow-roll
    effects, so we generally assume
    modes are in this regime only when all
    ultra-slow-roll phases have ended.}
In this regime
we can model their behaviour
as stochastic
fields with classical time dependence.
Meanwhile, subhorizon modes
(and possibly others)
supply a bath of short-scale
oscillators that must be treated
quantum-mechanically.
The loop correction
must aggregate the influence
from fluctuations of both types.
An analogous situation occurs in
finite temperature field theory,
where loops
measure the aggregate effect of thermal
\emph{and} vacuum fluctuations~\cite{Kapusta:2006pm}.
For example,
it is well-known
that
when coupled to a spin-0 field,
the effect of thermal
fluctuations is to generate a confining potential
that drives the spin-0 vev to zero
at sufficiently high temperature.
This is a form of back-reaction in which
the cold, highly ordered,
long-wavelength
spin-0 condensate is disordered by
incoherent,
short-wavelength thermal effects.
It is possible that
the CMB-scale
density field
could be similiarly disordered
due to
short-scale fluctuations.
Related examples in a
cosmological context
are the loop corrections
that appear in the effective field
theory of large scale
structure~\cite{Carrasco:2012cv,Carrasco:2015uma,Pajer:2013jj,Abolhasani:2015mra}.

The conclusion drawn
by Kristiano \& Yokoyama
was that
one-loop back-reaction may be significant,
and possibly already
large enough (even without higher loop orders)
to invalidate the tree-level analysis~\cite{Kristiano:2022maq}.
Their
calculation
primarily depended on
approximate analytical estimates,
and
it is not yet entirely clear whether
these adequately capture the relevant physics. 
As a result, the calculations of
Ref.~\cite{Kristiano:2022maq}
were soon refined by
many other authors~\cite{Riotto:2023hoz, Choudhury:2023vuj,
Choudhury:2023jlt, Kristiano:2023scm, Riotto:2023gpm,
Firouzjahi:2023aum, Motohashi:2023syh, Choudhury:2023rks,
Choudhury:2023hvf, Firouzjahi:2023ahg, Firouzjahi:2023btw, Franciolini:2023lgy, Tasinato:2023ukp, 
Cheng:2021lif, Cheng:2023ikq,
Fumagalli:2023hpa, Maity:2023qzw, Tada:2023rgp,
Firouzjahi:2023bkt, Davies:2023hhn}.%
    \footnote{For an analysis of the contribution of enhanced small-scale
    modes to the one-loop power spectrum evaluated at the peak scales and at near infrared scales,
    see Refs.~\cite{Inomata:2022yte, Iacconi:2023slv} and Ref.~\cite{ Fumagalli:2023loc} respectively.}
This literature
is already too extensive to review in detail here.
However, the current position can be summarized
by saying there is not
yet agreement
on the size of the one-loop effect.
Neither is there agreement
on the way
regularization and
renormalization of the divergent answer
should be handled.
We discuss some of these issues in~\S\ref{sec:discussion}.

In this paper we revisit the calculation of
one-loop corrections.
Our primary tool is the separate universe
framework~\cite{Starobinsky:1985ibc,Sasaki:1995aw,Lyth:2005fi}.
Many previous analyses have used the \emph{in--in}
(or `Schwinger--Keldysh') formulation of nonequilibrium
quantum field theory, beginning with the early analyses
of Inomata {\etal} and
Kristiano \& Yokoyama~\cite{Inomata:2022yte,Kristiano:2022maq}.
Once calculations progressed beyond linear order,
it was soon established that
both approaches
yielded identical results
for the tree-level $\zeta$
three-point function~\cite{Seery:2005gb,Lyth:2005fi}.
A more general demonstration of the
equivalence appeared in Refs.~\cite{Mulryne:2013uka,Seery:2012vj},
although still restricted to tree-level.
At loop-level,
the precise relationship between these two approaches
does not yet appear to be settled.

In particular,
Kristiano \& Yokoyama focused on a loop-level
contribution proportional to the commutator
$[ \zeta, \zeta' ]$, which temporarily grows during
ultra-slow-roll inflation.
This growth signals amplification of a previously
decaying mode.
Kristiano \& Yokoyama characterized this effect
as intrinsically quantum-mechanical.
Once the ultra-slow-roll stage has passed,
the $[ \zeta, \zeta' ]$ commutator
decays normally.
We expect that this effect will
change the statistics of particle creation
around the horizon scale, where the
$[ \zeta, \zeta' ]$ commutator is not yet
strongly suppressed.
This would be relevant for
modes near the peak of the power spectrum.
On the other hand, once particle creation is complete
and the $[ \zeta, \zeta' ]$ commutator
begins to decay,
we expect that the subsequent
evolution of these modes can be described classically.
Hence, for scales near the peak,
the separate universe framework would then apply
in the usual way.

Meanwhile, on large scales where we wish to
estimate the one-loop back-reaction effect,
the commutator $[ \zeta, \zeta' ]$ is strongly suppressed
outside the horizon, even if it does receive a small
amplification during the ultra-slow-roll period.
(This argument parallels the discussion of the
decaying mode during ultra-slow-roll;
see~\S\ref{sec: delta N CMB mode from hc}
and~\S\ref{sec: delta N CMB mode from USR time}.)
We therefore conclude, on physical grounds,
that the separate universe framework should
provide an adequate description of the
time dependence of
modes between the peak scale and the
CMB scale,
provided we wait until all
inflationary particle production
effects have ceased and relevant commutators
$[ \zeta, \zeta' ]$ have begun to be suppressed.
This is not very different to the usual stipulation
that we must wait until all relevant scales are outside
the horizon before invoking the separate universe framework.

Accepting this approach,
we argue that the separate universe framework has
at least two concrete advantages.
First, on the basis of what has already been
said, the back-reaction effect to be computed
involves only modes that can be modelled by
a stochastic field with classical time dependence.
This is a considerable simplification
compared to the full
in--in approach,
and makes it easier
to separate important aspects of the physics.
It yields a relatively transparent physical
interpretation
that can be framed in terms of a
classical back-reaction model.
We introduce this model in~\S\ref{sec:toy-model}.

Second, the loop effects under discussion
involve a large hierarchy of scales---%
between the early horizon-exit time of the CMB
modes, and the late horizon-exit time of the enhanced peak scales.
Inflationary correlation functions
(just like correlation functions that appear in many other applications
of quantum field theory)
develop a very rich structure
in the presence of such hierarchies~\cite{Dias:2012qy,Burrage:2011hd}.
In many cases, the separate universe
framework
explains how
large effects associated with these hierarchies
may be factorized into products of
lower-order correlation functions~\cite{Dias:2012qy,Kenton:2015lxa}.
Maldacena's famous consistency condition is one
example of a factorization result of this type~\cite{Maldacena:2002vr}.
Another is the squeezed limit of the three-point
correlation function in phase space studied by
Kenton \& Mulryne~\cite{Kenton:2015lxa}.
Clearly,
it must be possible
to obtain these factorization formulae
within
the framework of in--in perturbation theory.
However,
experience has shown that the factorization results
require considerably more work to derive there.
The separate universe framework
encodes these results
in a technically simple way.

The separate universe framework (in the context of the
$\delta N$ formula for the curvature perturbation,
$\zeta$, on uniform
density slices)
has already been used to compute loop corrections by
Firouzjahi \& Riotto~\cite{Firouzjahi:2023ahg}.
While our implementation is similar to theirs, 
they differ in the application of the framework.
We give a more detailed discussion in~\S\ref{sec:delta-N-back-reaction}. 

\subsection{A classical, stochastic back-reaction model}
\label{sec:toy-model}
As explained above,
in this paper our primary focus is the contribution
from outside-the-horizon modes that can be treated
as a stochastic source.
We will return to the question of vacuum
contributions in~\S\ref{sec:discussion-wilsonian-eft}.

\para{Correlation function between patches}
To calibrate our expectations,
suppose that we divide some region
of the universe into large superhorizon-sized patches
of characteristic size $\bigBox \sim 2\pi / \kBig$.
We spatially average over these regions
to produce a smoothed value for each perturbation.
To keep the discussion simple, we
usually work
only with the
curvature perturbation on a uniform density
slicing,
$\zeta$,
and denote its smoothed value by $\zetaBig$.
(However, our methods are general and will
apply to other species of perturbation.)
It is obtained from
the spatial average
\begin{equation}
    \label{smoothed zeta_L}
    \zetaBig(\vect{x}) =
    \frac{1}{V} \int \d^3 x' \;
    W(L^{-1}|\vect{x}-\vect{x}'|)
    \zeta(\vect{x}') , 
\end{equation}
where $W(z)$ is a window function that falls to zero
rapidly when $z \gtrsim 1$, and
$V$ is its volume.
The coarse-grained
coordinate $\vect{x}$ labels the spatial position
of each $\bigBox$-sized box.
The familiar mechanism of inflationary perturbations will
disorder the field,
so that $\zetaBig$ takes a different value in
each patch.
Its distribution over the ensemble of patches will be almost
Gaussian.

\begin{figure}
    \centering
    \captionsetup[subfigure]{justification=centering}
    \begin{subfigure}[b]{0.49\textwidth}
    \includegraphics[width=\textwidth]{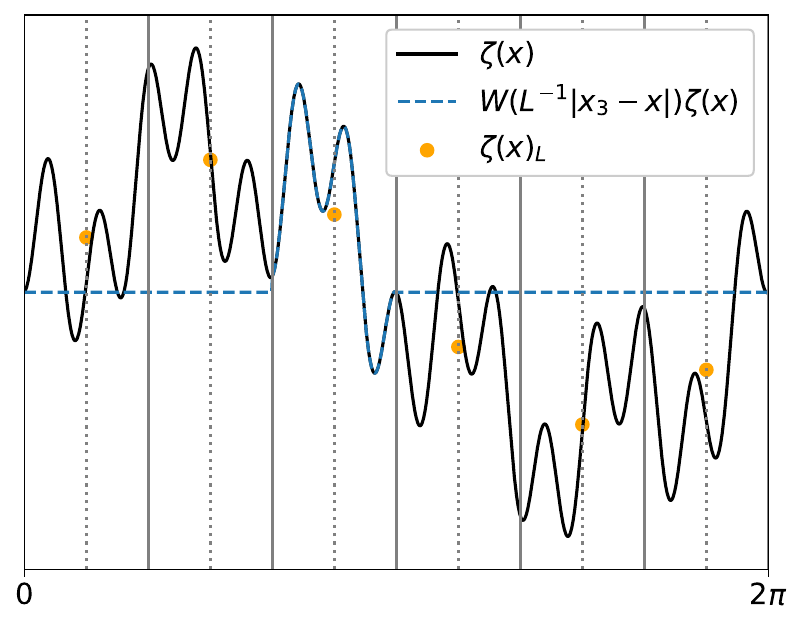}
    \end{subfigure}
    \begin{subfigure}[b]{0.49\textwidth}
    \includegraphics[width=\textwidth]{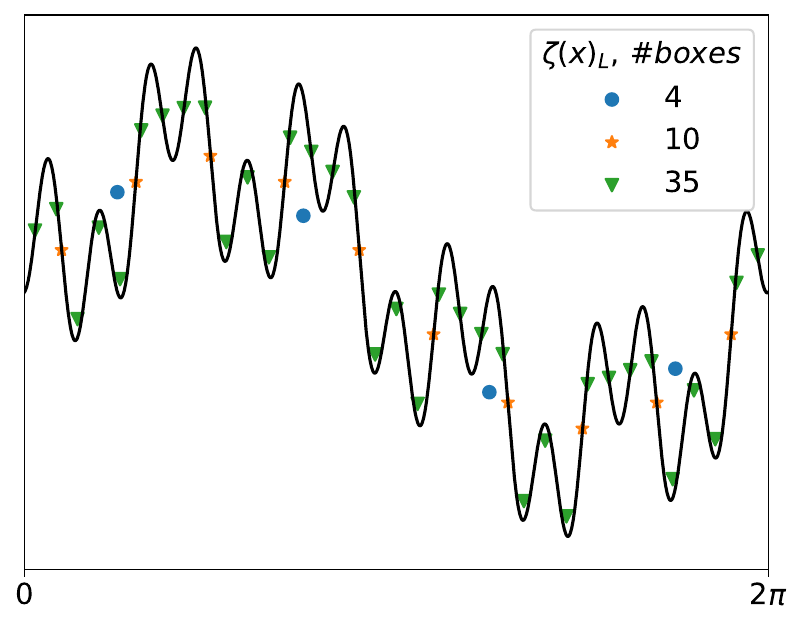}
    \end{subfigure}
    \caption{\label{fig:spatial averaging}\semibold{Left panel}:
    Spatial averaging of a one-dimensional signal,
    $\zeta(x)$ with $x\in [0,2\pi]$,
    over patches with size $\bigBox=2\pi/6$.
    \semibold{Right panel}:
    Effect of changing the size of the
    $\bigBox$-patches, where $\bigBox=2\pi/ (\# \text{boxes})$.}
\end{figure}
[For comparison,
we show an example of spatial averaging in
Fig.~\ref{fig:spatial averaging}.
We apply Eq.~\eqref{smoothed zeta_L} to a one-dimensional signal,
and demonstrate the effect of changing the size of the $\bigBox$-patch.
As $\bigBox$ decreases, an increasing
number of high-wavenumber modes
are retained after smoothing.]

Using Eq.~\eqref{smoothed zeta_L} we find
the correlation of the smoothed field between patches
satisfies
\begin{equation}
    \label{eq:patch-correlation}
    \langle \zetaBig(\vect{x}) \zetaBig(\vect{x} + \vect{r}) \rangle
    =
    \int_0^\infty \frac{\d k}{k} \;
    \Big[ L^3 W(2\pi k/\kBig) \Big]^2
    \dimlessP(k)
    \sinc kr ,
\end{equation}
where $\vect{r}$ is the coarse-grained displacement
between patches and
$W(k)$ is the Fourier transform of the normalized window function $W(z)/V$.
Notice that $\zetaBig$ and $\dimlessP(k)$
may each be time dependent, although we suppress
this in the notation.

For $r \lesssim L$, the locations $\vect{x}$ and $\vect{x} + \vect{r}$
label the same coarse-grained volume.
The integral in~\eqref{eq:patch-correlation}
receives contributions from $k$
smaller than the
window-function cutoff
$\sim \kBig$. For these wavenumbers we can
roughly approximate
$\sinc kr \approx 1$,
and therefore
$\langle \zetaBig(\vect{x}) \zetaBig(\vect{x}+\vect{r}) \rangle
\approx \langle \zetaBig(\vect{x})^2 \rangle = \sigmaBig^2$,
where%
    \footnote{\label{footnote:ir-convergence}Here and in the remainder of this paper
    we ignore questions of convergence in the infrared limit
    $k \rightarrow 0$,
    which concern the distribution of field values on unobservably
    large scales
    and do not significantly
    affect the discussion of back-reaction on the scale $L$.
    For example,
    if $\dimlessP(k)$ is red-tilted we assume an infrared cutoff
    is implicitly present by computing conditional expectations within
    our locally observable region, which has a defined value of the ultra-large scale
    background. See, e.g., Lyth~\cite{Lyth:2007jh} and also Ref.~\cite{Seery:2010kh}.}
\begin{equation}
    \label{eq:sigmaBig-def}
    \sigma_L^2 \approx \int_0^{\kBig} \frac{\d k}{k} \; \dimlessP(k) .
\end{equation}
We identify
$\sigmaBig^2$ as
the variance of the smoothed field $\zeta_L$ among the $L$-sized
patches.

If instead
$r \gg L$,
the oscillations of $\sinc kr$ effectively cut off the integral
at $k \sim 1/r \ll \kBig$.
In this region
the window function satisfies $W \approx 1$.
On these scales we have
$\langle \zetaBig(\vect{x}) \zetaBig(\vect{x}+\vect{r}) \rangle \approx \sigma^2_{r}$.
If the power spectrum is scale invariant,
or grows towards low wavenumbers, then spatial correlations
persist between patches.
On the other hand, if the power spectrum drops significantly
for $k \lesssim \kBig$
then the correlation function behaves like a smoothed $\delta$-function.
If the drop is sufficiently sharp then it may be a good approximation to
neglect correlations over scales larger than a few $L$ patches.%
    \footnote{To make a more precise statement
    one can develop asymptotics for the $\sinc$
    integral~\eqref{eq:patch-correlation}
    using the Barnes representation
    \begin{equation}
        \label{eq:barnes-sinc}
        \sinc x =
        \frac{1}{2\pi \im}
        \int_{c-\im \infty}^{c + \im \infty}
        \Gamma(s) x^{-s-1} \sin \frac{\pi s}{2} \, \d s ,
    \end{equation}
    for $x > 0$ and $0 < \Re(s) < 1/2$.
    Using~\eqref{eq:barnes-sinc},
    it is possible to
    evaluate
    Eq.~\eqref{eq:patch-correlation}
    for a prescribed functional form
    of $\dimlessP(k)$.
    The large $r$
    behaviour can
    be extracted using standard asymptotic
    methods for Mellin--Barnes integrals,
    which have been widely deployed in the
    analysis of Feynman
    diagrams~\cite{paris2001asymptotics,Friot:2005cu,Dubovyk:2022obc}.
    However, for blue spectra with suitable growth to represent
    a spike, the conclusion agrees with the simpler analysis
    presented here.
    A similar procedure can be used to estimate gradient corrections,
    discussed below,
    to the approximation of independent,
    identically distributed small-scale realizations $S_i$.}

\para{Smoothing over substructure}
Now suppose that
at some later stage, fluctuations associated with
a much smaller physical scale
$\smallBox \sim 2\pi/\kSmall$ exit the horizon.
To describe these fluctuations we
subdivide each $L$-patch into smoothed regions
of characteristic size
$\smallBox \ll \bigBox$, and suppose that each
region is populated by one or more
perturbations
$\SSmall$ that contribute additively
to $\zeta$.%
    \footnote{By `additive', we mean that
    when $\SSmall \rightarrow \SSmall + \Delta$,
    we also have $\zetaSmall \rightarrow \zetaSmall + \Delta$.}
We will sometimes describe these regions as `boxes'.
Each
$\SSmall$ is a random variable, but we are not yet being
specific about its distribution.
We apply this analysis to explicit examples
below.%
    \footnote{In~\S\ref{sec: SR to USR to SR} we introduce a
    3-phase ultra-slow-roll model, and evaluate its loop corrections
    in~\S\ref{sec:delta N loops calculation}.
    For this example, $S_\smallBox$ might be the quadratic
    operator $\delta \phi^2$.
    However, we emphasize
    that the discussion presented here is
    general, and applies beyond this scenario.}

\begin{figure}
    \begin{center}
        \includegraphics[scale=0.23]{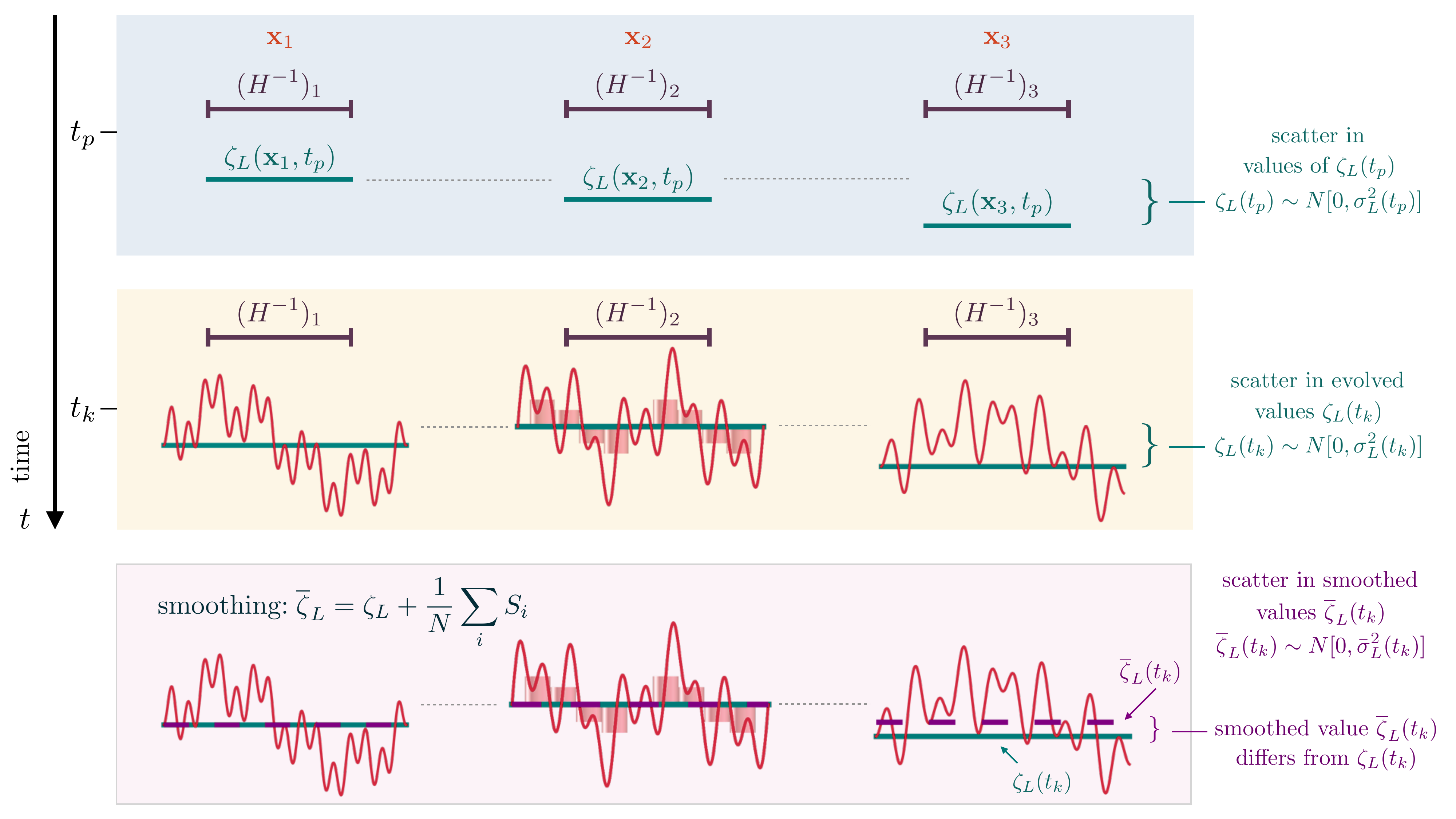}
    \end{center}
    \caption{\label{fig:box-evolve}Evolution of superhorizon regions.
    \semibold{Top row}: Three regions of size $L \sim 2\pi/p$
    just after horizon exit at time $t_p$. At this time, each
    region has no coherent substructure. The average value
    of $\zeta$ in each patch is $\zeta_L(\vect{x}_i, t_p)$.
    The distribution of $\zeta_L$ has variance $\sigma_L^2(t_p)$.
    \semibold{Middle row}: At a much later time, the scale
    $\ell \sim 2\pi/k$ exits the horizon.
    At this time, the value of
    $\zeta_L(\vect{x}_i, t_k)$ may have evolved due to isocurvature
    effects between $t_p$ and $t_k$, yielding a modified
    variance $\sigma_L^2(t_k)$. Long-lived coherent
    substructure associated
    with the scale $\ell$ is now present.
    \semibold{Bottom row}: In many regions (here $\vect{x}_1$, $\vect{x}_2$),
    averaging over the substructure
    does not appreciably alter the long-wavelength field.
    But in some patches (here $\vect{x}_3)$ the substructure does not
    average to zero. The spatial average $\overline{\zeta}_L(t_k)$
    therefore differs from $\zeta_L(t_k)$ and may have
    differing variance $\bar{\sigma}^2_L(t_k)$.}
\end{figure}

The net effect is to overlay each
$\bigBox$-patch
with a mosaic of many $\smallBox$-patches;
see Fig.~\ref{fig:box-evolve}.
In many realizations the mosaic will contain
roughly equal positive and negative fluctuations,
and its spatial average will be nearly zero.%
    \footnote{Notice that this discussion applies
    regardless of the statistical distribution of the random
    variable $S_\smallBox$. Indeed, if $S_\smallBox$ has a
    non-zero average, $\mu$, this will be independent
    of the $\bigBox$-patch and therefore can be subtracted
    by redefining $\zetaBig \rightarrow \zetaBig - \mu$.
    On the other hand, the contribution to
    $\sigmaBigSmooth^2$
    due to correlations
    with the long mode $\zetaBig$ cannot be
    subtracted, as $\langle \zetaBig S_\smallBox\rangle$
    depends on the value of $\zetaBig$,
    which changes across different $\bigBox$-patches.}

For these realizations the long wavelength field
$\zetaBig$ will be almost unaffected.
However, for some realizations the spatial
average over the moasic will not be zero.
This may happen by chance,
but is more likely where non-Gaussianity systematically
skews the distribution to produce more
positive or negative regions.
In this case, the spatial average over the mosaic
may significantly shift $\zetaBig$.
This is the back-reaction effect we wish to analyse.

Now focus on a single $\bigBox$-patch.
It contains roughly
$N = (\bigBox/\smallBox)^3$
independent small boxes in the overlaid mosaic.
We label the $\SSmall$ perturbation in each small box
by $S_i$, where $i \in \{ 1, 2, \ldots, N \}$.
The autocorrelation of the $S_i$
is the variance
$\langle S_i S_i \rangle = \sigmaSmall^2$,
whereas the cross-correlation between different boxes
is $\langle S_i S_j \rangle$.
If the spectrum used to generate the
perturbation within each small box
is sufficiently steep,
the discussion above shows that we can take the $S_i$
to be approximately independent, identically distributed (`IID')
random variables.
The leading corrections to this picture will be
gradient suppressed of order $\sim (\ell/r)^{2}$.
Here, $r$ is the displacement appearing in Eq.~\eqref{eq:patch-correlation}.

Each $S_i$ contributes additively to the curvature perturbation,
so the corrected
long-wavelength field in the $\bigBox$-patch,
denoted $\zetaBigSmooth$,
can be written
\begin{equation}
    \label{eq:corrected-zetaL-field}
    \zetaBig(\vect{x}) = \zetaBig +
    \sum_i S_i \IndicatorFunction_i(\vect{x}) .
\end{equation}
where $\IndicatorFunction_i(\vect{x})$
is an indicator function that is unity when
$\vect{x}$ is within the $i^{\mathrm{th}}$
box and zero otherwise.
The field $\zetaBig(\vect{x})$ traces the original
background field $\zetaBig$ (which is constant over the $\bigBox$-patch),
but is dressed by the small-scale fluctuations.
Its
volume-weighted
average over the $\bigBox$-patch
can be written
\begin{equation}
    \label{eq:corrected-average-zetaL-field}
    \zetaBigSmooth = \zetaBig + \frac{1}{N} \sum_i S_i .
\end{equation}
Eq.~\eqref{eq:corrected-average-zetaL-field}
is our desired formula for the back-reaction of
small-scale fluctuations on the long-wavelength mode,
with
the factor $1/N$ accounting for the
volume average over the mosaic.
The realization of
both the long-wavelength field $\zetaBig$
and the short-wavelength fluctuations $S_i$
vary between $\bigBox$-sized regions.
They can therefore be regarded as stochastic variables.
Meanwhile,
the effect of the short-wavelength fluctuations
can be regarded as a sort of shot noise:
they arise from a discreteness
effect, because we can fit only
$\sim N$ independent samples into each
$\bigBox$-patch. 
They formally disappear
in the limit $N \rightarrow \infty$,
corresponding to an infinite hierarchy of scale $\bigBox/\smallBox$.
A similar discussion was
already given by Riotto~\cite{Riotto:2023hoz}.

We now ask how the short-scale fluctuations contribute to
the variance $\langle \zetaBigSmooth \; \zetaBigSmooth \rangle \approx \sigmaBigSmooth^2$.
The simplest situation
occurs when the short-scale fluctuations do not
correlate with the long-wavelength mode $\zetaBig$.
In this case only correlations
among the $S_i$ are present,
generating a contribution
\begin{equation}
    \label{eq:Si-correlation}
    \sigmaBigSmooth^2
    \supseteq
    \frac{1}{N^2} \left(
        \sum_i \sum_j \delta_{ij} \langle S_i S_j \rangle
        + \sum_i \sum_{j\neq i} \langle S_i S_j \rangle
    \right)
    = \frac{1}{N} \sigmaSmall^2
    + \frac{1}{N^2} \sum_i \sum_{j\neq i} \langle S_i S_j \rangle .
\end{equation}
The symbol `$\supseteq$' is used to indicate that
$\langle \zetaBigSmooth \; \zetaBigSmooth \rangle$
contains this contribution among others.
For independent fluctuations,
Eq.~\eqref{eq:Si-correlation}
reduces to
\begin{equation}
    \label{eq:Si-auto-correlation}
    \sigmaBigSmooth^2 \supseteq
    \frac{1}{N} \sigmaSmall^2
    \sim \frac{\smallBox^3}{\bigBox^3} \sigmaSmall^2 .
\end{equation}
The physical picture is that, although short-scale power
is distributed evenly throughout the entire
$\bigBox$-patch,
the realization in each $\smallBox$-patch is an independent
sample.
Therefore,
the contributions from distinct $\smallBox$-patches
cannot add up coherently.
Although there are $N$ small patches, the absence of correlation
means that their cumulative effect is insufficient to overcome
the $1/N$ suppression from spatial averaging
over the $\bigBox$-patch. This is simply the
usual argument that leads to the central limit theorem.
Riotto had earlier made the same observation,
that this scaling
should be associated with
incoherent addition of small volumes
populated by a Poisson process~\cite{Riotto:2023hoz}.

\para{Corrections to the IID scenario}
If the fluctuations in each $\smallBox$-patch are
not independent then we should expect
this conclusion to be modified.
We have seen that this will happen if the spectrum
populating the $\smallBox$-patches is not
sufficiently steep,
which would allow long-range cross-correlation
between patches.
In this case the second contribution in
Eq.~\eqref{eq:Si-correlation} is not zero. 

Alternatively, the $\smallBox$-patches could correlate with the
long-wavelength $\zetaBig$ mode.
We then find
\begin{equation}
    \label{eq:Si-cross-correlation}
    \sigmaBigSmooth^2
    \supseteq
    \bigg\langle
        \zetaBig
        \Big( \frac{1}{N} \sum_i S_i \Big)
    \bigg\rangle
    =\langle \zetaBig \SSmall \rangle \equiv
    \sigmaSmall \sigmaBig R(\SSmall, \zetaBig) ,
\end{equation}
where we have assumed that all $S_i$ correlate
with $\zetaBig$ in the same way.
In the last step we have introduced
the correlation coefficient
$R(\SSmall, \zetaBig)$
between $\SSmall$ and $\zetaBig$.
Notice there is no longer a volume suppression factor $1/N$.
This happens
because---%
although each $\smallBox$-patch is still an independent
sample of the short-scale noise---%
each patch now correlates with
the \emph{same} long-wavelength mode $\zetaBig$ and therefore can
add up coherently over the $\bigBox$-patch.
Whether this effect is more or less important than the
$S$ autocorrelation~\eqref{eq:Si-auto-correlation}
depends on a balance between the amplitudes $\sigmaSmall$,
$\sigmaBig$,
the scale hierarchy $\smallBox/\bigBox$ and the
long--short coupling $R(\SSmall, \zetaBig)$.
In some cases, we will see that $R$ can be
interpreted as the reduced bispectrum $\fNL$
evaluated on a squeezed momentum configuration with sides
$\sim 1/\ell$, $1/L$~\cite{Riotto:2023hoz,Riotto:2023gpm,Firouzjahi:2023bkt}.

Note that the form of~\eqref{eq:Si-cross-correlation}
does not require that back-reaction
persists even in the limit of an infinite scale hierarchy
$\smallBox/\bigBox \rightarrow 0$, which would be difficult
to interpret physically.
In practice the
correlation coefficient $R$ will typically become
negligible
(at least formally)
in this limit.
In cases
where $R$ can be associated with the reduced bispectrum,
this corresponds to $\fNL$ approaching zero in the ultra-squeezed limit.

\subsection{Outline and summary}
In this paper our primary aim is to
develop a more sophisticated version
of the toy model described in~\S\ref{sec:toy-model},
suitable for the accurate evauation of
cosmological correlation functions.
Our main tool is the separate universe framework.
We carefully demonstrate how this should be deployed
to capture back-reaction from small-scale
substructure.
Along the way, we introduce simplified
models that help clarify the meaning
of each loop,
and the physical processes that contribute
to it.

In~\S\ref{sec:separate-universe-deltaN}
we briefly introduce and review the
separate universe framework.
This is a very general tool.
It
is most familiar
when applied to the curvature perturbation
$\zeta$,
where it leads to the
`$\delta N$' formula.
This is widely used
to evaluate inflationary correlation
functions
on superhorizon scales.
In~\S\ref{sec:deltaN-review}
we summarize the main ideas.
In~\S\ref{sec:delta-N-back-reaction}
we carefully explain how to set up the
separate universe framework
to be used for calculations of back-reaction.
This does not require the introduction
of new elements,
but the necessary scales and times
must be carefully distinguished.
We relate the $\delta N$
formula to the toy model of~\S\ref{sec:intro},
which provides an extremely useful framework in which to build
geometrical intuition.
In particular, the key features
of volume-suppressed auto-correlation,
as in Eq.~\eqref{eq:Si-auto-correlation},
and unsuppressed cross-correlation,
as in Eq.~\eqref{eq:Si-cross-correlation},
emerge naturally
from the more detailed $\delta N$ picture.

In~\S\S\ref{sec:back-reaction-particle-creation}--\ref{sec:back-reaction-quasiparticles}
we discuss the different physical processes
that drive back-reaction in the separate universe framework.
In~\S\ref{sec:back-reaction-particle-creation}
we discuss the effect of incoherent particle creation
at the horizon scale.
One approach would be to account for this
using
a stochastic framework, such as
the original Starobinsky model~\cite{Starobinsky:1986fx}.
In such frameworks
the aim is usually to compute the one-point
probability distribution of the
fields smoothed on the horizon scale.
(For a more general discussion, see Ref.~\cite{Tada:2021zzj}.)
In our application
we require the fields smoothed on
much larger scales, so the stochastic formalism
cannot be applied directly.
In~\S\ref{sec:back-reaction-quasiparticles}
we discuss a different effect,
caused by a cascade of power towards the infrared.
This is driven by nonlinear interactions among
the different particle species that contribute to $\zeta$.
To interpret the contributing processes,
and relate them to S-matrix elements
describing spacetime scattering processes,
we draw an analogy with
a system consisting of a Bose--Einstein
condensate interacting with a cloud of quasiparticles.

Up to this point our argument is very general.
In~\S\ref{sec: SR to USR to SR}
we introduce a 3-phase ultra-slow-roll model,
which is used in~\S\ref{sec:delta N loops calculation}
to illustrate our framework by explicit calculations.
The 3-phase model is very similar to
models previously introduced by
Cai {\etal}~\cite{Cai:2018dkf}
and (especially) Firouzjahi \& Riotto~\cite{Firouzjahi:2023ahg}.
Readers familiar with the discussion given by
Firouzjahi \& Riotto will find
no new material in this section and may wish to skip
directly to~\S\ref{sec:delta N loops calculation}.
However,
we discuss the model carefully
because (as explained above)
there are differences of detail in our
application of the framework.
Although our model is the same,
our final expressions for
the $\delta N$ coefficients
needed 
in the back-reaction calculation differ from those reported
in Ref.~\cite{Firouzjahi:2023ahg}.

In~\S\ref{sec:delta N loops calculation} we evaluate
the 12-, 22- and 13-type loops in
the 3-phase ultra-slow-roll model.
We report values for both instantaneous and smooth transitions
between the ultra-slow-roll phase and the second
slow-roll phase,
and emphasize the role of different types of nonlinear coupling.
We also emphasize that
loop contributions that are \emph{not} volume-suppressed
generally depend on soft limits of
correlation functions.
These limits can frequently
be evaluated by another
application of the separate universe framework,
but they may have gauge ambiguities,
and
suitable subtractions may be needed
to isolate the \emph{physical} correlations.
In this paper we do not pursue these
subtractions in detail.
Taken together,
our detailed outcomes differ
from previously published results.
We find that---even if taken at face value, without subtractions---%
the 13-type loop vanishes,
and the 12- and 22-type loops
are too small
to invalidate the tree-level prediction.
This applies
no matter how the we choose to
transition between the different slow-roll
and ultra-slow-roll phases.
In comparison with
the results of Firouzjahi \& Riotto~\cite{Firouzjahi:2023ahg},
these differences can be attributed partly to
differences in our
application of the separate universe framework.

In~\S\ref{sec:wilsonian-counterterms}
we consider the effect of changing the separate universe
smoothing scale.
We argue that this functions exactly like a Wilsonian cutoff.
As the cutoff changes, quanta are included or
excluded from the loop, depending whether the cutoff is increased
or decreased, respectively.
This procedure is simply a reorganization of the calculation,
and can not change the outcome for any correlation function.
To absorb changes from the cutoff-dependence of the loop
integrals will require counterterms, as in any application
of effective field theory methods.
In this section we show explicitly that the 12- and 13-type
loops can be absorbed into a renormalization of
local operators in the $\zeta$ Lagrangian.
We argue that the necessity to introduce counterterms
means that contributions to the loops
of~\S\ref{sec:delta N loops calculation}
that are analytic in the CMB-scale wavenumber
must be combined with unknown ultraviolet-sensitive
terms before they are be used
to evaluate observables.
We comment briefly on the prospects for finding measures
of the loop contribution that are not sensitive
to the unknown ultraviolet completion.

In~\S\ref{sec:discussion}
we conclude with an extended discussion.
Appendix~\ref{sec: numerical example}
collects a number of numerical results
for the ultra-slow-roll model of~\S\ref{sec: SR to USR to SR}.

\para{Conventions}
We work in natural units where $c = \hbar = 1$.
The reduced Planck mass is defined by $\Mp = (8\pi G)^{-1/2}$,
where $G$ is Newton's gravitational constant.
Numerically, $\Mp = 2.435 \times 10^{18} \, \GeV$.
Latin indices $I$, $J$, \ldots, label the coordinates
needed to cover the phase-space manifold
of fields. In the $N$-field model, they run from $1$
up to $2N$.
For the single-field models
considered in detail below, they range
over a field fluctuation $\delta\phi$ and a momentum fluctuation
$\delta\pi$.
These fluctuations are defined in Section~\ref{sec:deltaN-review}.

Our Fourier transform convention is
\begin{equation*}
    f(\vect{k}) = \int \d^3 x \; f(\vect{x})
        \e{-\im \vect{k}\cdot\vect{x}}
    \quad
    \text{and}
    \quad
    f(\vect{x}) = \int \frac{\d^3 k}{(2\pi)^3} \; f(\vect{k})
        \e{\im \vect{k}\cdot\vect{x}} ,
\end{equation*}
where $f(\vect{x})$ and $f(\vect{k})$ represent an arbitrary
function and its Fourier transform, respectively.
We sometimes use the shorter notation
$f_{\vect{k}} = f(\vect{k})$,
or $[ \OpO ]_{\vect{k}}$
for the Fourier mode of a composite quantity $\OpO$.

We use the notation $\langle \cdots \rangle'$ to indicate
a correlation function with removal of the momentum-conservation
$\delta$-function and the accompany factor of $(2\pi)^3$.

\section[\texorpdfstring{The separate universe framework and the $\bm{\delta N}$ formula}{The separate universe framework and the delta N formula}]{The separate universe framework and the $\bm{\delta N}$ formula}
\label{sec:separate-universe-deltaN}

The separate universe framework is a tool
used
to compute
the nonlinear evolution of perturbations
on superhorizon scales.
Let $\vect{p}$ be a comoving wavenumber
associated with some
perturbation whose evolution we wish to compute.
Outside the horizon, if
$p = |\vect{p}|$ and $H$ are the only relevant
scales, spatial gradients of this Fourier mode
will typically be suppressed by at least
$(p/a)/H = p/(aH) \ll 1$
in any local model.%
    \footnote{We use the term `local' to mean
    that the model remains finite when all
    wavenumbers go to zero together.
    Although suppression of gradients
    by the horizon scale $1/H$
    is usually assumed in presentations of
    the separate universe technique,
    this situation can be
    different
    in models with more than one
    relevant scale~\cite{Jackson:2023obv}.
    In our scenario, this manifests as gradient
    corrections that are suppressed by $\sim \smallBox/\bigBox$.
    In~\S\ref{sec:wilsonian-counterterms}
    and~\S\ref{sec:discussion-wilsonian-eft}
    we show that this can be usefully interpreted
    in the context of an effective field theory description.}
Where this estimate is valid,
a smoothed patch of the universe of size $L \sim 1/p$
will evolve like a ``separate'' unperturbed
universe
up to corrections of order $p^2/(aH)^2$.

\subsection[\texorpdfstring{Review: the ${\delta N}$ formula in phase space}{Review: the delta N formula in phase space}]{Review: the $\bm {\delta N}$ formula in phase space}
\label{sec:deltaN-review}
\para{Trajectory description}
Dropping gradients,
each smoothed patch evolves along an available
trajectory in the \emph{background} phase space $M$.
This trajectory is
selected (at least in the absence
of back-reaction) by its initial conditions,
which identify a particular point in $M$.
To be concrete,
let us suppose that the background model
is described by some number of scalar fields,
although
our discussion is more general than
this scenario.
We assume the corresponding phase space
to be labelled by
the value and conjugate momentum
for each field,
which can be
collected into a set of coordinates $X^I$.

To build statistical quantities we require an ensemble
of smoothed patches. The time evolution of this ensemble can
be expressed in terms of
the trajectories
followed by its members.
Consider two regions whose initial conditions
(specified at some early time $t_\ast$)
differ by an amount $\delta X^I(t_\ast)$.
We should regard $\delta X^I(t)$, for $t > t_\ast$,
as a connecting vector linking the
distinct
trajectories followed by these regions.
At some later time $t$
this connecting vector
can be written
\begin{multline}
    \label{eq:basic-separate-universe}
    \smoothBig{\delta X^I(t, \vect{x})} =
    \frac{\delta X^I(t)}{\delta X^J(t_\ast)}
    \smoothBig{\delta X^J(t_\ast, \vect{x})}
    \\
    +
    \frac{1}{2}
    \frac{\delta^2 X^I(t)}{\delta X^J(t_\ast) \delta X^K(t_\ast)}
    \smoothBig{\delta X^J(t_\ast, \vect{x})}
    \smoothBig{\delta X^K(t_\ast, \vect{x})}
    + \cdots .
\end{multline}
The brackets $\smoothBig{\cdots}$ represent
smoothing by spatial averaging over the scale $L$,
and
we have added a label $\vect{x}$
to distinguish the regions
linked by $\delta X^I(t)$.%
    \footnote{The coordinate label $\vect{x}$
    identifies only one of these regions.
    The other is taken to be a fixed fiducial region.
    The choice of fiducial region drops out of connected
    correlation functions.}
In the quadratic term---%
and its higher-order counterparts, not written here---%
it follows from the discussion of initial conditions
above that
each factor $\delta X^I$ is smoothed
\emph{before} forming the product.
As a point of principle
the operations of multiplication and smoothing do not commute. 

The variational derivatives
$\delta X^I(t)/\delta X^J(t_\ast)$,
$\delta^2 X^I(t) / \delta X^J(t_\ast) \delta X^K(t_\ast)$,
and their higher-order analogues,
measure changes in the background trajectory
$X^I(t)$ due to changes in its initial conditions.
They are the essential building blocks of the
separate universe method.
Each derivative
can be computed from explicit knowledge of
the trajectories $X^I = X^I[t, X^J(t_\ast)]$
as functions of their initial
conditions~\cite{Starobinsky:1985ibc,Lyth:2005fi},
or by integrating
a Jacobi-like equation
along the phase-space flow~\cite{Seery:2012vj}.

For typical models in which isocurvature modes
decay by the time initial conditions are
set for the CMB anisotropy,
the primary quantity of interest is the curvature
perturbation $\zeta$.
This
represents a locally defined scale factor 
$a(t, \vect{x}) = a_0(t) \e{\zeta(t, \vect{x})}$
for each smoothed region.
It can be shown that $\zeta(t,\vect{x})$
satisfies a separate-universe equation analogous
to~\eqref{eq:basic-separate-universe}, 
\begin{multline}
    \label{eq:zeta-separate-universe}
    \smoothBig{\zeta (t, \vect{x})}
    = \frac{\delta N(t)}{\delta X^J(t_\ast)}
    \smoothBig{\delta X^J(t_\ast, \vect{x})}
    \\
    + \frac{1}{2}
    \frac{\delta^2 N(t)}{\delta X^J(t_\ast) \delta X^K(t_\ast)}
    \smoothBig{\delta X^J(t_\ast, \vect{x})}
    \smoothBig{\delta X^K(t_\ast, \vect{x})}
    + \cdots ,
\end{multline}
where $N(t)$ represents the number of e-folds elapsed
along a given trajectory between a spatially flat hypersurface
at time $t_\ast$
with initial conditions set by $\delta X^I(t_\ast)$,
and a uniform-density hypersurface at time $t$.
As before, $\smoothBig{\cdots}$ indicates
smoothing on the scale $L$.
For details, see Refs.~\cite{Dias:2014msa,Dias:2015rca}.
Eq.~\eqref{eq:zeta-separate-universe} is
the `$\delta N$ formula', introduced
by Starobinsky~\cite{Starobinsky:1985ibc}
and extended nonlinearly by Lyth \& Rodr\'{i}guez~\cite{Lyth:2005fi}.

In practical calculations we typically
wish to work in Fourier space
and extract a superhorizon mode $\vect{p}$
from~\eqref{eq:basic-separate-universe}
or~\eqref{eq:zeta-separate-universe}.
For this purpose we can (almost) neglect
the window function $W$,
which
is nearly unity
if $1/p$ is at least a little larger than the
smoothing scale. 
It has negligible effect except to specify
how the convolution integrals implied by
the quadratic term in~\eqref{eq:zeta-separate-universe},
and higher-order terms, should be handled.
Accordingly, when working in Fourier space,
we will often drop an explicit smoothing label
on the field.
Notwithstanding this notational convenience,
however,
we will see that
in a back-reaction calculation---%
which involves averaging over
small-scale structure---%
the cutoff on each convolution integral
must be treated with due care.

\para{Low-order correlation functions of $\zeta$}
In the later discussion we will need
explicit forms for some $\zeta$ correlation functions.
The main quantities of interest
are the 2-, 3- and 4-point functions,
\begin{subequations}
\begin{align}
    \label{Power spectrum def}
    \langle \zeta_{\vect{k}_1} \zeta_{\vect{k}_2} \rangle
        & \equiv  (2\pi)^3 \delta (\vect{k}_1 + \vect{k}_2) P_\zeta(k_1) , \\
    \label{bispectrum def}
    \langle\zeta_{\vect{k}_1} \zeta_{\vect{k}_2} \zeta_{\vect{k}_3} \rangle 
        & \equiv (2\pi)^3 \, \delta(\vect{k}_1+\vect{k}_2+\vect{k}_3)
        \, B_\zeta(k_1, k_2, k_3) , \\
    \label{trispectrum def}
    \langle\zeta_{\vect{k}_1} \zeta_{\vect{k}_2} \zeta_{\vect{k}_3} \zeta_{\vect{k}_4} \rangle
        & \equiv (2\pi)^3 \, \delta(\vect{k}_1+\vect{k}_2+\vect{k}_3+\vect{k}_4)
        \, T_\zeta(\vect{k}_1, \vect{k}_2, \vect{k}_3, \vect{k}_4) .
\end{align}
\end{subequations}
The functions $P_\zeta$, $B_\zeta$ and $T_\zeta$ are the power spectrum,
bispectrum, and trispectrum, respectively.
In order to deal with dimensionless amplitudes it is
often preferable to work with the
dimensionless power spectrum $\dimlessP_\zeta(k)$
and the reduced bispectrum $\fNL(k_1, k_2, k_3)$,
which are defined to satisfy
\begin{subequations}
\begin{align}
    \label{dimensionless Pz def}
    P_\zeta(k)
        & \equiv \frac{2\pi^2}{k^3} \dimlessP(k) , \\
    \label{f NL def}
    B_\zeta(k_1, k_2, k_3)
        & \equiv \frac{6}{5} \fNL(k_1, k_2, k_3)
        \Big[ P_\zeta(k_1) P_\zeta(k_2) + \text{2 perms} \Big] .
\end{align}
\end{subequations}
Note that $\fNL$ defined in this way depends on the momentum
configuration $\{ k_1, k_2, k_3 \}$, and therefore can vary
as a function of both its
scale (measured by the perimeter $k_t = k_1 + k_2 + k_3$)
and shape (measured by the ratios
$k_i/k_t$).

There is no simple parametrization for the amplitude of the
trispectrum, but it is conventional to recognize
two distinct `shapes'
with amplitude $\tauNL$ and $\gNL$,
\begin{multline}
    \label{t NL and g NL def}
    T_\zeta(\vect{k}_1, \vect{k}_2, \vect{k}_3, \vect{k}_4)
    \equiv
    \tauNL \Big[
            P_\zeta(k_{13}) P_\zeta(k_3)  P_\zeta(k_{4}) +\text{11 perms}
    \Big] \\
    + \frac{54}{25} \gNL \Big[
        P_\zeta(k_{2}) P_\zeta(k_3)  P_\zeta(k_{4}) +\text{3 perms}
    \Big] ,
\end{multline}
where $\vect{k}_{ij} = \vect{k}_i + \vect{k}_j$ and $k_{ij} = |\vect{k}_{ij}|$.
Eq.~\eqref{t NL and g NL def}
is exact within the local model, in which the nonlinear $\zeta$ is
\emph{defined}
to satisfy
\begin{equation}
    \label{eq:local-model}
    \zeta(t, \vect{x})
    = \zetaGaussian(t, \vect{x})
    + \frac{3}{5} \fNLloc \Big( \zetaGaussian(t, \vect{x})^2 - \langle \zetaGaussian(t, \vect{x})^2 \rangle \Big)
    + \frac{9}{25} \gNLloc \zetaGaussian(t, \vect{x})^3 ,
\end{equation}
where $\tauNL = (6\fNLloc/5)^2$ and $\gNLloc$
are constant,
and $\zetaGaussian$ is a Gaussian random field.
In a realistic model $\tauNL$ and $\gNL$
will develop scale- and shape-dependence, but it is then difficult to
separate them unambiguously.

\subsection[\texorpdfstring{The ${\delta N}$ formula with back-reaction}{The delta N formula with back-reaction}]{The $\bm {\delta N}$ formula with back-reaction}
\label{sec:delta-N-back-reaction}

Separate universe expressions such
as Eqs.~\eqref{eq:basic-separate-universe} and~\eqref{eq:zeta-separate-universe}
invoke a Taylor expansion in the initial conditions
for each patch.
This yields a correct result
only when the time evolution
can be predicted uniquely
from knowledge of
the initial conditions $\delta X^I(t_\ast)$.
If back-reaction is sufficiently strong it will spoil
the critical property of uniquely predictable evolution.

In this situation the picture becomes
more complicated.
At one extreme, the trajectories followed by
smoothed regions within the ensemble may be disturbed continuously.
For example,
this is the case if we repeatedly adjust the size of the smoothed
patches to follow the horizon scale.
In any time interval,
the value of $\zeta$ within each (adjusted)
patch is coherently disturbed
due to inflationary particle creation
at the horizon.
Eqs.~\eqref{eq:basic-separate-universe}
and~\eqref{eq:zeta-separate-universe}
are then invalidated,
with the outcome that
$\delta X^I(t)$ and $N(t)$ should be replaced
by a statistical distribution representing the range of 
phase space coordinates accessible from a given starting position
$X^I(t_\ast)$.
This approach leads to the Starobinsky formulation of stochastic
inflation~\cite{Starobinsky:1985ibc},
and more recent variants including the `stochastic $\delta N$' method (see, e.g. Ref.~\cite{Vennin:2015hra}).
Our analysis will not apply to this regime.

At the other extreme, each patch may be disturbed just once
during its evolution.
This is the scenario needed to study back-reaction from
a relatively narrow spike over a well-defined range
of wavenumbers,
and
is the main focus of the analysis in this paper.
It is somewhat different to the conventional stochastic
formalism.
First,
the intention is to follow the evolution of an ensemble
of patches of \emph{fixed} scale as they are inflated far beyond
the horizon.
Second,
when the back-reaction event occurs, it involves the eruption
of substructure in many comparatively small horizon-scale
patches, which add incoherently as explained
in~\S\ref{sec:intro}.
For both these reasons we cannot immediately
compute the back-reaction
using a Starobinsky-type approach.

\para{A $\delta N$ formula for back-reaction}
Instead, we now aim to use
Eq.~\eqref{eq:zeta-separate-universe}
to build an analogue of the toy
model~\eqref{eq:corrected-zetaL-field}
and its smoothed counterpart~\eqref{eq:corrected-average-zetaL-field}.
To fix ideas,
we let $\vect{p}$ label a soft mode
with $p = |\vect{p}| \sim \bigBox^{-1}$.
In our later applications this will represent
a typical CMB scale.
As explained in~\S\ref{sec:intro}
we are primarily interested in models
where the short-scale
modes $\sim \smallBox^{-1}$
receive an enhanced amplitude.
We use
$\vect{k}$
to label typical Fourier modes in this enhanced region.
There will now be three types of gradient correction to the
separate universe formula---those suppressed by
$p/(aH)$, $k/(aH)$, and those suppressed by $\smallBox/\bigBox \sim p/k$. 

Allow the ensemble of
large $\bigBox$-sized patches to exit the horizon.
Up to the point where the smaller
$\smallBox$-sized boxes emerge
from the horizon,
the field value distribution within each patch
can be evolved using any convenient method, including the
$\delta N$ formula~\eqref{eq:zeta-separate-universe}.
However,
once the $\smallBox$-boxes have emerged,
we re-apply~\eqref{eq:zeta-separate-universe}
\emph{with smoothing scale set by $\smallBox$}.
The result is $\smoothSmall{\zeta(\vect{x})}$.
It represents the full spatial dependence of the
$\zeta$ field, smoothed on the scale of the
$\smallBox$-sized patches,
and is
the exact
analogue of the
field $\zetaBig(\vect{x})$
in the
toy model~\eqref{eq:corrected-zetaL-field}.
Specifically,
$\smoothSmall{\zeta(\vect{x})}$
traces the distribution of
field values over the large $\bigBox$-patches,
but is dressed by small $\smallBox$-scale fluctuations.
In Fourier space,
this dressing includes
contributions from enhanced peak-scale
modes $k \sim \smallBox^{-1}$.

The necessity to re-apply the formula~\eqref{eq:zeta-separate-universe}
after the $\smallBox$-scale boxes emerge
is a restatement of the well-known
requirement that, to apply separate-universe arguments, 
one must wait until
\emph{all} relevant scales have left the horizon.
As explained in~\S\ref{sec:intro} we must also wait until
the subsequent time evolution is classical.
The necessary conditions in this case are $p/(aH) \ll 1$
and $k/(aH) \ll 1$.
The natural time
at which
to base the $\delta N$ calculation is therefore
(a few e-folds after)
the horizon-crossing time of the peak scales, $t_k$.

In~\S\ref{sec:delta N loops calculation} we
specialize the general treatment developed
here to a 3-phase model with ultra-slow-roll,
to be introduced in~\S\ref{sec: SR to USR to SR}. 
This is the same scenario described by
Firouzjahi \& Riotto in Ref.~\cite{Firouzjahi:2023ahg}.
However, in their calculation, the $\delta N$
formula was applied from the horizon exit time $t_p$ 
for the large-scale mode $\vect{p}$.
We argue that this approach does not correctly account
for the influence of the short-scale modes,
which are still deep inside the horizon at $t_p$.%
    \footnote{In principle, there is another
    reason why it is necessary to restart
    the calculation at $t_k$.
    After the short-scale boxes exit the horizon,
    it is possible for their aggregate effect to
    displace larger smoothed regions
    onto a new trajectory in the background phase
    space.
    See Eq.~\eqref{eq:fourier-p-back-reaction}.
    Therefore, if the $\delta N$ calculation is based at $t_p$,
    to account for this effect there must be a step 
    in which we re-evaluate the initial
    condition for each smoothed region
    in light of our knowledge of the newly-emerged
    short-scale modes.
    This is exactly how stochastic
    approaches work, and is needed here for the same
    reason.
    We estimate this effect
    in~\S\ref{sec:back-reaction-particle-creation}
    and find it to be negligible when there is
    appreciable separation of scales,
    because the small-scale boxes add incoherently.
    There is a possible coherent effect that
    may be less
    suppressed by the scale hierarchy,
    and which is not
    included in~\S\ref{sec:back-reaction-particle-creation}.
    This is model-dependent.
    It would be interesting to obtain a more accurate
    estimate of its possible importance.}
    
In the prescription given above,
we choose to smooth over the scale $\smallBox$
when the $\delta N$ calculation is restarted at $t_k$.
However, one might instead consider
smoothing immediately on the scale $\bigBox$.
In comparison with our prescription,
this choice
represents a different way to organize the calculation,
but must give a compatible outcome.
We comment on the relation
between these two computational schemes,
and their respective merits,
in~\S\ref{sec:wilsonian-counterterms} below.

Because we assume there is only one back-reaction event,
Eq.~\eqref{eq:zeta-separate-universe}
(with the replacement $\bigBox \rightarrow \smallBox$)
can be used to evaluate $\smoothSmall{\zeta(\vect{x})}$
at any late time of interest, $t$.
Once we have arrived at this time,
the next step conceptually
is to smooth on the
scale $\bigBox$ associated with the original large patches.
In practice, however, we simply extract a Fourier mode $\vect{p}$.
As explained above, this effectively entails a spatial average.
The resulting expression for $\zeta_{\vect{p}}$
is
\begin{multline}
    \zeta_{\vect{p}}(t)
    = \frac{\delta N(t)}{\delta X^J(t_\ast)}
        \Big[
            \smoothSmall{\delta X^J(\vect{x}, t_\ast)}
        \Big]_{\vect{p}}
    \\
    + \frac{1}{2}
    \frac{\delta^2 N(t)}{\delta X^J(t_\ast) \delta X^K(t_\ast)}
    \int\limits_{q \lesssim \smallBox^{-1}}
    \frac{\d^3 q}{(2\pi)^3}
    \Big[
        \smoothSmall{\delta X^J(\vect{x}, t_\ast)}
    \Big]_{\vect{p} - \vect{q}}
    \Big[
        \smoothSmall{\delta X^K(\vect{x}, t_\ast)}
    \Big]_{\vect{q}}
    + \cdots ,
    \label{eq:zeta-separate-universe-fourier}
\end{multline}
where
`$\dots$' denotes terms
of cubic order and higher in
$\smoothSmall{\delta X}$ that we have not written
explicitly.

We could obtain
equivalent results by choosing any smoothing lengthscale
smaller than $\smallBox$.
Whichever time we pick,
the base time $t_\ast$ can be chosen at any time after horizon exit of the
smoothing scale.
The particular choice we are making here,
i.e. $t_\ast = t_k$,
is simply the most convenient
for our purpose.
There is no advantage in choosing a base time $t_\ast$ significantly
later than horizon exit of modes with wavelength
comparable to the smoothing scale, because
one must then account
for the time evolution of the $\smallBox$-size boxes between
their horizon exit
and $t_\ast$.
Likewise, there is no advantage in choosing a smoothing scale that is significantly
smaller then $\smallBox$, because one must then wait for it to exit the horizon.

As explained above, we have dropped the smoothing label
on $\zeta_{\vect{p}}$.
However, it appears on the right-hand side in two
locations.
First, we have written
$[ \smoothSmall{\delta X^J(\vect{x}, t_\ast)} ]_{\vect{p}}$
to denote the
$\vect{p}^{\mathrm{th}}$ Fourier mode
of the field $\smoothSmall{\delta X^J}$.
Since $p \ll \smallBox^{-1}\sim k$
it must be remembered that many $\smallBox$-size
patches will fit within a box of size $\bigBox \sim 1/p$.
Therefore one must include the scatter among these boxes
when performing the Fourier transform, as in
the toy model Eq.~\eqref{eq:corrected-average-zetaL-field}.
This effect
is discussed in~\S\ref{sec:back-reaction-particle-creation} below.
Second,
the smoothing scale appears in the cutoff $q \lesssim \smallBox^{-1}$
applied to the convolution integral.
Since $p \lesssim \bigBox^{-1} \ll \smallBox^{-1}$,
both Fourier modes in the convolution lie approximately
below the cutoff.
Back-reaction from the convolution is discussed
in~\S\ref{sec:back-reaction-quasiparticles}.

\subsubsection{Inflationary particle creation at the horizon}
\label{sec:back-reaction-particle-creation}

Our first task is to
estimate the Fourier transform
$[\smoothSmall{\delta X^J(\vect{x}, t_\ast)}]_{\vect{p}}$
when $p \sim \bigBox^{-1} \ll \smallBox^{-1}$.
This is needed in the linear term of
Eq.~\eqref{eq:zeta-separate-universe-fourier}.
In contrast,
for our scenario, the convolution integral is dominated
by Fourier modes close to the peak in short-scale
power.
We do not need to include back-reaction
from particle creation when evaluating
these convolution modes.
When included in a one loop correction to the power spectrum
its effect would be higher order in the peak amplitude than the contributions that we keep.

As we have explained,
once the $\smallBox$-size boxes have exited the horizon,
$[\smoothSmall{\delta X^J(\vect{x}, t_\ast)}]_{\vect{p}}$
differs from the background
Fourier mode
$\delta X^J_{\vect{p}}$
because of the stochastic $\smallBox$-scale modes.
The effect can be estimated using
Eq.~\eqref{eq:corrected-average-zetaL-field}.
After translation to
the present context, this would suggest
\begin{equation}
    \label{eq:fourier-p-back-reaction}
    [ \smoothSmall{\delta X^J(\vect{x}, t_\ast)} ]_{\vect{p}}
    \approx
    \delta X^J_{\vect{p}}(t_\ast)
    +
    \frac{1}{N} \sum_i S^J_i ,
\end{equation}
where $\delta X^J_{\vect{p}}(t_\ast)$ is the original
background Fourier mode.
To specify $S^J_i$, take $(\sigma^2)^{IJ}_{\smallBox}$
to be the variance
in $\delta X^I(t_\ast)$,
computed over the narrow band of wavenumbers
that support the peak,
\begin{equation}
    (\sigma^2)^{IJ}_{\smallBox}
    \approx
    \int\limits_{\substack{\text{peak} \\ \text{scales}}} \frac{\d q}{q} \;
    \dimlessP^{IJ}(q) .
\end{equation}
The (cross-)power spectrum $\dimlessP^{IJ}(q) = q^3 P^{IJ}(q) / (2\pi^2)$
is determined by the correlator
\begin{equation}
    \label{eq:short-mode-power-spectrum}
    \langle \delta X^I_{\vect{k}_1}( t_\ast) \delta X^J_{\vect{k}_2}( t_\ast) \rangle
    =
    (2\pi)^3 \delta(\vect{k}_1 + \vect{k}_2) P^{IJ}(k;t_\ast) ,
    \quad
    \text{where} \;
    |\vect{k}_1| = |\vect{k}_2| = k .
\end{equation}
Then
$S^J_i$ is a random variable
whose covariance should be chosen to satisfy
\begin{equation}
    \langle S^I_i S^J_j \rangle \approx (\sigma^2)^{IJ}_{\smallBox} \delta_{ij} .
\end{equation}
The correlation~\eqref{eq:short-mode-power-spectrum} is evaluated just after
horizon exit for the $\smallBox$-scale modes and can be estimated
using the usual methods of in--in quantum field theory.

The contribution of Eq.~\eqref{eq:fourier-p-back-reaction}
to the two-point correlation function of $\zeta_{\vect{p}}$ is 
\begin{equation}
    \label{eq:2pt-zeta-particle-creation}
    \langle
        \zeta_{\vect{p}} \zeta_{-\vect{p}}
    \rangle
    \supseteq
    \frac{\delta N(t)}{\delta X^I(t_\ast)}
    \frac{\delta N(t)}{\delta X^J(t_\ast)}
    \left[
        \langle \delta X^I_{\vect{p}}(t_\ast) \delta X^J_{-\vect{p}}(t_\ast) \rangle
        + \frac{1}{N} (\sigma^2)^{IJ}_{\smallBox}
        + \langle \delta X^I_{\vect{p}}(t_\ast) S_\ell^J  \rangle
    \right] \;,
\end{equation}
where the first term encodes the linear piece
(that is, without back-reaction).

In a na\"{\i}ve estimate where one
counts only powers
of the short-scale amplitude $(\sigma^2)^{IJ}_{\smallBox}$, the second term in Eq.~\eqref{eq:2pt-zeta-particle-creation} would be the \emph{leading}
contribution to the back-reaction.
It involves only a single power
of $(\sigma^2)^{IJ}_{\smallBox}$.
In practice, Eq.~\eqref{eq:fourier-p-back-reaction}
shows that the effect is suppressed by a central-limit-like
volume factor $N^{-1} \sim (\smallBox / \bigBox)^3$.
For applications to formation of primordial black holes,
$\bigBox$ is a CMB scale and $\smallBox$
might be a scale associated with formation of solar-mass
objects or smaller. This makes $N^{-1}$ exponentially small,
and the correction~\eqref{eq:fourier-p-back-reaction}
is negligible.
The outcome is that other sources of back-reaction can
dominate, if they are not similarly volume-suppressed,
even if they are formally higher order in the power spectrum amplitude.

Although it is the leading contribution
when counting powers
of the amplitude,
the effect
described by~\eqref{eq:fourier-p-back-reaction}
has not previously been considered (in this context).
One reason is that it is not visible directly
from Eq.~\eqref{eq:zeta-separate-universe-fourier}
when this is used to construct a loop expansion for
correlation functions,
of the form to be described
in~\S\ref{sec:back-reaction-quasiparticles}.
Nor is it visible directly from
the loop expansion in the in--in formalism,
because any such loop must involve at least one
$n$-point vertex for $n \geq 3$.
In contrast, Eq.~\eqref{eq:fourier-p-back-reaction}
depends only on particle creation from coupling
to the background geometry
and is present even for a massless free field.

Instead,
to compute this type
of back-reaction within the in--in formalism,
one should coarse-grain the fields in the Wilsonian
sense by integrating out fluctuations
in momentum modes $k \gtrsim \smallBox^{-1}$.
This approach was used
by Gell-Mann \& Hartle to obtain effective equations
of motion for the coarse-grained field~\cite{Gell-Mann:1992wkv},
generalizing
earlier work by
Feynman \& Vernon~\cite{Feynman:1963fq,feynman2010quantum} and
Caldeira \& Leggett~\cite{Caldeira:1982iu,Caldeira:1982uj}.
Their analysis was reformulated in terms
of the in--in path integral by
Calzetta \& Hu, who emphasized the application
to effective field theories~\cite{Calzetta:1996sy,Calzetta:1999xh}.
After coarse-graining, the path integral yields
an influence functional of Feynman--Vernon
type
that describes energy exchange between the high- and low-energy
sectors in terms of noise and dissipation kernels~\cite{Feynman:1963fq}.
An analysis of this type could be used, if required,
to provide a more precise estimate of the
back-reaction effect~\eqref{eq:fourier-p-back-reaction}.

\para{Long--short correlations}
This discussion leaves open the question of what happens
when the short-scale amplitude $(\sigma^2)^{IJ}_{\smallBox}$
correlates with the $\bigBox$-scale mode
$\delta X^{J}_{\vect{p}}$. This effect is encoded in the third term in Eq.~\eqref{eq:2pt-zeta-particle-creation}.
The outcome in that case
depends on the long--short correlation,
but is not \emph{explicitly} suppressed by the 
central-limit volume factor $N^{-1} = (\smallBox / \bigBox)^3$.
In practice it will
typically inherit at least mild
dependence on the ratio $\smallBox/\bigBox$
from the scaling of the long--short correlation.

In this situation the magnitude of the back-reaction
requires a detailed evaluation in each
model of interest.
Here, we assume that the physics of the spike
is sufficiently insensitive to the background
that long--short correlations are suppressed.
Under this assumption
the loops to be described
in~\S\ref{sec:back-reaction-quasiparticles}
would remain the dominant source of
back-reaction.
However, this should clearly be checked.
It certainly appears possible
to imagine scenarios in which
the long--short correlation from
inflationary particle creation
contributes significantly.

\subsubsection{Energy exchange between quasiparticles and condensate}
\label{sec:back-reaction-quasiparticles}

The conclusion of~\S\ref{sec:back-reaction-particle-creation}
is that, under our assumptions,
$\bigBox$-scale modes
$[\smoothSmall{\delta X^J(\vect{x}, t_\ast)}]_{\vect{p}}$
with $p \sim \bigBox^{-1}$
are undisturbed by the effect of particle creation
at the (much smaller) horizon scale.
The physical reason is that an exponentially large number
of horizon volumes fit within each $\bigBox$-sized box,
allowing their fluctuations to balance each other statistically
to very high accuracy.
Hence, at lowest order
and in typical regions,
the corresponding $\zeta$
mode $\zeta_{\vect{p}}$
will
inherit its amplitude
from the background
Fourier mode $\delta X_{\vect{p}}^J(t_\ast)$
with negligible back-reaction.
However, it is still possible for
$\zeta_{\vect{p}}$
to be
corrected by the convolution
term in~\eqref{eq:zeta-separate-universe-fourier}
and its higher-order analogues.

The convolution integral
is dominated by modes with enhanced amplitude.%
    \footnote{Recall that the smoothing scale sets an effective cutoff
    of order $\sim \smallBox^{-1}$ on wavenumbers that participate
    in the integral.}
These occur
near the spike in short-scale
power and have wavenumbers
$k \sim \smallBox^{-1}$.
Let $\vect{k}$ label a Fourier mode
near the peak scale with $k = |\vect{k}| \sim \smallBox^{-1}$,
and let $\vect{p}$ label a soft mode
with $p = |\vect{p}| \sim \bigBox^{-1}$.
The convolution allows
a pair of
excitations
with
wavenumbers $\vect{k}$, $\vect{p}-\vect{k}$
to combine,
producing
a disturbance in the long-wavelength
mode $\vect{p}$.

\para{Noise from short-scale interactions}
What physical processes are described by such combinations?
The particle content of an inflating patch of spacetime can be
described as a dense condensate of zero-momentum particles
with a comparatively
dilute cloud of excitations scattering over it.
These excitations are described as \emph{quasiparticles}
in the condensed matter literature, to emphasize that
their properties and interactions are dressed by
the condensate
and differ from their vacuum values.
In our situation the background field values
$X^I$ describe one or more of these scalar condensates,
and the fluctuations
$\delta X^I_{\vect{p}}$
describe the quasiparticle cloud.
The spike in power is produced by an abundant
population of excitations in the quasiparticle
cloud with wavenumbers $\sim \smallBox^{-1}$.
We describe these as `fast' or `peak' excitations.
Meanwhile, the longer-wavelength degrees of freedom
of wavenumber $\lesssim \bigBox^{-1}$
can be coarse-grained to form an effective
value for the background condensate.
We describe modes in this region as `slow'.

Using this language we can give a spacetime interpretation
of the convolution in~\eqref{eq:zeta-separate-universe-fourier}.
A peak-scale
quasiparticle with momentum $\vect{k}$
may strike the condensate, ejecting two
particles with momenta $\vect{k} - \vect{p}$ and $\vect{p}$.
(Here, we are continuing
to use the convention introduced above
that $\vect{k}$ and $\vect{p}$ label the fast and
slow momenta, respectively.)
The reverse process will also occur,
in which
two peak-scale quasiparticles with
nearly opposite momenta $\vect{k}$, $\vect{p} - \vect{k}$
collide, producing a slow excitation that
joins the coarse-grained
condensate.
Three-body exchange processes
of this type
typically
describe the growth or decay
of Bose--Einstein condensates in
laboratory conditions;
see, e.g., the book by
Kamenev~\cite{kamenev2023field}.

The properties of the newly-created
condensate mode
depend on the statistical distribution
of the fast progenitors,
which are (mostly)
decorrelated from the
coarse-grained background.
The net effect is
as if the long-wavelength condensate
were coupled to a stochastic source of
noise.
We will see that the loops
evaluate the auto-
and cross-correlations of this
noise source.
They can be calculated explicitly within
our framework, precisely because
we prescribed that
the
$\delta N$ formula including back reaction---%
Eq.~\eqref{eq:zeta-separate-universe-fourier}---%
should be smoothed on the scale $\smallBox$.
Had we instead chosen to smooth on the
scale $\bigBox$, most information about
the $\smallBox$-scale modes would have been erased
except for a few aggregate statistics.
This is just the usual Wilsonian
effect of integrating out
ultraviolet modes.
The missing information about the $\smallBox$-scale
modes would then have to be provided via counterterms,
including stochastic counterterms to replicate the
effect of the decorelated noise.
When properly defined and renormalized, these two approaches
would agree on the final outcome.
However, the $\smallBox$-smoothed approach allows a much
more explicit analysis to be given.
We defer a discussion of whether it is actually more predictive
than the $\bigBox$-smoothed approach
to~\S\ref{sec:wilsonian-counterterms}.

The conclusion is that spacetime
processes described by the convolution
in Eq.~\eqref{eq:zeta-separate-universe-fourier}
correspond to
redistribution of excitations
within the quasiparticle cloud.
The degree to which 3-body processes
contribute to this redistribution is measured
by the
time dependent prefactor of the quadratic terms
in an expression such as~\eqref{eq:basic-separate-universe}
or~\eqref{eq:zeta-separate-universe}.
There are similar contributions involving higher
$n$-body processes which arise from the cubic and
higher terms in these equations.
The interpretation of these contributions
parallels the discussion of 3-body interactions given here.
Meanwhile, injection of new excitations into the
cloud is described by the
particle creation process
discussed in~\S\ref{sec:back-reaction-particle-creation}.

Each available
3-body channel is described by tree-level
$S$-matrix elements
shown in Fig.~\ref{fig:three-body-S-matrix}.
\begin{figure}
    \begin{equation*}
        \eqgraph{1ex}{0ex}{
            \begin{fmfgraph*}(70,50)
                \fmfleft{l}
                \fmfright{r1,r2}
                \fmf{plain,label={\small $\vect{p}$}}{l,v}
                \fmf{plain,label={\small $\vect{k}$}}{v,r1}
                \fmf{plain,label={\small $\vect{p}-\vect{k}$}}{r2,v}
                \fmfv{label={\footnotesize $|\text{condensate} \rangle$}}{l}
                \fmfv{label={\footnotesize $|\text{cloud} \rangle$}}{r1}
                \fmfv{label={\footnotesize $|\text{cloud} \rangle$}}{r2}
            \end{fmfgraph*}
        }
    \end{equation*}
    \caption{\label{fig:three-body-S-matrix}Three-body $S$-matrix processes
    describing exchange between the condensate and the quasiparticle cloud.
    From the perspective of the fast mode $\vect{k}$,
    the soft mode $\vect{p}$ can be regarded as part
    of the local condensate.}
\end{figure}
\begin{figure}
    \begin{equation*}
        \begin{split}
            \langle \delta X^I_{\vect{p}} \delta X^J_{-\vect{p}} \rangle
            &
            \sim
            \sum_{\substack{\footnotesize \text{cloud} \\ \text{states}}}
            \Bigg|
                \mspace{5mu}
                \eqgraph{2ex}{2ex}{
                    \begin{fmfgraph*}(50,40)
                        \fmfleft{l}
                        \fmfright{r1,r2}
                        \fmf{plain,label={\small $\vect{p}$},label.side=right}{l,v}
                        \fmf{dashes,label={\small $\vect{k}$},label.side=right}{v,r1}
                        \fmf{dashes,label={\small $\vect{p}-\vect{k}$},label.side=left}{v,r2}
                    \end{fmfgraph*}
                }
            \Bigg|^2
            =
            \sum_{\substack{\footnotesize \text{cloud} \\ \text{states}}}
            \Bigg(
                \mspace{5mu}
                \eqgraph{2ex}{2ex}{
                    \begin{fmfgraph*}(50,40)
                        \fmfleft{l}
                        \fmfright{r1,r2}
                        \fmf{plain,label={\small $\vect{p}$},label.side=right}{l,v}
                        \fmf{dashes,label={\small $\vect{k}$},label.side=right}{v,r1}
                        \fmf{dashes,label={\small $\vect{p}-\vect{k}$},label.side=left}{v,r2}
                    \end{fmfgraph*}
                }
            \Bigg)
            \Bigg(
                \eqgraph{2ex}{2ex}{
                    \begin{fmfgraph*}(50,40)
                        \fmfleft{l1,l2}
                        \fmfright{r}
                        \fmf{plain,label={\small $\vect{p}$},label.side=left}{r,v}
                        \fmf{dashes,label={\small $\vect{k}$},label.side=left}{v,l1}
                        \fmf{dashes,label={\small $\vect{p}-\vect{k}$},label.side=right}{v,l2}
                    \end{fmfgraph*}
                }
                \mspace{5mu}
            \Bigg)^\dag
            \\
            &
            \sim
            \int \frac{\d^3 k}{(2\pi)^3}
            \mspace{30mu}
            \eqgraph{4ex}{4ex}{
                \begin{fmfgraph*}(60,40)
                    \fmfleft{l}
                    \fmfright{r}
                    \fmf{plain}{l,v1}
                    \fmf{dashes,left,tension=0.3,label={\small $\vect{p}-\vect{k}$}}{v1,v2}
                    \fmf{dashes,left,tension=0.3,label={\small $\vect{k}$}}{v2,v1}
                    \fmf{plain}{r,v2}
                    \fmfv{label={\small $\vect{p}$}}{l}
                    \fmfv{label={\small $-\vect{p}$}}{r}
                \end{fmfgraph*}
            }
        \end{split}
    \end{equation*}
    \caption{\label{fig:s-matrix-to-power-spectrum}The tree-level
    $S$-matrix process in
    Fig.~\ref{fig:three-body-S-matrix} sets up correlations in the condensate.
    To evaluate these, we require the expectation
    of the modulus-squared of the $S$-matrix element, averaged over the
    unobserved fast excitations in the quasiparticle cloud.
    In these diagrams and in Fig.~\ref{fig:one-loop-delta-N}
    we show cloud excitations as dashed lines and
    condensate modes as solid lines.
    (Elsewhere in this paper, the meaning of these lines is
    as shown in the key to Fig.~\ref{fig:deltaN-diagrams}.)
    Averaging over the cloud excitations
    glues the two S-matrix elements together to produce
    an in--in loop diagram
    (or at least its short-wavelength part).}
\end{figure}
To evaluate the spacetime correlations
they produce (up to normalization), we should compute the
modulus-squared of the $S$-matrix element
and sum over the bath of unobserved
cloud states
that source the interaction.
The sum over cloud states implies that this is
a loop-level calculation within the in--in formalism,
even though each individual
$S$-matrix process is at tree level.
An exactly analogous loop appears when calculating
the spectrum of gravitational waves
produced by a model of this type with a peak in
short-scale
power~\cite{Ananda:2006af,Baumann:2007zm,Kohri:2018awv,Domenech:2021ztg}.%
    \footnote{See, for example, Refs.~\cite{Ota:2022xni, Chen:2022dah} for
    an application of the in--in formalism to the calculation of gravitational
    waves at one-loop.}
However,
in this case one is normally interested in
configurations where there is not
a large hierarchy among the wavenumbers.

Similar reasoning leads to the formalism
of \emph{cut diagrams} in hadronic physics,
which are closely related to
Cutkosky-type rules.
In these diagrams one partitions the vertices
in a Feynman diagram
into two parts using a \emph{final state cut},
which implies summation over states crossing the cut.
The two partitions correspond to
different groupings of vertices
on the Schwinger--Keldysh closed time contour.
There are differences in the cosmological case,
because
the final state cut becomes an \emph{initial} state cut.
It would be valuable to
understand whether there is
an analogous
statement involving cosmological cutting
rules~\cite{Melville:2021lst,Goodhew:2021oqg},
but we do not pursue that question here.
For a presentation of cut diagrams in the context of QCD,
see, e.g., the book by Collins~\cite{Collins:2011zzd}.

In the language of the effective field theory of large-scale
structure, the in--in loop of Fig.~\ref{fig:s-matrix-to-power-spectrum}
is of `22-type'.
This analysis shows that it can be interpreted
using the toy model of
Eqs.~\eqref{eq:corrected-zetaL-field}--\eqref{eq:corrected-average-zetaL-field}.
Specifically,
we may write~\eqref{eq:zeta-separate-universe-fourier}
in a form
that parallels~\eqref{eq:corrected-average-zetaL-field},
\begin{equation}
    \zeta_{\vect{p}}
    = \zeta^{(1)}_{L,\vect{p}}
    + \OpS_{\vect{p}} .
\end{equation}
Each term has a counterpart in Eq.~\eqref{eq:corrected-average-zetaL-field}.
The $\vect{p}^{\mathrm{th}}$ Fourier mode
$\zeta_{\vect{p}}$
plays the role of $\zetaBigSmooth$;
the Gaussian field
$\zeta^{(1)}_{L,\vect{p}}$
plays the role of the
long-wavelength background $\zetaBig$,
\begin{subequations}
\begin{equation}
    \zeta^{(1)}_{L,\vect{p}}(t)
    \equiv
    \frac{\delta N(t)}{\delta X^J(t_\ast)}
    \Big[ \smoothSmall{\delta X^J(\vect{x}, t_\ast)} \Big]_{\vect{p}} ;
\end{equation}
and
$\OpS_{\vect{p}}$ is a composite operator
that should be regarded as
an avatar for the sum $N^{-1} \sum_i S_i$,
\begin{equation}
    \OpS_{\vect{p}}
    \equiv \frac{1}{2} \frac{\delta^2 N(t)}{\delta X^J(t_\ast) \delta X^K(t_\ast)}
    \int\limits_{q \lesssim \smallBox^{-1}}
    \frac{\d^3 q}{(2\pi)^3}
    \,
    \Big[ \smoothSmall{\delta X^J(\vect{x}, t_\ast)} \Big]_{\vect{p}-\vect{q}}
    \Big[ \smoothSmall{\delta X^K(\vect{x}, t_\ast)} \Big]_{\vect{q}} .
\end{equation}
\end{subequations}
Source excitations in the quasiparticle cloud
that contribute to $\OpS_{\vect{p}}$
are decorrelated
from the long-wavelength condensate
and can be localized
on a scale $\sim \smallBox$.
In each of these quantities,
as explained in the discussion following~\eqref{eq:zeta-separate-universe-fourier},
extracting the Fourier mode $\vect{p}$ is already tantamount to spatial averaging
on the scale $\bigBox \sim p^{-1}$.
The loop of Fig.~\ref{fig:s-matrix-to-power-spectrum}
is therefore nothing more than
an estimate of the
$\OpS_{\vect{p}}$ autocorrelation
$\langle \OpS_{\vect{p}} \OpS_{-\vect{p}} \rangle$,
with the approximation of Gaussian statistics for the source excitations.
It does not account for any inter-box correlation
captured by the off-diagonal covariance
$\langle S_i S_j \rangle$ for $i \neq j$
in Eq.~\eqref{eq:Si-correlation}.
One could do so, if required, by including 
non-Gaussian effects.
In summary, we expect the contribution from the 22-loop
to be suppressed by
the central-limit-like factor volume $N^{-1} = (\smallBox/\bigBox)^3$,
just as its counterpart~\eqref{eq:Si-auto-correlation}.

\subsubsection[\texorpdfstring{Loops in ${\delta N}$ perturbation theory}{Loops in delta N perturbation theory}]{Loops in $\bm{\delta N}$ perturbation theory}
\label{sec: loops in delta N}
Let us now systematically consider the types of loop that
can appear.
Each loop bears an interpretation analogous to the sum over
cloud states in Fig.~\ref{fig:s-matrix-to-power-spectrum}.
In particular, the use of the term ``loop'' should not be taken
to indicate that the calculation involves quantum corrections,
but rather an average over unobserved
fluctuations regardless of their character.%
    \footnote{In this context one may compare with the discussion
    of Ref.~\cite{Lyth:2006qz}.
    The discussion in {\S}4 of that reference
    distinguished between 
    what were called ``C-Feynman''
    and ``Q-Feynman'' diagrams,
    with the C-Feynman diagrams roughly corresponding to
    what we describe as
    averages over classical stochastic structure.
    As pointed out there, the epithet \emph{classical}
    strictly refers only to the time dependence of these modes;
    they do in fact arise from a vacuum fluctuation
    and carry factors of $\hbar$ allowing the whole diagrammatic
    expansion to match with the in--in formalism.}

Correlation functions of $\zeta_{\vect{p}}$
are computed
by taking expectation values of products
of~\eqref{eq:zeta-separate-universe-fourier},
with due care regarding the loop
order at which these are calculated.
(See, e.g., Refs.~\cite{Seery:2007we,Seery:2007wf}.)
We assign an order in the loop expansion to each
product of $\delta X^I$,
determined by the number of unconstrained
momentum integrals they contain.
It follows from combinatorial considerations
that we can represent
such products using Feynman-like diagrams
constructed using the language of Fig.~\ref{fig:s-matrix-to-power-spectrum}.
As usual, there is an unconstrained momentum
integral associated with each closed loop.
These
represent independent sums over
excitations in the quasiparticle cloud.
Notice that
(as for vertices in any type of Feynman diagram)
a given term in~\eqref{eq:zeta-separate-universe-fourier}
may contribute at different loop orders
in different correlation functions.
For example, the quadratic term
generates tree-level 112-type contributions
to the bispectrum, but
one-loop 22-type
contributions to the power spectrum.

\begin{figure}
    \begin{equation*}
        \underbracket{
            \eqgraph{1ex}{0ex}{
                \begin{fmfgraph*}(100,50)
                    \fmfleft{l}
                    \fmfright{r}
                    \fmf{plain}{l,v1}
                    \fmf{dashes}{v1,v2}
                    \fmf{dashes,left,tension=0.3}{v2,v3}
                    \fmf{dashes,left,tension=0.3}{v3,v2}
                    \fmf{plain}{v3,r}
                    \fmfv{label=$N_{I}^{(t_{k}\comma t)}$,label.angle=90,
                        decor.shape=circle,decor.filled=full,decor.size=2thick}{v1}
                    \fmfv{label=$N_{JK}^{(t_{k}\comma t)}$,label.angle=45,
                        decor.shape=circle,decor.filled=full,decor.size=2thick}{v3}
                \end{fmfgraph*}
            }
        }_{\text{12-type diagrams}}
        \mspace{20mu}
        +
        \mspace{12mu}
        \underbracket{
            \mspace{5mu}
            \eqgraph{1ex}{0ex}{
                \begin{fmfgraph*}(100,50)
                    \fmfleft{l}
                    \fmfright{r}
                    \fmf{plain}{l,v1}
                    \fmf{dashes,left,tension=0.45}{v1,v2}
                    \fmf{dashes,left,tension=0.45}{v2,v1}
                    \fmf{plain}{v2,r}
                    \fmfv{label=$N_{IJ}^{(t_{k}\comma t)}$,label.angle=135,
                        decor.shape=circle,decor.filled=full,decor.size=2thick}{v1}
                    \fmfv{label=$N_{KL}^{(t_{k}\comma t)}$,label.angle=45,
                        decor.shape=circle,decor.filled=full,decor.size=2thick}{v2}
                \end{fmfgraph*}
            }
        }_{\text{22-type diagrams}}
        \mspace{20mu}
        +
        \mspace{10mu}
        \underbracket{
            \mspace{5mu}
            \eqgraph{1ex}{0ex}{
                \begin{fmfgraph*}(100,50)
                    \fmfleft{l}
                    \fmfright{r}
                    \fmf{plain}{l,v1}
                    \fmf{dashes}{v1,v2}
                    \fmf{dashes,right,tension=0.75}{v2,v2}
                    \fmf{plain}{v2,r}
                    \fmfv{label=$N_{I}^{(t_{k}\comma t)}$,label.angle=90,
                        decor.shape=circle,decor.filled=full,decor.size=2thick}{v1}
                    \fmfv{label=$N_{JKL}^{(t_{k}\comma t)}$,label.angle=-45,
                        decor.shape=circle,decor.filled=full,decor.size=2thick}{v2}
                \end{fmfgraph*}
            }
        }_{\text{13-type diagrams}}
    \end{equation*}
    \caption{\label{fig:one-loop-delta-N}$\delta N$ diagrams,
    in the language of Fig.~\ref{fig:s-matrix-to-power-spectrum},
    for one-loop corrections to the
    $\bigBox$-scale power spectrum due to
    $\smallBox$-scale cloud modes.
    The unconstrained momentum integration implicit in
    each diagram, representing an average over unobserved
    cloud states,
    runs over scales near the short-scale
    power spectrum peak.
    The notation $N_I^{(t_\ast, t)}$, $N_{IJ}^{(t_\ast, t)}$ 
    is introduced in Eqs.~\eqref{eq:NI-def}--\eqref{eq:NIJ-def}.
    In the diagrams we have set the initial time $t_\ast$ to equal $t_k$;
    see the discussion below Eq.~\eqref{eq:prototype-12}.}
\end{figure}
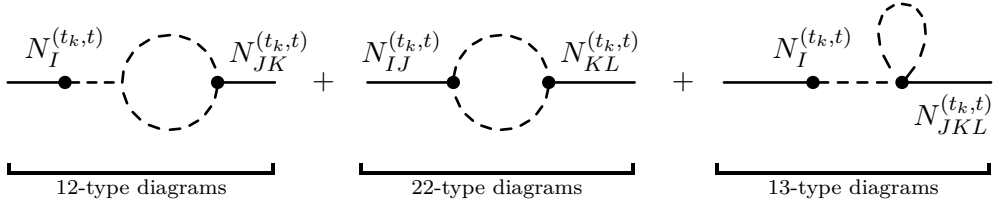

At one loop,
the three
possibilities are shown in Fig.~\ref{fig:one-loop-delta-N}.
In these diagrams we have introduced a compact notation
\begin{subequations}
\begin{align}
    \label{eq:NI-def}
    N_I^{(t_\ast, t)}
    & =
    \frac{\delta N(t)}{\delta X^I(t_\ast)} ,
    \\
    \label{eq:NIJ-def}
    N_{IJ}^{(t_\ast, t)}
    & =
    \frac{\delta^2 N(t)}{\delta X^I(t_\ast) \delta X^J(t_\ast)} ,
\end{align}
\end{subequations}
with similar definitions for $N_{IJK}^{(t_\ast, t)}$ (and so on), as needed.

\para{$\delta X^I$ correlators}
Each diagram in Fig.~\ref{fig:one-loop-delta-N}
involves correlators of the $\delta X^I$ fields.
In linear perturbation theory these are free, Gaussian fields.
Higher-order interactions and nonlinear evolution generate
non-Gaussian contributions. We have already written down the
two-point function for $\delta X^I_{\vect{k}_1}( t_\ast)$ in
Eq.~\eqref{eq:short-mode-power-spectrum}.
We will also use the corresponding three-point function,
\begin{equation}
    \label{def alpha_ijk}
    \langle
        \delta X^I_{\vect{k}_1}(t_\ast)
        \delta X^J_{\vect{k}_2}(t_\ast)
        \delta X^K_{\vect{k}_3}( t_\ast)
    \rangle
    \equiv
    (2\pi)^3 \delta (\vect{k}_1 + \vect{k}_2 + \vect{k}_3)
    \alpha^{IJK}(k_1, k_2, k_3;t_\ast) .
\end{equation}
$P^{IJ}(k_1;t_\ast)$ and $\alpha^{IJK}(k_1, k_2, k_3;t_\ast)$ are, respectively,
second-order and fourth-order
in powers of $H/\Mp$.
For free, massless fields in a nearly de Sitter spacetime we have,
after dropping decaying
modes,
\begin{equation}
     \label{massless correlator}
     P^{\phi\phi}(k_1;t_{k_1}) = \frac{H(t_{k_1})^2}{2 {k_1}^3} ,
\end{equation}
where $t_{k_1}$ is the horizon-crossing time for the wavenumber $k_1$.
In this paper we assume that Eq.~\eqref{massless correlator} is a good
approximation for the field fluctuations at horizon-crossing.
We discuss this choice in the context of an explicit
numerical study in Appendix~\ref{sec: numerical example}.

\para{Tree-level}
The $\delta N$ tree-level contribution comes from the product of linear
terms in Eq.~\eqref{eq:zeta-separate-universe-fourier},
\begin{equation}
    \label{eq:prototype-tree}
    \langle \zeta_{\vect{p}} \zeta_{-\vect{p}} \rangle_{\text{tree}}
    \sim
    N_I^{(t_k ,t)}
    N_J^{(t_k, t)}
    \langle
        \delta X^I_{\vect{p}}(t_k)
        \delta X^J_{-\vect{p}}(t_k)
    \rangle_{\text{tree} + \text{1-loop}} .
\end{equation}
There is a subtlety with the $\delta X$ two-point function appearing here.
This term enters at tree-level within the $\delta N$ expansion.
However, in order to obtain a result that contains all contributions
up to one-loop,
the $\delta X$ correlator
should contain \emph{both} tree-level and one-loop contributions.

One option is to evaluate these loops using a second application
of the separate universe picture. This could be a
fruitful approach if
large contributions
are expected from substructure that emerges between the horizon exit time
$t_p$ for the scale $\bigBox \sim 1/p$,
and the corresponding exit time
$t_k$ for the scale $\smallBox \sim 1/k$.
Specifically,
it is at the time $t = t_k$
that
we need to evaluate the correlation function.

In our present scenario, however, there are no large effects of this kind.
At time $t_k$ the enhanced modes on the scale $\smallBox$
have only just emerged from the horizon.
There has not yet been time for scattering
processes
to have redistributed
any power from small scales to large scales,
so we do not expect a significant loop effect.
Instead, the only source of back reaction is the particle creation
event itself. This was considered in~\S\ref{sec:back-reaction-particle-creation}
and found to be negligible provided there is
at least a modest hierarchy between the scales $p$ and $k$.
On this basis (here and in the remainder of this paper), we retain only the
tree-level contribution to the $\delta X$ two-point function
in~\eqref{eq:prototype-tree}.

To close this section we give
general expressions for each one-loop term
in Fig.~\ref{fig:one-loop-delta-N}.
In principle these can be applied to \emph{any}
model of inflation with an arbitrary number
of fields.
In~\S\S\ref{sec: SR to USR to SR}--\ref{sec:delta N loops calculation}
we specialize these expressions to the
case of a concrete 3-phase ultra-slow-roll model.

\para{12-type loops}
Using a standard notation
we label the
three loop topologies
as contributions of type 12, 22 (already introduced above),
and 13.

First, consider the 12 loops.
These arise from a product of the linear term
in Eq.~\eqref{eq:zeta-separate-universe-fourier}
with the quadratic term,
\begin{equation}
    \label{eq:prototype-12}
    \langle \zeta_{\vect{p}} \zeta_{-\vect{p}} \rangle_{12}
    \sim
    N^{(t_k, t)}_I
    N^{(t_k, t)}_{JK}
    \int \frac{\d^3 q}{(2\pi)^3}
    \big\langle
        \delta X^I_{\vect{p}}(t_k)
        \delta X^J_{-\vect{p}-\vect{q}}(t_k)
        \delta X^K_{\vect{q}}(t_k)
    \big\rangle_{\text{tree}} .
\end{equation}
We have dropped the smoothing label $\smoothSmall{\cdots}$
on the perturbations $\delta X^I_{\vect{k}}$,
partly to reduce clutter,
and partly
because none of the fields in this product include
back-reaction from $\smallBox$-scale effects.
For the slow mode $\vect{p}$ this follows from the discussion
of~\S\ref{sec:back-reaction-particle-creation}.
Meanwhile, the convolution is dominated by fast modes
$\vect{k}$ that are themselves associated with the scale
$\smallBox$.
The convolution should still be cut off at $q \sim \smallBox^{-1}$,
as in Eq.~\eqref{eq:zeta-separate-universe-fourier}.
Finally, in Eq.~\eqref{eq:prototype-12}
and Fig.~\ref{fig:one-loop-delta-N}
we have set $t_\ast = t_k$, where $t_k$
is the horizon exit time for modes contributing to the
short-scale
peak in the power spectrum.

Notice that (as explained above)
it is important to set
the smoothing lengthscale 
for the separate universe
formula~\eqref{eq:zeta-separate-universe-fourier}
to be no larger than the scale $\smallBox$
associated with the peak modes $\sim \vect{k}$.
For this reason, one must choose the base time
to be at least as late as the horizon exit time $t_k$,
in order that these modes begin to evolve according to the
classical superhorizon time dependence.
If one chooses a larger smoothing
length
then the modes whose back-reaction we intend
to compute will be erased.

The
choice of smoothing scale
$t_\ast = t_k$ has the consequence that we must
\emph{separately}
account for the
evolution of the $\delta X_{\vect{p}}$
mode
between its horizon exit time and the time
$t_k$.
We will see explicit examples
in~\S\ref{sec:delta N loops calculation}.
The soft limit for the $\delta X$
three-point function
(and other soft limits needed for the remaining
loop topologies)
can generally be estimated using powerful
factorization methods~\cite{Kenton:2015lxa};
see Eq.~\eqref{alpha generic}.

The $\delta X^I$ correlation function
on the right-hand side of~\eqref{eq:prototype-12}
would be exactly zero in a Gaussian approximation.
It is present only when the cloud modes have long--short correlations
that persist between the modes $p \sim \bigBox^{-1}$
and $k \sim \smallBox^{-1}$.
The amplitude of these correlations
is measured by the reduced bispectrum
$\fNL$
in a squeezed configuration, roughly $\sim \fNL(p, k, k)$.
This evidently corresponds to the
long--short correlation~\eqref{eq:Si-cross-correlation}
in the toy model of~\S\ref{sec:toy-model},
with $\fNL$ playing the role of the correlation coefficient $R$.
It follows that
the short-scale power may
add coherently over $\bigBox$-sized regions.
The contribution from
these modes is therefore
not suppressed by
the central-limit-like volume factor $N^{-1} = (\smallBox/\bigBox)^3
\sim (p/k)^3$.

If this is the case, how should we expect the correlation
to behave when $p/k \rightarrow 0$?
An estimate of this behaviour is critical,
for two reasons.
First, it will
determine whether the back-reaction is significant.
Second, on the basis of the decoupling principle,
Fumagalli argued that one should expect
the back-reaction (and therefore the loop)
to vanish in the limit of an infinite hierarchy
$p/k \rightarrow 0$~\cite{Fumagalli:2023hpa}.
A similar argument was given by Tada {\etal}~\cite{Tada:2023rgp}; see
also Firouzjahi~\cite{Firouzjahi:2023bkt}.
In a single-field adiabatic model,
the asymptotic squeezed limit of the $\zeta$
bispectrum
is controlled by Maldacena's consistency
condition~\cite{Maldacena:2002vr}.
It has been argued that this is a gauge
artefact
that should be subtracted by moving to physically defined
coordinates appropriate for a local
observer~\cite{Tanaka:2011aj,Creminelli:2011rh,Pajer:2013ana,dePutter:2015vga}.
This amounts to fixing our small-scale
rulers so that, in each region, the
peak is produced on a fixed physical scale,
rather than a fixed comoving
scale~\cite{Pajer:2013ana,Tada:2016pmk}.
If so, then
when
working on such a physically-defined hypersurface,
the asymptotic squeezed limit of the reduced
$\zeta$-bispectrum would be zero,
with
the leading correction from gradients of order
$(p/k)^2$~\cite{Pajer:2013ana,Creminelli:2011rh},
at least if there is a massless mode in the
spectrum.
(If all modes are massive then we expect stronger suppression.)
This suggests
that Eq.~\eqref{eq:prototype-12}
may require corrections in order that the integral
represents a
sum over a suitable, physically-defined squeezed limit.

A general recipe to compute the necessary subtraction
for any correlation function was given by Pajer {\etal}~\cite{Pajer:2013ana}.
Their prescription involved an infrared regulator scale
that was used to divide modes into a `soft' part,
which is effectively resummed into the new physical coordinates,
and a `hard' part.
There does not yet appear to be a practical prescription
for this subtraction that yields results manifestly independent
of this infrared regulator.
Nevertheless, it would seem plausible that,
in a single-field adiabatic model, Eq.~\eqref{eq:prototype-12}
scales like $(p/k)^2$ (or a steeper function of $p/k$)
when $p/k \rightarrow 0$ at fixed $k$.
If so, we should apparently conclude that
the physical back-reaction on CMB-scale modes due to
a spike of power associated with PBH-scale modes is
exponentially small.%
    \footnote{There are some ambiguities.
    For example, we should be free to compute the loop correction on
    any well-defined hypersurface and later translate to the physical
    hypersurface.
    The outcome of this calculation should be equivalent to
    performing the loop correction on the physical hypersurface from
    the outset by making subtractions to the three-point
    function in Eq.~\eqref{eq:prototype-12}.
    (This neglects the effect of back-reaction on the definition of the
    physical hypersurface, which may be reasonable if the loop
    correction is smaller than the tree-level.)
    It is not clear that these operations commute.}

In this situation
it would be unclear whether the separate universe framework
is adequate to evaluate the loop,
because we already expect gradient
corrections to formulae such as~\eqref{eq:prototype-12}.
If we base the separate universe expansion
near the horizon exit time for $k$
then we would expect these gradients to be suppressed
by $(p/k)^2$. They would therefore be the same
order in gradients as the surviving terms from~\eqref{eq:prototype-12}.
Effects of this type at order $(p/k)^2$
have been reported by Tasinato
using a large $|\eta|$ expansion~\cite{Tasinato:2023ukp},
which could possibly be identified
with these surviving contributions.
However,
the situation here is quite unclear.

On the other hand, if the model is \emph{not} adiabatic,
long--short correlations may survive,
at least over some range of scales.
For example, this may happen in inflationary models with
multiple fields, if their isocurvature modes do not decay rapidly.
The 12-loop~\eqref{eq:prototype-12}
could then receive contributions from squeezed
configurations.
On physical grounds, one would still expect
the soft limit of the correlation function to decay
at least mildly as a function of $p/k$.
This decay would be inherited by the loop,
generating the expected decoupling
as $p/k \rightarrow 0$.

In this paper we work with the separate universe
formula~\eqref{eq:zeta-separate-universe-fourier},
and the resulting 12-type loop,
as they stand---%
with the understanding that
the integrand of~\eqref{eq:prototype-12}
(and similar loop contributions that depend on squeezed configurations of
higher $n$-point functions)
may require subtractions in order to define a physical squeezed limit.
However,
we defer a detailed analysis of how these squeezed limits should be handled
to future work.

Just as for the 22-loops,
the 12-loops have a particle interpretation.
The topology depicted in Fig.~\ref{fig:one-loop-delta-N}
shows this must be related to the 22-loop;
see Fig.~\ref{fig:s-matrix-to-power-spectrum}.
The major difference
is that the 22-loop has the same interaction at both vertices,
associated with conversion of two short-wavelength field
fluctuations into a slow $\zeta$ mode.
It correlates the volume-averaged field produced by
this process with itself.
In contrast, the 12-loop has different vertices.
One of these is the same as
the vertices occurring in the 22-loop,
and is used to tie together two
fast legs
of the $\delta X^I$ three-point function to produce a slow $\zeta$.
The other describes production of a slow $\delta X^I$ mode
which is re-interpreted directly as a slow $\zeta$.
It is this cross-correlation effect that allows it to evade
volume suppression, as we have already seen in
Eq.~\eqref{eq:Si-cross-correlation}.
The same conclusions would apply
to a $\zeta$-gauge calculation using the in--in formalism.
Specifically,
the 12- and 12-loops discussed here would be
both appear to be part of the in--in 22-loop (there being no 12-loop
in the interaction picture).

\para{22-type loops}
Some aspects of these loops have already been discussed
in~\S\ref{sec:back-reaction-quasiparticles}.
They arise from the product of two quadratic terms
in Eq.~\eqref{eq:zeta-separate-universe-fourier},
\begin{equation}
    \label{eq:prototype-22}
    \langle \zeta_{\vect{p}} \zeta_{-\vect{p}} \rangle_{22}
    \sim
    N_{IJ}^{(t_k, t)}
    N_{KL}^{(t_k, t)}
    \int \frac{\d^3 q}{(2\pi)^3}
    \int \frac{\d^3 r}{(2\pi)^3}
    \langle
        \delta X^I_{\vect{q}}(t_k)
        \delta X^J_{\vect{p}-\vect{q}}(t_k)
        \delta X^K_{\vect{r}}(t_k)
        \delta X^L_{-\vect{p}-\vect{r}}(t_k)
    \rangle_{\text{tree}} .
\end{equation}
The leading contribution comes from the disconnected
part of this four-point correlation,
representing a Gaussian approximation.
We will see in~\S\ref{sec:delta N loops calculation}
that it reproduces the expected
shot-noise-like behaviour $\sim (p/k)^3$,
described above.
If desired, the connected contribution could also be kept,
which would enter at two-loop order.

\para{13-type loops}
These are generated by a product between the linear
term in Eq.~\eqref{eq:zeta-separate-universe-fourier}
and the cubic term, which until now we have not written explicitly,
\begin{equation}
    \label{eq:prototype-13}
    \langle \zeta_{\vect{p}} \zeta_{-\vect{p}} \rangle_{13}
    \sim
    N_{I}^{(t_k, t)}
    N_{JKL}^{(t_k, t)}
    \int \frac{\d^3 q}{(2\pi)^3}
    \int \frac{\d^3 r}{(2\pi)^3}
    \langle
        \delta X^I_{\vect{p}}(t_k)
        \delta X^J_{\vect{q}}(t_k)
        \delta X^K_{\vect{r}}(t_k)
        \delta X^L_{-\vect{p}-\vect{q}-\vect{r}}(t_k)
    \rangle_{\text{tree}}
\end{equation}
The corresponding diagram (see Fig.~\ref{fig:one-loop-delta-N})
has a different topology to the 12- and 22-type loops.
It constitutes a multiplicative renormalization of the
power spectrum at wavenumber $p$,
rather than a stochastic contribution from a decorrelated
noise-like source.
It represents a sum over $S$-matrix processes of the form
\begin{equation}
    \langle \zeta_{\vect{p}} \zeta_{-\vect{p}} \rangle
    \sim
    \sum_{|\text{cloud}\rangle}
    \Bigg(
        \mspace{5mu}
        \eqgraph{2ex}{2ex}{
            \begin{fmfgraph*}(50,40)
                \fmfleft{l}
                \fmfright{r}
                \fmf{plain,label={\small $\vect{p}$}}{l,r}
                \fmffreeze
                \fmfforce{(0.5w, 0.5h)}{v}
            \end{fmfgraph*}
        }
        \mspace{5mu}
    \Bigg)
    \Bigg(
        \mspace{50mu}
        \eqgraph{2ex}{2ex}{
            \begin{fmfgraph*}(50,40)
                \fmfleft{l}
                \fmfright{r}
                \fmftop{t1,t2}
                \fmf{dashes}{l,v}
                \fmf{dashes}{t1,v}
                \fmf{dashes}{t2,v}
                \fmf{plain,label={\small $\vect{p}$},tension=1.3}{v,r}
                \fmffreeze
                \fmfforce{(0.1w, 0.85h)}{t1}
                \fmfforce{(0.5w, 0.95h)}{t2}
                \fmfforce{(0.5w, 0.5h)}{v}
                \fmfv{label={\small $|\text{cloud}\rangle$}}{t1}
                \fmfv{label={\small $|\text{cloud}\rangle$}}{t2}
            \end{fmfgraph*}
        }
        \mspace{5mu}
    \Bigg)^\dag
    +
    \text{h.c.} ,
\end{equation}
where `$+$ h.c.' indicates that we should add the Hermitian conjugate
of the preceding term.
The sum over cloud states $|\text{cloud}\rangle$ ties together two
of the dashed legs in the diagram contaning a four-point vertex,
producing a loop.
Roughly, the $\vect{p}$ Fourier mode passes straight through the
diagram;
none of the decorrelated cloud excitations propagate
into the final state.
 
In the $\zeta$-gauge calculation performed using the in--in formalism, the 13-type diagram would correspond to the (topologically equivalent) diagram due to quartic-order $\zeta$-interactions. This contribution has been calculated by Firouzjahi in Ref.~\cite{Firouzjahi:2023aum} for a 3-phase ultra-slow-roll model with instantaneous transitions; see also the discussion in Ref.~\cite{Firouzjahi:2023bkt}.

\section{The 3-phase ultra-slow-roll model}
\label{sec: SR to USR to SR}
The framework elaborated in~\S\S\ref{sec:intro}--\ref{sec:separate-universe-deltaN}
can be applied to a wide range of scenarios.
As an illustration,
we consider a single-field model
including
a transient phase of ultra-slow-roll
inflation~\cite{Kinney:2005vj, Dimopoulos:2017ged, Pattison:2018bct}.
This phase is both preceded and succeeded by
conventional periods of slow-roll inflation.
In this section we briefly
review details of the ultra-slow-roll
model and show how to describe the time evolution of its perturbations
using the separate universe framework.

Scalar fluctuations
are amplified during the ultra-slow-roll era.
This produces a peak in the scalar power spectrum.
By choosing the ultra-slow-roll phase to occur at the right
time, the peak can be positioned on scales smaller than those constrained
by CMB experiments.
After horizon re-entry in the radiation era, these large perturbations
may collapse to form primordial black holes~\cite{10.1093/mnras/168.2.399};
for example,
see the review by Sasaki {\etal}~\cite{Sasaki:2018dmp}.
Such models are of interest because there is still a small mass window in which
primoridal black holes could account for a fraction
(or even all) of the dark matter~\cite{Bird:2016dcv,Bertone:2018krk, Bartolo:2018evs}.
To trigger sufficient black hole formation, the small-scale power spectrum would need
to be enhanced by roughly $10^7$ relative to its value on CMB scales.

Similar
multi-phase USR $+$ SR
scenarios have been discussed by Cai {\etal}~\cite{Cai:2018dkf}, Kristiano \& Yokoyama~\cite{Kristiano:2022maq}
and Firouzjahi \& Riotto~\cite{Firouzjahi:2023ahg}.

\para{Background evolution}
We define the first three Hubble slow-roll parameters, 
\begin{equation}
    \label{slow roll paramters}
    \epsilon
    \equiv
    -\frac{H'}{H}
    ,
    \quad
    \eta
    \equiv
    \frac{\epsilon'}{\epsilon}
    ,
    \quad
    \xi
    \equiv
    \frac{\eta'}{\eta} ,
\end{equation}
where a prime indicates a derivative with respect
to the elapsed e-folding number,
$N = \int^t H(t') \, \d t'$.
The inflationary phase is assumed to be supported
by the potential energy $V(\phi)$ associated with a
single scalar
field $\phi$.
We also introduce
$V$-type
slow-roll parameters defined with respect to the potential,
\begin{equation}
    \label{potential slow roll paramters}
    \epsilon_V
    \equiv
    \frac{\Mp^2}{2}
    \left(
        \frac{V_\phi}{V}
    \right)^2
    ,
    \quad
    \eta_V
    \equiv
    \Mp^2
    \frac{V_{\phi\phi}}{V} .
\end{equation}
Here, $V_\phi$ and $V_{\phi\phi}$ are the first and second derivatives
of $V(\phi)$ with respect to $\phi$.
The $V$-type slow-roll parameters are not the
same as the Hubble-type parameters
defined in~\eqref{slow roll paramters},
but their order of magnitude is comparable when
the slow-roll approximation is valid.

We suppose that $V(\phi)$ supports a period of slow-roll inflation
while the largest observable scales leave the horizon.
A slow-roll epoch (`SR') is characterized by
the conditions $\epsilon = \dot{\phi}^2/(2H^2 \Mp^2) \ll 1$,
and usually also $|\eta| = |\ddot{\phi} / (H \dot{\phi})| \ll 1$.
At some later time we assume $V(\phi)$ flattens abruptly.
In this phase there is no potential
gradient to support
the inflaton velocity, which therefore decays rapidly
with $\eta \approx -6$.
This epoch is described as \emph{ultra-slow-roll inflation}
(`USR').
Finally, when the velocity has
relaxed
to a suitable value,
there is a second phase of slow-roll inflation.
We distinguish the two slow-roll phases
using the labels `$1$' and `$2$'.
In particular,
$\epsilon_1$ and $\epsilon_2$
denote the $V$-type $\epsilon$
parameter in the first and second slow-roll epochs,
\begin{equation}
    \epsilon_1 = ( \epsilon_V )_1,
    \quad
    \epsilon_2 = ( \epsilon_V )_2 .
\end{equation}
We assume the second slow-roll epoch to be (eventually)
terminated
with $\epsilon \sim 1$, leading to a graceful exit.

The transitions between these phases are of particular
importance. 
Kristiano \& Yokoyama assumed an instant transition between
ultra-slow-roll and the final slow-roll phase~\cite{Kristiano:2022maq}.
They used in--in methods
to calculate the one-loop correction
due to a (bulk) cubic
interaction proportional to $\eta'$, which
has a large $\delta$-function spike in the
limit of an instantaneous transition.
Under certain assumptions,
they found that the one-loop correction
might become comparable to the tree-level.
The same scenario featuring an instantaneous transition
was also used by Firouzjahi~\cite{Firouzjahi:2023aum, Firouzjahi:2023bkt},
leading to a similar conclusion.
Riotto~\cite{Riotto:2023gpm} and Firouzjahi~\cite{Firouzjahi:2023aum,Firouzjahi:2023bkt}
suggested that the correction might be suppressed when the transition is smooth.
(See also Firouzjahi \& Riotto, Ref.~\cite{Firouzjahi:2023ahg}. We make a detailed
comparison with this reference in~\S\ref{sec:delta-N-back-reaction} and~\S\ref{sec:delta N loops calculation}.)

Franciolini {\etal} modelled the time evolution of $\eta$
using an analytic function that allows for a smooth
transition~\cite{Franciolini:2023lgy}.
They found that, after fixing the final abundance of primordial
black holes, the size of the one-loop correction is not significantly reduced by considering smooth evolution.
Nevertheless, these numerical results show that the loop estimate given by Kristiano \& Yokoyama is too large, even for models with instantaneous transitions.
In Ref.~\cite{Davies:2023hhn}, Davies {\etal} reached similar conclusions by numerically evaluating the one-loop term directly from a single-field potential yielding ultra-slow-roll.

In the following, we adopt the framework
introduced by Cai {\etal}~\cite{Cai:2018dkf} 
to describe the transition from ultra-slow-roll to
slow-roll, which we
summarize below in~\S\ref{sec: delta N peak mode from hc}.
Our treatment will apply to instantaneous and smooth transitions.
We discuss how the resulting one-loop corrections depend the
character of the transition in~\S\ref{sec:delta N loops calculation}.

To proceed, we label the e-folding time
at the start of the ultra-slow-roll phase
by $N = N_s$,
and
use $t = t_s$ and $\tau = \tau_s$
to denote the same time in cosmic time and conformal time,
respectively.
We write the wavenumber leaving the horizon
at this time as $k_s = (aH)|_{t = t_s}$.
We take
the
transition
at the end of ultra-slow-roll
to occur at time $N = N_e$
(or $t = t_e$, $\tau = \tau_e$),
and write the wavenumber leaving the horizon as $k_e$.%
    \footnote{
    \label{footnote on criterion to define USR}
    In practice, when dealing with a smooth transition, it is necessary
    to introduce an additional criterion to identify the ultra-slow-roll phase.
    See the discussion in Appendix~\ref{sec: numerical example}.}

In the following sections we use the $\delta N$
formula to compute the tree-level $\zeta$
power spectrum
for modes of order the CMB scale
(\S\ref{sec: delta N CMB mode from hc}),
and
those of order the peak scale
(\S\ref{sec: delta N peak mode from hc}).
In each case,
we must select an initial time $t_\ast$
at which to base
the separate universe formula~\eqref{eq:zeta-separate-universe}.
In~\S\S\ref{sec: delta N CMB mode from hc}--\ref{sec: delta N peak mode from hc}
we take this to be the natural initial time,
a little after horizon exit for the corresponding mode.
However, when calculating loop corrections
in~\S\ref{sec:delta N loops calculation}
it will be necessary to
set $t_\ast$ equal to the horizon-crossing
time for the peak modes,
even when computing the power spectrum on CMB scales.
In \S\ref{sec: delta N CMB mode from USR time}
we demonstrate that
this calculation
yields the same result as~\S\ref{sec: delta N CMB mode from hc}.
This version of the
argument will be
critical for the
discussion in~\S\ref{sec:delta N loops calculation}.

\subsection{CMB scales based at horizon exit}
\label{sec: delta N CMB mode from hc}
First, consider fluctuations on a scale
that will be imprinted on the CMB, assumed to exit the horizon
during the first slow-roll
epoch.
Throughout this section
we continue to use the notation introduced
in~\S\ref{sec:back-reaction-quasiparticles},
in which $\vect{p}$ labels the wavevector for a large-scale
mode (here representing the CMB scale),
and $\vect{k}$ labels a wavevector near the short-scale
peak.
The corresponding horizon exit times are written
$t_p$ and $t_k$, respectively.

In this section the power spectrum will be computed
up to tree level using
the separate universe formula~\eqref{eq:zeta-separate-universe}
based at the horizon exit time $t_p$.%
    \footnote{In presentations of the separate universe method
    that invoke classical arguments, it is sometimes
    suggested that one should wait until a few e-folds
    \emph{after} horizon exit before applying a formula
    such as Eq.~\eqref{eq:zeta-separate-universe}.
    Ref.~\cite{Dias:2012qy} argued, based on a direct
    analysis of the evolving correlation functions and
    \emph{without}
    invoking a classical model,
    that to obtain the most accurate
    estimate one should begin the
    separate universe calculation exactly at horizon exit
    (here $t = t_p$), but drop decaying modes in the
    correlation functions of phase-space variables.}
The analysis is standard,
and we briefly recall only the major steps.
During a slow-roll phase, the inflaton velocity is locked
to the gradient of the potential.
Introducing the momentum $\pi \equiv \d \phi / \d N$, this condition
can be written
\begin{equation}
    \label{eom SR}
    \pi = - \Mp \sqrt{2 \epsilon_1} .
\end{equation}
We have used the relationship $3H^2 \Mp^2 \approx V$.
If we further assume
that $\epsilon_1$ is approximately constant
during the first slow-roll epoch then
Eq.~\eqref{eom SR} can be integrated
to obtain a formula
for $N^{(t_p, t)}$, 
\begin{equation}
    \label{eq:N-CMB-hc-def}
    N^{(t_p, t)} \approx \frac{1}{\sqrt{2 \epsilon_1}}
    \frac{\phi(t_p) - \phi(t)}{\Mp} \;, 
\end{equation}
where $t>t_p$ is a time during the first slow-roll era. 
Note that the notation used here
matches Eqs.~\eqref{eq:NI-def}--\eqref{eq:NIJ-def}. 
We are assuming $\phi$ rolls from large to small
expectation values.
If the opposite is true, a sign flip is needed
in this expression.

\para{Evolution up to end of first slow-roll epoch}
Eq.~\eqref{eq:N-CMB-hc-def}
measures the number of e-folds that elapse between
two hypersurfaces on which
$\phi$ assumes prescribed values
$\phi(t_p)$ and $\phi(t)$.
This is not strictly the quantity that appears in
the $\delta N$ formula~\eqref{eq:zeta-separate-universe},
in which the final surface should be taken to
have uniform density. In general $\rho = \dot{\phi}^2/2 + V$,
and therefore hypersurfaces of uniform density do not coincide
with hypersurfaces of fixed $\phi$.
However, during slow-roll inflation
$\dot{\phi}^2/2$ is small compared with $V$.
Therefore the difference is $\Or(\epsilon)$ and can be neglected if we aim only for expressions
valid to leading order in the slow-roll parameters.

Outside the horizon,
the perturbations begin to evolve like
the background field.
Variation of Eq.~\eqref{eom SR} shows that
$\delta\pi$
becomes locked to the
field fluctuation
$\delta\phi$, with its amplitude suppressed
by a slow-roll factor,
\begin{equation}
    \label{eq:pi-SR-attractor}
    \delta \pi = 
    \frac{\eta}{2} \delta \phi + \text{decaying} ,
\end{equation}
where
$\eta$ is given in Eq.~\eqref{slow roll paramters},
and
`decaying' represents a decaying mode that is
exponentially suppressed on the slow-roll attractor.

We conclude that only the $\langle \phi \phi \rangle$
correlator is needed in the tree-level
formula~\eqref{eq:prototype-tree}.
The coefficient $N_\phi^{(t_p, t)}$
can be obtained from variation of~\eqref{eq:N-CMB-hc-def}
with respect to its initial condition
$\phi(t_p)$,
taking the final density---and hence $\phi(t)$---to be fixed,
\begin{equation}
    \label{N phi SR}
    N_\phi^{(t_p, t)} \equiv \frac{\delta N(t)}{\delta \phi(t_p)} \approx \frac{1}{\sqrt{2\epsilon_1}} \frac{1}{\Mp} .
\end{equation}
One can reproduce the expected result by
combining
Eq.~\eqref{massless correlator}
for the spectrum of $\phi$,
Eq.~\eqref{eq:prototype-tree}
for the $\langle \zeta \zeta \rangle$
correlation function,
and
Eq.~\eqref{N phi SR} for $N_\phi^{(t_p, t)}$, viz.,
\begin{equation}
    \label{Pz large scales}
    \dimlessP_\zeta(p;t)_\text{tree} = \frac{1}{8\pi^2 \epsilon_1} \frac{H(t_p)^2}{\Mp^2} . 
\end{equation}

\para{Evolution during ultra-slow-roll epoch}
After horizon-exit, $\dimlessP_\zeta$
becomes time-independent if the evolution
is adiabatic.
Alternatively,
if isocurvature modes are present, it may
evolve on a slow-roll timescale
until the onset of the ultra-slow-roll
phase.
At that time a different analysis is required
to determine its evolution.

During ultra-slow-roll the potential gradient is very shallow,
so that $(\epsilon_V)_{\text{USR}}$ is extremely small.
This can be occur,
for example, if the potential
includes almost-stationary point of
inflection (see, e.g.,~\cite{Garcia-Bellido:2017mdw,Germani:2017bcs,Ballesteros:2017fsr}), or a bump or dip~\cite{Atal:2019cdz, Mishra:2019pzq}.
At the beginning of the ultra-slow-roll phase,
the inflaton velocity is (comparatively) large,
since it is controlled by $\epsilon_1 \gg (\epsilon_V)_{\text{USR}}$.
The potential is no longer
sufficient to support this velocity,
so $\dot{\phi}$
must be balanced by the acceleration
term rather than $V'$,
\begin{equation}
    \label{eq:USR-field-equation}
    \ddot{\phi} + 3 H \dot{\phi} \approx 0 .
\end{equation}
Notice that the transition must be reasonably
abrupt in order that $\dot{\phi}$ does not
have time to smoothly relax
to a value that can be supported by the new shallow
gradient $(\epsilon_V)_{\text{USR}}$.
If that were to happen, Eq.~\eqref{eq:USR-field-equation}
would no longer apply
and we would merely have a phase of very shallow
slow-roll inflation.

The solution to Eq.~\eqref{eq:USR-field-equation}
can be written
\begin{equation}
    \label{eq:USR-field-solution}
    \phi \approx A + B \e{-3 H (t - t_0)}
    \approx A + B \e{-3 (N - N_0)}
    \approx A + B' \sqrt{2\epsilon}
\end{equation}
where $A$, $B$ (and $B'$) are constants of integration,
and $t_0$, $N_0$ label an arbitrary reference time.
In the second approximate equality we have used
$\Delta N \approx H \Delta t$,
which follows because $H$ is nearly constant.
The final approximate equality uses
$\epsilon \propto \e{-6Ht}$,
which can be demonstrated using
the first relation in~\eqref{eq:USR-field-solution}.

The transition between slow-roll and
ultra-slow-roll
occurs on a hypersurface
$\phi(t_s) = \text{const}$,
determined by the position on the potential
where the gradient rapidly decreases.
Up to the transition point, the slow-roll
attractor operates to lock the momentum fluctuation
to the field fluctuation,
as in~\eqref{eq:pi-SR-attractor}.
At the transition point, this
typically makes $\delta \pi$ a few orders of magnitude
smaller than $\delta \phi$.
Meanwhile,
if it is not extinguished,
the decaying term is responsible for different regions
of the universe
entering the ultra-slow-roll phase with very
slightly different
values of the field momentum.
However, it rapidly becomes
negligible
provided the horizon-exit time $t_p$
occurs appreciably before the transition.

During ultra-slow-roll, Eq.~\eqref{eq:USR-field-solution}
requires the momentum to decay exponentially,
$\pi \propto \e{-3N}$.
Inverting this relation,
it follows that the number of e-folds that elapse
during the ultra-slow-roll phase can be written
\begin{equation}
    \label{eq:N-USR}
    N^{(t_s, t)} \approx \frac{1}{3} \ln \frac{\pi_s}{\pi(t)} ,
\end{equation}
where, as explained above,
$t_s$ labels the start of the ultra-slow-roll phase
and $\pi_s$ is the value of the momentum
at this time.
If the decaying mode is present, 
it will contribute a small $\delta N$
through variation in the initial
and final momenta.
To apply Eq.~\eqref{eq:N-USR}
to compute $\zeta$, we should choose the final
time $t$ to lie on a hypersurface
of fixed $\phi$.
These hypersurfaces practically coincide with
slices of fixed density, because the kinetic
energy is decaying exponentially.

We can now determine how
$\dimlessP_\zeta$ evolves at the transition
and during the ultra-slow-roll epoch.
The number of elapsed e-folds
$N^{(t_p, t)}$ should be broken into
a slow-roll contribution
$N^{(t_p, t_s)}$ from Eq.~\eqref{eq:N-CMB-hc-def},
and an ultra-slow-roll contribution
$N^{(t_s, t)}$ from Eq.~\eqref{eq:N-USR}.
These components join on the fixed surface $\phi(t_s) = \text{const}$.
Therefore, the only contribution
to $\delta N$ from the slow-roll part
is due to variation with respect to
$\phi(t_p)$.
Next, consider variation of the ultra-slow-roll part $N^{(t_s, t)}$.
We first ignore the decaying mode of the momentum,
so at the transition $\delta \pi_s$
would be determined by
$\delta \phi_s$.
During the ultra-slow-roll phase
$\delta \pi$
would continue to be determined by Eq.~\eqref{eq:pi-SR-attractor}.
It follows that there is no late-time contribution
from either
the initial or final slices in~\eqref{eq:N-USR}.
(Recall that $\phi(t_s)$ is fixed, and $\phi(t)$ is
also fixed if the final slice has constant energy density.)

We now restore the decaying mode,
which
generates a contribution to $\delta N$
from variation of~\eqref{eq:N-USR}
with respect to $\pi$,
on both the initial and final 
hypersurface.
This yields a \emph{growing}
contribution
to
$\langle \zeta \zeta \rangle$,
which is
proportional to $\pi^{-2}(t) = \epsilon^{-1}$.
During an ultra-slow-roll phase this grows like
$\e{6N}$.
This growing contribution is very significant for modes
that exit near the transition, or during the ultra-slow-roll
phase~\cite{Cai:2018dkf}.
(See \S\ref{sec: delta N peak mode from hc}.)
However,
for modes such as $\vect{p}$
that exit significantly before the transition,
suppression of the decaying mode is very strong.
In principle, its influence could be recovered
if the ultra-slow-roll phase lasts for a sufficiently
long time.
However,
in a model intended to produce a realistic
abundance of primordial black holes,
we
typically expect only
between $2$ and $2.5$ e-folds
of ultra-slow-roll inflation~\cite{Franciolini:2023lgy}.

The conclusion is that
provided the horizon exit time $t_p$
occurs more than a few e-folds
before the transition,
the decaying mode will be sufficiently
extinguished that even the rapid $\e{6N}$ enhancement
during ultra-slow-roll cannot compensate
for it.
Therefore $\dimlessP_\zeta$ will practically not evolve during the
whole ultra-slow-roll phase.
It will also not evolve during the final slow-roll epoch,
provided the evolution there is adiabatic.
The conclusion is that $\dimlessP_\zeta(p)$
for the CMB mode is given by~\eqref{Pz large scales},
with $t$ now labelling a time during the
second slow-roll epoch.
Up to this point our
analysis agrees with the discussion given
in Cai {\etal}~\cite{Cai:2018dkf}
and Firouzjahi \& Riotto~\cite{Firouzjahi:2023ahg}.

\subsection{Peak scales based at horizon exit}
\label{sec: delta N peak mode from hc}
Next, we calculate the tree-level power spectrum
for a short-scale mode, $\vect{k}$,
in the vicinity of the peak
in the power spectrum.
This mode is taken to cross during the ultra-slow-roll
phase.
The number of e-folds the elapse
from horizon exit
at $t = t_k$ up to some later time $t$---still assumed to be
within the ultra-slow-roll epoch---%
is determined by Eq.~\eqref{eq:N-USR}
with the replacement $t_s \rightarrow t_k$,
\begin{equation}
    \label{duration USR}
    N^{(t_k, t)} = \frac{1}{3} \ln \frac{\pi(t_k)}{\pi(t)} .
\end{equation}

Unlike the CMB-scale mode $\vect{p}$,
the $\delta \pi$ fluctuation
at $\vect{k}$ is independent
of the field fluctuation.
Therefore
Eq.~\eqref{duration USR}
yields a significant growing mode
during the ultra-slow-roll phase,
arising from
variation in~\eqref{duration USR}
between smoothed regions.
To determine the final value of
$\dimlessP_\zeta(k)$
we should follow the evolution of this
mode until the dynamics become
adiabatic.
This requires a detailed description
of the transition into the second slow-roll epoch.
Following Ref.~\cite{Cai:2018dkf} we write
the scalar field solution~\eqref{eq:USR-field-solution}
during ultra-slow-roll in terms of the
field and momentum
expectation values $\phi_e$, $\pi_e$
(respectively)
attained at time $t = t_e$,
at the end of ultra-slow-roll.
That yields
\begin{equation}
    \label{pi and phi USR}
    \phi(N) = \phi_e + \frac{\pi_e - \pi(N)}{3}
    \quad
    \text{and}
    \quad
    \pi(N) = \pi_e \e{-3 (N - N_e)} .
\end{equation}
We are not necessarily taking the transition into the
second slow-roll phase to be
abrupt, so it need not be meaningful
to talk about a precise moment where the transition
happens. For our purposes in this paper
we assume it is possible to
write down a condition that marks the division
between ultra-slow-roll and slow-roll, at least to a reasonable
approximation. 
(See the footnote \ref{footnote on criterion to define USR} on p.~\pageref{footnote on criterion to define USR}.)
Considering the exponential decay of the velocity during
ultra-slow-roll, this condition must identify a practically
unique value $\phi_e$
that does
not vary from region to region.
(If the phase of ultra-slow-roll inflation is extremely short,
however, it may be necessary to allow for this possibility.)

To describe the subsequent slow-roll epoch
we require information about the
potential
that supports it.
Our analysis again follows Cai {\etal}~\cite{Cai:2018dkf}.
Since $V(\phi)$ is assumed
to be fairly smooth,
it can be represented
near $\phi_e$ as a series,
\begin{equation}
    \label{V end USR}
    V(\phi)
    \approx
    V_e
    \Bigg[
        1
        + (2 \epsilon_2)^{1/2} \frac{\phi-\phi_e}{\Mp}
        + \frac{1}{2} \eta_2 \bigg( \frac{\phi-\phi_e}{\Mp} \bigg)^2
        + \cdots
    \Bigg]
    ,
\end{equation}
where $V_e \equiv V(\phi_e)$ is the fixed value of the potential
on the transition hypersurface $\phi = \phi_e$.
In this expression, $\epsilon_2 = (\epsilon_V)_2$
and $\eta_2 = (\eta_V)_2$
are taken to be roughly constant
during the second slow-roll epoch,
and to coincide with their values at
$\phi = \phi_e$.
We will use Eq.~\eqref{V end USR}
only on the slow-roll side of the transition.
The smoothness of this transition
is quantified by the dimensionless
parameter $h$~\cite{Cai:2018dkf}
\begin{equation}
    \label{h def}
    h \equiv 6 (2\epsilon_2)^{1/2} \frac{\Mp}{\pi_e} .
\end{equation}
This compares the actual inflation velocity
at the end of the ultra-slow-roll epoch
$\sim \pi_e$
to the natural value $\sim \epsilon_2^{1/2} \Mp$
in the slow-roll phase.
In the limit $h\rightarrow 0$,
the transition becomes increasingly smooth.
If the inflaton approaches the second slow-roll plateau with a velocity that matches the one on the second slow-roll attractor, $\pi_e=-(2\epsilon_2)^{1/2}\Mp$, this corresponds to an instantaneous transition, described by $h=-6$.

The solutions for $\phi$ and $\pi$
\emph{after} the transition
can be found by solving
the slow-roll equation~\eqref{eom SR}
using Eq.~\eqref{V end USR}.
The solutions matching $\phi_e$, $\pi_e$, respectively,
at the transition time $N = N_e$
can be written%
    \footnote{There is a typo in Eq.~(2.17) of
    Ref.~\cite{Cai:2018dkf},
    which
    is corrected in our Eq.~\eqref{pi N tr}.}
\begin{subequations}
\begin{align}
    \label{phi N tr}
    \phi(N)
    & =
    \begin{multlined}[t]
        \phi_e
        + \frac{2h\pi_e}{s^2-9}
        + \pi_e \frac{s-3-h}{s(s-3)}
            \exp\bigg[
                \frac{s-3}{2} (N-N_e)
            \bigg] \\
        - \pi_e \frac{s+3+h}{s(s+3)}
            \exp\bigg[
                {-\frac{s+3}{2}} (N-N_e)
            \bigg] ,
    \end{multlined}
    \\
    \label{pi N tr}  
    \pi(N)
    & =
    \pi_e
    \exp\bigg[
        {-\frac{3}{2}(N-N_e)}
    \bigg]
    \bigg[
        \cosh{\frac{s(N- N_e)}{2}}
        - \frac{3+h}{s} \sinh{\frac{s(N-N_e)}{2}}
    \bigg] ,
\end{align}
\end{subequations}
where $s$ is defined
to satisfy $s^2 \equiv 9-12 \eta_2$.
The final
term in Eq.~\eqref{phi N tr} is subleading
when $N$ is at least modestly larger than $N_e$.
Dropping this subleading contribution,
the number of e-folds $N^{(t_e, t)}$
that elapse
between hypersurfaces
corresponding to $\phi = \phi_e$
and $\phi = \phi(t)$,
can be found by inverting~\eqref{phi N tr},
\begin{equation}
    \label{duration tr SR 1}
    N^{(t_e, t)}
    \approx
    \frac{2}{s-3}
    \ln \Bigg[
        \bigg(
            \phi(t)-\phi_e - \frac{12 \Mp (2\epsilon_2)^{1/2}}{s^2-9}
        \bigg)
        \frac{s(s-3)}{\pi_e (s-3-h)}
    \Bigg] .
\end{equation}
The number of e-folds $N^{(t_k, t)}$
that elapse
from horizon exit of the mode $\vect{k}$
at time $t_k$,
up to some practically uniform-density slice
$\phi = \phi(t)$
during the second slow-roll era,
should be obtained by
summing~\eqref{duration USR}
(with the replacement $t \rightarrow t_e$)
and~\eqref{duration tr SR 1}.

To use~\eqref{eq:zeta-separate-universe}
to determine the time evolution
of $\zeta_{\vect{k}}$,
we vary
$N^{(t_k, t)}$ with respect
to the initial conditions
at $t = t_k$.
In doing so, we must keep
$\phi_e$ and $\phi(t)$ fixed.
Also, we are taking $\epsilon_2$
and $\eta_2$ to be determined by $\phi_e$,
so these slow-roll parameters are also fixed.
Only $\pi_e$ will vary;
it should be regarded as a function of the
initial values $\phi(N_k)$
and $\pi(N_k)$
via the solution for $\phi(N)$
in Eq.~\eqref{pi and phi USR}.
This implies
$\delta \pi_e/\delta \phi(N_k) = 3$,
$\delta \pi_e/\delta \pi(N_k) = 1$.
Taken together, these conditions
mean that
the only contribution
from Eq.~\eqref{duration tr SR 1}
comes from
the factor
$\pi_e (s - 3 - h)$
inside the logarithm.
Hence we may write~\cite{Cai:2018dkf,Firouzjahi:2023ahg}
\begin{equation}
    \label{eq:full-N-3phase-model}
    N^{(t_k, t)}
    \asymp
    \frac{1}{3} \ln \frac{\pi(N_k)}{\pi_e}
    - \frac{2}{s-3} \ln \bigg|
        \frac{\pi_e}{\Mp} (s - 3 - h)
    \bigg|
    \approx
    \frac{1}{3} \ln \frac{\pi(N_k)}{\pi_e}
    + \frac{1}{\eta_2} \ln \bigg[
        2 \eta_2 \frac{\pi_e}{\Mp} + 6 (2\epsilon_2)^{1/2}
    \bigg] .
\end{equation}
The symbol `$\asymp$'
is used to mean that this expression
produces the same \emph{derivatives}
as the full formula for $N^{(t_k, t)}$,
although it is not numerically equal.
The second approximate equality
follows by expanding $s$ to lowest order in $\eta_2$.
Eq.~\eqref{eq:full-N-3phase-model}
agrees with the expression Eq.~(II.21)
reported by Firouzjahi \& Riotto~\cite{Firouzjahi:2023ahg},
except that we do not have their first term
representing the number of e-folds accumulated
during the first slow-roll phase.
This was discussed above in~\S\ref{sec: delta N CMB mode from hc}.
However, it is not needed here:
Eq.~\eqref{eq:zeta-separate-universe-fourier}
requires that we base our separate universe
formula at a time $t_\ast \sim t_k$,
just a little later than the horizon exit time of the peak modes,
which is already within the ultra-slow-roll phase. See~\S\ref{sec:delta-N-back-reaction} for more details. 

Taking variations with respect to the $\phi$ and $\pi$
fields at this time, we conclude
\begin{subequations}
\begin{align}
    \label{N_phi}
    N^{(t_k, t)}_{\phi}
    & =
    -\frac{1}{\pi_e} + \frac{3}{3 \Mp (2 \epsilon_{2})^{1/2} + \eta_{2} \pi_e}
    =
    \frac{1}{\pi_e} \bigg(
        - 1 + \frac{6}{h + 2\eta_2}
    \bigg) ,
    \\
    \label{N_pi}
    N^{(t_k, t)}_{\pi}
    & = \frac{1}{3 \pi(N_k)} -\frac{1}{3 \pi_e} + \frac{1}{3 \Mp (2 \epsilon_{2})^{1/2} + \eta_{2}\pi_e}
    =
    \frac{1}{3\pi_e} \bigg(
        \frac{\pi_e}{\pi(N_k)} - 1
        + \frac{6}{h + 2\eta_2}
    \bigg)
    .
\end{align}
\end{subequations}
For both $N_\phi$ and $N_\pi$,
we have re-introduced the smoothness parameter $h$
in the second equality.
This makes the dependence of the qualitative nature
of the transition explicit.

Notice that, although our background model~\eqref{eq:full-N-3phase-model}
agrees with Firouzjahi \& Riotto~\cite{Firouzjahi:2023ahg},
the derivatives Eq.~\eqref{N_phi}--\eqref{N_pi}
do not agree with the
derivatives reported by these authors.
Firouzjahi \& Riotto reported $N_\pi \approx 0$
and gave an expression for $N_\phi$---Eq.~(II.24) of
Ref.~\cite{Firouzjahi:2023ahg}---%
that agrees with our
formula~\eqref{N phi SR}.
This is because Ref.~\cite{Firouzjahi:2023ahg}
always elected to base the separate universe
formula at a time $t_\ast = t_p$
close to horizon exit for the CMB-scale $\vect{p}$-mode.
(See~\S\ref{sec:delta-N-back-reaction}.)
At this time the $\delta \pi$ mode can be neglected,
as discussed above.
The $\delta N$ coefficients needed
for our Eq.~\eqref{eq:zeta-separate-universe-fourier}
are those of Eqs.~\eqref{N_phi}--\eqref{N_pi}.

If the transition is at least modestly smooth,
the denominator of $6/(h+2\eta_2)$
is small.
Therefore this term gives the dominant
contribution to both derivatives.
If $h$ is not significantly smaller
that $\eta_2$,%
    \footnote{For a discussion of the $h\ll1$ expansion,
    see footnote \ref{footnote on h small expansion}
    on p.~\pageref{footnote on h small expansion}.}
we find
\begin{equation}
    \label{simplify N_phi}
    N_\phi^{(t_k, t)} \approx
    \frac{6}{h \pi_e}
    \approx \frac{1}{\Mp} \frac{1}{(2\epsilon_2)^{1/2}} .
\end{equation}
A similar expression can be given for $N_\pi{(t_k, t)}$,
but to calculate $\dimlessP_\zeta(k; t)$
for modes that cross the horizon during ultra-slow-roll
we do not need it.
For these modes,
the field perturbation
primarily populates the constant $A$-mode
of Eq.~\eqref{eq:USR-field-solution}.
Accordingly, the momentum perturbation
decays outside the horizon.
It follows that the tree-level power spectrum
Eq.~\eqref{eq:prototype-tree}
can be written
\begin{equation}
    \label{Pzeta peak scale}
    \dimlessP_\zeta(k; t)_\text{tree}
    =
    \Big[ N^{(t_k, t)}_\phi \Big]^2
    \dimlessP_{\phi\phi}(k; t_k) = \frac{1}{8\pi^2 \epsilon_2} \frac{H(t_k)^2}{\Mp^2}
    \quad
    \text{(assuming $t > t_e$)} , 
\end{equation}
where we have used~\eqref{simplify N_phi}.

We emphasize that~\eqref{Pzeta peak scale}
applies only after the transition into the second
slow-roll epoch.
At earlier times, $\dimlessP_\zeta$
is roughly given by
$\dimlessP_\zeta \approx \dimlessP_{\phi\phi} / (2 \Mp^2 \epsilon)$.
This is rapidly growing, because the factor
$\epsilon$ is decaying like $\e{-6N}$.
The late-time value therefore
should be evaluated when the
dynamics become adiabatic.
In this model, that occurs after the transition
to the second slow-roll epoch,
when $\epsilon \approx \epsilon_2$.
This provides a qualitative explanation
for the result of Eq.~\eqref{Pzeta peak scale}.

\subsection{CMB scales based at exit time for peak scales}
\label{sec: delta N CMB mode from USR time}

Finally, we demonstrate how to compute the tree-level power spectrum
for the CMB-scale mode $\vect{p}$,
but now using the separate universe
formula~\eqref{eq:zeta-separate-universe}
based at $t_\ast = t_k$,
the horizon-crossing time for the peak-scale modes.
According to~\eqref{eq:zeta-separate-universe-fourier}
it is this arrangement that is required
to compute the back-reaction from enhanced modes~$\vect{k}$.
Clearly, the answer must agree with our previous
explicit calculation~\eqref{Pz large scales}.

Eq.~\eqref{N phi SR}
shows that,
for a mode $\vect{p}$
that exits during the first slow-roll
phase,
$\zeta_{\vect{p}}
= \delta \phi / (2 \Mp^2 \epsilon)^{1/2}$.
Since $\zeta_{\vect{p}}$ is conserved on superhorizon
scales during this epoch,
it follows that (to linear order) the growing mode
of $\delta\phi$,
evaluated at the transition $t=t_s$
into the
ultra-slow-roll phase, can be written%
    \footnote{Recall that in~\S\ref{sec: delta N CMB mode from hc}
    we concluded the decaying mode could be safely dropped
    for modes that exit the horizon at least a few e-folds
    before the transition.}
\begin{equation}
    \label{delta phi large scale SR}
    \delta \phi_{\vect{p}}(t_s)
    =
    \Bigg(
        \frac{\epsilon(t_s)}{\epsilon(t_p)}
    \Bigg)^{1/2}
    \delta \phi_{\vect{p}}(t_p) .
\end{equation}
After the transition,
$\delta \phi$ evolves
according to the solution~\eqref{eq:USR-field-solution}.
Assuming $\epsilon$ varies continuously, and
matching smoothly across
the transition,
it follows that $\delta \phi$
will
populate only the $B$-mode of this solution.
Hence, at some time $t$
during the ultra-slow-roll phase, we have
\begin{subequations}
\begin{align}
    \label{delta phi CMB mode later time}
    \delta \phi_{\vect{p}}(t)
    & =
    \Bigg(
        \frac{\epsilon(t)}{\epsilon(t_p)}
    \Bigg)^{1/2}
    \delta \phi_{\vect{p}}(t_p) \\
    \label{delta pi CMB mode later time}
    \delta \pi_{\vect{p}}(t)
    & =
    \frac{\eta(t)}{2} \delta \phi_{\vect{p}}(t)
    \approx -3 \delta \phi_{\vect{p}}(t) .
\end{align}
\end{subequations}
In the final step we have used $\eta \approx -6$,
which is valid during ultra-slow-roll.
This is very different to
the behaviour of perturbations
that exit in the ultra-slow-roll phase,
described in~\S\ref{sec: delta N peak mode from hc}.
For an explicit numerical example, see Fig.~\ref{fig:correlators CMB and peak}.
For those $\vect{k}$-scale
modes, the field perturbation
is primarily in the constant $A$-mode,
and $\delta \pi$ rapidly becomes small.
In contrast,
for the $\vect{p}$-scale modes,
the field perturbation is in the $B$-mode
and
$\delta \pi$ rapidly \emph{grows}
during the transition to ultra-slow-roll,
becoming
comparable to (in fact, slightly
larger than) $\delta \phi$.
It follows that we \emph{cannot} neglect the
momentum perturbation
when the separate universe formula is based
at $t = t_k$.

The power spectrum is readily computed
from Eqs.~\eqref{N_phi}--\eqref{N_pi}.
Working from
Eq.~\eqref{eq:prototype-tree}
and substituting for $\delta\phi$,
$\delta\pi$ using
Eqs.~\eqref{delta phi CMB mode later time}--\eqref{delta pi CMB mode later time},
\begin{equation}
    \dimlessP_\zeta(p;t)_\text{tree}
    =
    \Big[
        N_\phi^{(t_k, t)}
    \Big]^2
    \dimlessP_{\phi\phi}(p, t_k)
    +
    2
    N_\phi^{(t_k, t)}
    N_\pi^{(t_k, t)}
    \dimlessP_{\phi\pi}(p, t_k)
    +
    \Big[
        N_\pi^{(t_k, t)}
    \Big]^2
    \dimlessP_{\pi\pi}(p, t_k) .
\end{equation}
We have used $\dimlessP_{\pi\phi}(p, t_k) = \dimlessP_{\phi\pi}(p, t_k)$,
which is
approximately true because $\phi$ and $\pi$ nearly commute
outside the horizon.
There is significant cancellation between the
$\delta\phi$ and $\delta\pi$
terms in this expression. We find
\begin{equation}
    \label{Pz CMB from later times}
    \dimlessP_\zeta(p;t)_{\text{tree}}
    =
    \frac{1}{\pi(N_k)^2} \dimlessP_{\phi\phi}(p, t_k)
    =
    \frac{1}{2 \epsilon(N_k) \Mp^2}
    \frac{\epsilon(N_k)}{\epsilon(t_p)}
    \frac{H(t_p)^2}{4\pi^2}
    =
    \frac{1}{8\pi^2 \epsilon(t_p)} \frac{H(t_p)^2}{\Mp^2} .
\end{equation}
In the final step we have used $\pi(N_k) = - \Mp [2\epsilon(N_k)]^{1/2}$,
which follows from~\eqref{eom SR}
and~\eqref{eq:USR-field-solution}.
Since $\epsilon(t_p) \approx \epsilon_1$,
this reproduces our earlier expression~\eqref{Pz large scales}.

The primary conclusion from this analysis is that
a correct outcome for the $\vect{p}$-scale
mode requires inclusion of the momentum
fluctuation
when the separate universe expansion is based
near the horizon exit time for the short modes.
This calculation also provides a sanity-check of the $\delta N$ expressions  
which we will use in~\S\ref{sec:delta N loops calculation} to calculate the one-loop corrections.

\section[\texorpdfstring{Case study: $\bm{\delta N}$ loops in the 3-phase ultra-slow-roll model}{Case study: delta N loops in the 3-phase ultra-slow-roll model}]{Case study: $\bm{\delta N}$ loops in the 3-phase ultra-slow-roll model}
\label{sec:delta N loops calculation}

In this section, we apply the general
formula~\eqref{eq:zeta-separate-universe-fourier} to
compute the one-loop correction for the
3-phase ultra-slow-roll model
of~\S\ref{sec: SR to USR to SR}. 
In~\S\ref{sec: 12 type loop}, \S\ref{sec: 22 type loop}
and \S\ref{sec: 13 type loop}
we give explicit expressions for the enhanced part
of the 12-loop, 22-loop,
and 13-loop,
respectively,
as defined in~\S\ref{sec: loops in delta N}. 

When
using~\eqref{eq:zeta-separate-universe-fourier}
to compute one-loop corrections, or correlation functions
in general, we are obliged to carry out
Wick contractions among the phase space fluctuations.
These are best visualized using a diagrammatic
expansion.
In the case of the 3-phase model,
the phase space coordinates $X^I$ comprise
a field value $\phi$ and its momentum $\pi$,
and
the corresponding diagrammatic representation is summarized in
Fig.~\ref{fig:deltaN-diagrams}.
In the following sections we use these diagrams
to enumerate the contractions that contribute to each type
of loop.
\begin{figure}
    \begin{equation*}
        \mspace{-100mu}
        \begin{split}
            \zeta_{\vect{p}}(t) =
            \mbox{}
            &
            \mspace{5mu}
            \eqgraph{1ex}{0ex}{
                \begin{fmfgraph*}(60,30)
                    \fmfleft{l}
                    \fmfright{r}
                    \fmf{plain}{l,v}
                    \fmf{dashes}{v,r}
                    \fmfv{label=$[\smoothSmall{\delta\phi^{(t_k)}}]_{\vect{p}}$}{r}
                    \fmfv{label=$N_\phi^{(t_k\comma t)}$,label.angle=90,
                        decor.shape=circle,decor.filled=full,decor.size=2thick}{v}
                \end{fmfgraph*}
            }
            \mspace{105mu}
            +
            \mspace{5mu}
            \eqgraph{1ex}{0ex}{
                \begin{fmfgraph*}(60,30)
                    \fmfleft{l}
                    \fmfright{r}
                    \fmf{plain}{l,v}
                    \fmf{dbl_dots}{v,r}
                    \fmfv{label=$[\smoothSmall{\delta\pi^{(t_k)}}]_{\vect{p}}$}{r}
                    \fmfv{label=$N_\pi^{(t_k\comma t)}$,label.angle=90,
                        decor.shape=circle,decor.filled=full,decor.size=2thick}{v}
                \end{fmfgraph*}
            }
            \\
            & +
            \frac{1}{2!} \int \frac{\d^3 q}{(2\pi)^3}
            \mspace{10mu}
            \eqgraph{3ex}{3ex}{
                \begin{fmfgraph*}(60,50)
                    \fmfleft{l}
                    \fmfright{r1,r2}
                    \fmf{plain}{l,v}
                    \fmf{dashes}{v,r1}
                    \fmf{dashes}{v,r2}
                    \fmfv{label=$[\smoothSmall{\delta\phi^{(t_k)}}]_{\vect{q}}$,label.angle=0}{r2}
                    \fmfv{label=$[\smoothSmall{\delta\phi^{(t_k)}}]_{\vect{p}-\vect{q}}$,label.angle=0}{r1}
                    \fmfv{label=$N_{\phi\phi}^{(t_k\comma t)}$,label.angle=120,
                        decor.shape=circle,decor.filled=full,decor.size=2thick}{v}
                \end{fmfgraph*}
            }
            \mspace{105mu}
            +
            \frac{1}{2!} \int \frac{\d^3 q}{(2\pi)^3}
            \mspace{10mu}
            \eqgraph{3ex}{3ex}{
                \begin{fmfgraph*}(60,50)
                    \fmfleft{l}
                    \fmfright{r1,r2}
                    \fmf{plain}{l,v}
                    \fmf{dbl_dots}{v,r1}
                    \fmf{dbl_dots}{v,r2}
                    \fmfv{label=$[\smoothSmall{\delta\pi^{(t_k)}}]_{\vect{q}}$,label.angle=0}{r2}
                    \fmfv{label=$[\smoothSmall{\delta\pi^{(t_k)}}]_{\vect{p}-\vect{q}}$,label.angle=0}{r1}
                    \fmfv{label=$N_{\pi\pi}^{(t_k\comma t)}$,label.angle=120,
                       decor.shape=circle,decor.filled=full,decor.size=2thick}{v}
                \end{fmfgraph*}
            }
            \mspace{50mu}
            +
            \cdots
            \\
            & +
            \frac{1}{3!}
            \int \frac{\d^3 q_1}{(2\pi)^3}
            \int \frac{\d^3 q_2}{(2\pi)^3}
            \mspace{10mu}
            \eqgraph{1ex}{0ex}{
                \begin{fmfgraph*}(60,50)
                    \fmfleft{l}
                    \fmfright{r1,r2,r3}
                    \fmf{plain}{l,v}
                    \fmf{dashes}{v,r1}
                    \fmf{dashes}{v,r2}
                    \fmf{dashes}{v,r3}
                    \fmfv{label=$[\smoothSmall{\delta\phi^{(t_k)}}]_{\vect{q}_1}$,label.angle=0}{r3}
                    \fmfv{label=$[\smoothSmall{\delta\phi^{(t_k)}}]_{\vect{q}_2}$,label.angle=0}{r2}
                    \fmfv{label=$[\smoothSmall{\delta\phi^{(t_k)}}]_{\vect{p}-\vect{q}_1-\vect{q}_2}$,label.angle=0}{r1}
                    \fmfv{label=$N_{\phi\phi\phi}^{(t_k\comma t)}$,label.angle=120,
                        decor.shape=circle,decor.filled=full,decor.size=2thick}{v}
                \end{fmfgraph*}
            }
            \mspace{110mu}
            +\cdots
        \end{split}
    \end{equation*}
    {\footnotesize
        \semibold{Key}
        \begin{flalign*}
            \eqgraph{0ex}{0ex}{\begin{fmfgraph*}(15,5)\fmfleft{l}\fmfright{r}\fmf{plain}{l,r}\end{fmfgraph*}}
                & \mspace{30mu} \text{long wavelength field $\zeta_{\vect{p}}$} \\
            \eqgraph{0ex}{0ex}{\begin{fmfgraph*}(15,5)\fmfleft{l}\fmfright{r}\fmf{dashes}{l,r}\end{fmfgraph*}}
                & \mspace{30mu} \text{field fluctuation $\delta\phi$} \\
            \eqgraph{0ex}{0ex}{\begin{fmfgraph*}(12,5)\fmfleft{l}\fmfright{r}\fmf{dbl_dots}{l,r}\end{fmfgraph*}}
                & \mspace{30mu} \text{momentum fluctuation $\delta\pi$} \\
            \eqgraph{0ex}{0ex}{\begin{fmfgraph*}(8,8)\fmfiv{decor.shape=circle,decor.filled=full,decor.size=2thick}{c}\end{fmfgraph*}}
                & \mspace{30mu} \text{contraction between the $\delta N$ coefficients $N_{I}$, $N_{IJ}$, \ldots,
                and fluctuations $X^I = \{ \delta\phi, \delta\pi \}$} \\
            N_\phi^{(t_k, t)}
                & \mspace{30mu} \delta N(t) / \delta \phi(t_k)
                \mspace{10mu} \text{(similarly for higher-order derivatives)} \\
            N_\pi^{(t_k, t)}
                & \mspace{30mu} \delta N(t) / \delta \pi(t_k)
                \mspace{10mu} \text{(similarly for higher-order derivatives)} &
        \end{flalign*}
    }
    \caption{\label{fig:deltaN-diagrams}
    Diagrammatic form of the separate universe formula for
    $\zeta_{\vect{p}}$, Eq.~\eqref{eq:zeta-separate-universe-fourier},
    for a single-field model with a transient phase of ultra-slow-roll.
    The $\delta N$ expansion has been based
    at the horizon-crossing time $t_k$ of the peak-scale modes.}
\end{figure} 

\subsection{12-loop}
\label{sec: 12 type loop}
The 12-type loop~\eqref{eq:prototype-12} was discussed
in~\S\ref{sec:delta-N-back-reaction}.
It corresponds to the first diagram in Fig.~\ref{fig:one-loop-delta-N}.
Here, we specialize it to
the 3-phase model.
In Eq.~\eqref{eq:prototype-12}, the large-scale
fluctuation $\delta X^I_{\vect{p}}(t_{k})$ can be
of either field or velocity type:
they are both relevant and comparable at time $t_{k}$.
(See Eqs.~\eqref{delta phi CMB mode later time}--\eqref{delta pi CMB mode later time},
which are valid for a large-scale mode during ultra-slow-roll.)
The convolution integral runs over momenta $q\lesssim k$,
where $k$
continues to identify the peak scales.
In this section,
when making explicit estimates, we integrate
loop momenta from the first mode
leaving the horizon
during ultra-slow-roll ($q = k_s$), up to the last such mode ($q = k_e$).
Hence, the generic scale $k$ can be interpreted as $k \sim k_e$,
and likewise $t_k \sim t_{k_e}$.

The hierarchy $p \ll q$
implies that the momenta
$\vect{q}$ and $-\vect{p}-\vect{q}$
appearing in
Eq.~\eqref{eq:prototype-12} correspond to peak scales. 
For these, the velocity
auto- and cross-correlations become negligible
soon after horizon crossing.
Therefore we can neglect the velocity fluctuations.
For an explicit numerical example, see the bottom panel
of Fig.~\ref{fig:correlators CMB and peak}.
It follows that the 12-type contribution
to the $\langle \zeta \zeta \rangle$
correlation function can be written 
\begin{multline}
    \label{type 1 one}
    \langle \zeta_{\vect{p}} \zeta_{-\vect{p}}\rangle_\text{12} =
    N_{\phi\phi}^{(t_k, t)}
    \int\limits_{k_s\lesssim q\lesssim k_e} \frac{\d^3q}{(2\pi)^{3}}\\
    \times \left(  N_\phi^{(t_k,t)}  \langle \delta \phi_{\vect{p}}{(t_{k})} \delta \phi_{\vect{q}}{(t_{k})} \delta \phi_{-\vect{p}-\vect{q}}{(t_{k})} \rangle + N_\pi^{(t_k,t)}\langle \delta \pi_{\vect{p}}{(t_{k})} \delta \phi_{\vect{q}}{(t_{k})} \delta \phi_{-\vect{p}-\vect{q}}{(t_{k})} \rangle \right) \;.
\end{multline}
In the language of Fig.~\ref{fig:deltaN-diagrams}, this can be represented diagrammatically as 
\begin{equation}
    \langle \zeta_{\vect{p}} \zeta_{-\vect{p}} \rangle_{\text{12}}
    =
    \eqgraph{1ex}{0ex}{
        \begin{fmfgraph*}(100,50)
            \fmfleft{l}
            \fmfright{r}
            \fmf{plain}{l,v1}
            \fmf{dashes}{v1,v2}
            \fmf{dashes,left,tension=0.3}{v2,v3}
            \fmf{dashes,left,tension=0.3}{v3,v2}
            \fmf{plain}{v3,r}
            \fmfv{label=$N_{\phi}^{(t_{{k}},,t)}$,label.angle=90,
                decor.shape=circle,decor.filled=full,decor.size=2thick}{v1}
            \fmfv{label=$N_{\phi\phi}^{(t_{{k}},,t)}$,label.angle=45,
                decor.shape=circle,decor.filled=full,decor.size=2thick}{v3}
        \end{fmfgraph*}
    }
    \mspace{20mu}
    +
    \mspace{5mu}
    \eqgraph{1ex}{0ex}{
        \begin{fmfgraph*}(100,50)
            \fmfleft{l}
            \fmfright{r}
            \fmf{plain}{l,v1}
            \fmf{dbl_dots}{v1,v2}
            \fmf{dashes,left,tension=0.3}{v2,v3}
            \fmf{dashes,left,tension=0.3}{v3,v2}
            \fmf{plain}{v3,r}
            \fmfv{label=$N_{\pi}^{(t_{{k}},,t)}$,label.angle=90,
                decor.shape=circle,decor.filled=full,decor.size=2thick}{v1}
            \fmfv{label=$N_{\phi\phi}^{(t_{{k}},,t)}$,label.angle=45,
                decor.shape=circle,decor.filled=full,decor.size=2thick}{v3}
        \end{fmfgraph*}
    }\quad.
\end{equation}
We can rewrite~\eqref{type 1 one} in a more compact form
by using Eq.~\eqref{delta pi CMB mode later time},
\begin{equation}
    \label{type 1 two}
    \langle \zeta_{\vect{p}} \zeta_{-\vect{p}}\rangle_\text{12} =N_{\phi\phi}^{(t_k, t)} \left(  N_\phi^{(t_k,t)} + \frac{\eta}{2} N_\pi^{(t_k,t)} \right) \int\limits_{k_s\lesssim q\lesssim k_e} \frac{\d^3q}{(2\pi)^{3}}\,
    \langle \delta \phi_{\vect{p}}{(t_{k})} \delta \phi_{\vect{q}}{(t_{k})} \delta \phi_{-\vect{p}-\vect{q}}{(t_{k})} \rangle  \; .
\end{equation}
The prefactor
$N_{\phi}^{(t_k, t)} + \eta N_{\pi}^{(t_k, t)}/2$
appears frequently
and is equal to the
\emph{total} field-space variation
$\delta N^{(t_k, t)}/\delta \phi$
assuming the slow-roll attractor condition.
As would be expected,
in a single-field model on this attractor,
we could (if we wished) eliminate
the phase-space
momentum $\delta\pi$.

The three-point correlation function
appearing in~\eqref{type 1 two}
is in a squeezed configuration
where
$p \ll q$ and $|\vect{p}+\vect{q}|\simeq q$.
It acts as a book-keeper of the hierarchy between large and small scales.
On physical grounds we expect it must (formally)
vanish in the limit of an infinite separation
where $p/k\to0$.
In this limit we expect that all back-reaction decouples and the 12-loop
should vanish.

In Ref.~\cite{Kenton:2015lxa}
this squeezed
three-point function
was estimated using
soft-limit arguments,
with the outcome%
    \footnote{Note that Ref.~\cite{Kenton:2015lxa}
    worked on the slow-roll attractor, so their
    Latin indices run over field-space labels only
    and exclude the momenta.
    However, their formulae are valid more generally.
    In the present case
    we are considering a single-field model, but with
    indices running over its field value and velocity.}
\begin{equation}
    \label{alpha generic}
    \alpha^{IJK}(p,q_1,q_2; t_{q}) \simeq P^{IM}(p; t_q) {P^{JK}}_{M}(q; t_q)
    \quad
    \text{where $|\vect{q}_1| \sim |\vect{q}_2| \sim q$ and $p \ll q$} .
\end{equation}
This result does not rely on the
slow-roll approximation~\cite{Kenton:2015lxa}. 
We expect that leading corrections to~\eqref{alpha generic}
are at relative order $(p/k)^2$ or higher.
The quantity $P^{IM}$ is the
cross-power spectrum between $X^I$ and $X^M$,
and
${P^{JK}}_M$ is defined to be the derivative
\begin{equation}
\label{def Sigma derivative}
    {P^{JK}}_{M}(q;t_q) \equiv \frac{\partial P^{JK}(q,t_q)}{\partial  X^M(t_q) } \;. 
\end{equation}
In order to apply Eq.~\eqref{alpha generic}
we must push \emph{back} the evaluation
time for each fluctuation
appearing in Eq.~\eqref{type 1 two} from time $t_k$ to
the earlier time $t_q < t_k$.
For the large-scale mode $\delta\phi_{\vect{p}}$,
the linear time evolution is described by
Eq.~\eqref{delta phi CMB mode later time}.
At the level to which we are now working, it should be supplemented
by a quadratic correction. However, we drop this quadratic
piece because it produces a result similar to the 22-loop and is therefore
volume-suppressed, as explained in~\S\ref{sec:back-reaction-quasiparticles},
and to be shown in detail in~\S\ref{sec: 22 type loop} below.
Meanwhile, the peak-scale modes $\delta \phi_{\vect{q}}$
and $\delta \phi_{-\vect{p}-\vect{q}}$
are dominated by the constant `A-mode'
(again up to possible quadratic corrections that we drop).
Therefore we may simply shift
$\delta \phi_{\vect{q}}(t_k) \to \delta \phi_{\vect{q}}(t_q)$.
It follows that
\begin{equation}
    \label{type 1 two at t_q}
    \langle \zeta_{\vect{p}} \zeta_{-\vect{p}}\rangle_\text{12} =N_{\phi\phi}^{(t_k, t)} \left(  N_\phi^{(t_k,t)} + \frac{\eta}{2} N_\pi^{(t_k,t)} \right)
    \mspace{-5mu}
    \int\limits_{k_s\lesssim q\lesssim k_e}
    \mspace{-5mu}
    \frac{\d^3q}{(2\pi)^{3}}
    \sqrt{\frac{\epsilon(t_k)}{\epsilon(t_q)}}
    \langle \delta \phi_{\vect{p}}{(t_{q})} \delta \phi_{\vect{q}}{(t_{q})} \delta \phi_{-\vect{p}-\vect{q}}{(t_{q})} \rangle \;. 
\end{equation}

Eq.~\eqref{type 1 two at t_q} clearly highlights the two
ingredients necessary to obtain a non-vanishing 12-type loop correction. 
First, the small-scale modes
$\delta\phi_{\vect{q}}$ and
$\delta\phi_{-\vect{p}-\vect{q}}$
must \emph{react} to the
presence of the long-wavelength curvature perturbation. 
The nature of this reaction is
encoded in the expression for the squeezed bispectrum.
Specifically,
the derivative~\eqref{def Sigma derivative}
is non-vanishing only when the small-scale power spectrum
depends on the background, which
absorbs the contribution of the long mode
$\zeta_{\vect{p}}$.

Second, one needs \emph{non-linearity} in the mapping between the
late-time, large-scale curvature perturbation $\zeta_{\vect{p}}(t)$
and the short scale field fluctuations.
This requires $N_{\phi\phi}^{(t_k, t)}\neq 0$,
which would generate a contribution to the \emph{equilateral}
bispectrum for momentum configurations of characteristic scale $k$.
The conclusion is that
the amplitude of the 12-type
diagram depends on \emph{two}
distinct types of non-Gaussianity:
one accounting for long--short mode coupling,
and one measuring
the non-linear interactions of the small-scale modes among themselves.

Combining Eq.~\eqref{alpha generic}
and Eq.~\eqref{type 1 two at t_q}
yields
\begin{multline}
    \label{type 1 three}
    P_\zeta(p;t)_\text{12} = N_{\phi\phi}^{(t_k, t)}
    \left(  N_\phi^{(t_k,t)} + \frac{\eta}{2} N_\pi^{(t_k,t)} \right)
    \mspace{-5mu}
    \int\limits_{k_s\lesssim q\lesssim k_e}
    \mspace{-5mu}
    \frac{\d^3q}{(2\pi)^{3}}
    \sqrt{\frac{\epsilon(t_k)}{\epsilon(t_q)}}
    P^{\phi M} (p;t_q) {P^{\phi\phi}}_{M}(q;t_q)\;.
\end{multline}
As explained above, we use Eq.~\eqref{massless correlator}
to express the small-scale power spectrum ${P^{\phi\phi}}(q;t_q)$
as a function of $H(t_q)$ only.
This is equivalent to the assumption that
modes exiting the horizon during
ultra-slow-roll were always in ultra-slow-roll,
and
will not reproduce the characteristic
oscillations in the tree-level power spectrum at peak scales.
These are
due to a spike in $\eta_V$
corresponding to the SR$\to$USR transition,
which displaces the mode functions.
(For an explicit numerical example see
Fig.~\ref{fig:Pz}.
A related discussion appears in Ref.~\cite{Jackson:2023obv}.)
Our treatment does not
include the effect
of these oscillations.
We anticipate that they
would typically boost the derivative~\eqref{def Sigma derivative}.
Therefore
our results should be regarded as a minimal estimate
for the one-loop correction produced within these models.

The estimates presented here,
and in~\S\S\ref{sec: 22 type loop}--\ref{sec: 13 type loop}
below,
such as Eq.~\eqref{type 1 three},
account for
the contribution to the loop
from a band of modes whose amplitude
is enhanced by ultra-slow-roll effects.
They are intended to be comparable
to estimates reported by other authors.
In a realistic model one would have to
understand how this enhanced
band is connected to the
broadband power spectrum
at its high- and low-momentum
boundaries.%
    \footnote{In our explicit numerical
    examples, such as Fig.~\ref{fig:Pz},
    this band forms a plateau
    and is not completed to a peak.
    This is because
    the dynamics of our 3-phase
    ultra-slow-roll model
    does not allow
    the power spectrum amplitude
    to reduce significantly
    even after the ultra-slow-roll
    phase has ended.
    However, our analytic formulae
    are more general than this scenario.}
In particular,
we shall not account for contributions from modes
rising towards, or falling away from,
a possible peak, which are
typically present in
realistic models.
These modes are potentially implicated in cancellations
that could reduce the size of the loop~\cite{Riotto:2023hoz,Tada:2023rgp,Firouzjahi:2023bkt}.
Accordingly, we caution that the results should be interpreted
with care.
 
Due to the flatness of the potential
during ultra-slow-roll, the derivative of ${P^{\phi\phi}}(q;t_q)$
with respect to $\phi$ is
suppressed by $|V'/V| \ll 1$,
and only the derivative with respect to $\pi$ will contribute. 
Hence,
\begin{equation} 
    \label{derivative of H wrt pi}
    \frac{\partial H(t_q)^2}{\partial \pi(t_q)}
    \approx H(t_q)^2 \frac{\pi(t_q)}{3 \Mp^2}
    =
    - H(t_q)^2 \frac{[2\epsilon(t_q)]^{1/2}}{3 \Mp} .
\end{equation} 
Substitution in Eq.~\eqref{type 1 three} yields
\begin{equation}
    \label{type 1 four}
    P_\zeta(p;t)_\text{12} = - N_{\phi\phi}^{(t_k, t)}
    \left(  N_\phi^{(t_k,t)} + \frac{\eta}{2} N_\pi^{(t_k,t)} \right)
    \frac{[2 \epsilon(t_k)]^{1/2}}{6 \Mp}
    \int\limits_{k_s\lesssim q\lesssim k_e} \frac{\d^3q}{(2\pi)^{3}}\,
    P^{\phi \pi} (p;t_q) \frac{H(t_q)^2}{q^3} 
     \;.
\end{equation}
Using Eqs.~\eqref{delta phi CMB mode later time}--\eqref{delta pi CMB mode later time},
one can derive the 2-point correlations for the large-scale fluctuations during the
ultra-slow-roll phase.
They are
\begin{equation}
\label{our evolved sigma}
    P^{\phi\phi}(p;t_q) = \frac{\epsilon(t_q)}{\epsilon(t_p)} \frac{H(t_p)^2}{2p^3} \;, \; P^{\pi\pi}(p;t_q) = \frac{\eta^2}{4} P^{\phi\phi}(p;t_q)\;, \; P^{\phi\pi}(p;t_q) = P^{\pi\phi}(p;t_q)= \frac{\eta}{2} P^{\phi\phi}(p;t_q) .
\end{equation}
We have left factors of $\eta$ explicit
in order that they may be counted,
but it should be remembered that
$\eta\approx -6$ during ultra-slow-roll.
We conclude that the amplitude of the 12-loop
relative to the tree-level can be written
\begin{equation}
\begin{split}
    \label{type 1 five}
    \Delta \dimlessP_\zeta(p;t)_\text{12} & \equiv \frac{\dimlessP_\zeta(p;t)_\text{12}}{\dimlessP_\zeta(p;t)_\text{tree}} \\
    &= - N_{\phi\phi}^{(t_k, t)}
    \left(  N_\phi^{(t_k,t)} + \frac{\eta}{2} N_\pi^{(t_k,t)} \right)
    \frac{\eta \Mp}{6} 
    [2\epsilon(t_k)]^{1/2}
    \frac{H^2}{2\pi^2} \int_{k_s}^{k_e} \frac{\d q}{q} \, \epsilon(t_q) \;,
\end{split}
\end{equation}
The quantity $H$ represents the Hubble parameter
during the ultra-slow-roll phase, which is
nearly time-independent.
From Eqs.~\eqref{N_phi}--\eqref{N_pi} 
we note that 
\begin{equation}
\label{Npi fct of Nphi}
    N_{\pi}^{(t_k,t)} =
    \frac{N_{\phi}^{(t_k,t)}}{3} + \frac{1}{3\pi(t_k)}
    =
    \frac{N_{\phi}^{(t_k,t)}}{3} - \frac{1}{3 \Mp [2\epsilon(t_k)]^{1/2}} . 
\end{equation}
When substituted in Eq.~\eqref{type 1 five} this yields
\begin{equation}
    \label{type 1 six}
    \Delta \dimlessP_\zeta(p;t)_\text{12} =
    - N_{\phi\phi}^{(t_k, t)}
    \left[
        N_\phi^{(t_k,t)}
        \left(1+\frac{\eta}{6} \right) [2 \epsilon(t_k)]^{1/2}
        - \frac{\eta}{6 \Mp} \right] \frac{\eta \Mp}{6}
        \frac{H^2}{2\pi^2} \int_{k_s}^{k_e} \frac{\d q}{q} \, \epsilon(t_q) \;.
\end{equation}
For ultra-slow-roll models the contribution proportional to
$N_{\phi}^{(t_k,t)}$ cancels
and can be discarded.
The remainder comes
from the second term
in square brackets.
For a smooth transition, this term is smaller than
the one associated with
$N_\phi^{(t_k,t)}$;
for an explicit numerical example, see
the right panel in Fig.~\ref{fig:background}.

To perform the momentum integral we note that,
during ultra-slow-roll,
the time-dependence of the $\epsilon$ parameter
has a simple expression when
written in terms of the momenta,
\begin{equation}
    \label{epsilon USR}
    \epsilon(t_q)= \epsilon(t_s) \left(\frac{k_s}{q} \right)^6 .
\end{equation}
This yields
\begin{equation}
    \label{type 1 seven}
    \Delta \dimlessP_\zeta(p;t)_\text{12} =
    N_{\phi\phi}^{(t_k, t)}\left(\frac{\eta}{6}\right)^2 \frac{\epsilon(t_s)}{12 \pi^2}
    H^2 . 
\end{equation}
As has already been explained,
$N_{\phi\phi}^{(t_k,t)}$
will contribute to
the
cubic non-linearity parameter~\eqref{f NL def}
evaluated on a peak-scale equilateral configuration,
$\fNLeq(k,k,k;t)$.
In certain circumstances it actually
provides the dominant contribution.
Specifically,
$\fNL$
receives 
two contributions:
one from 
the intrinsic bispectrum
$\alpha^{IJK}$
of the phase-space fluctuations,
and another from non-linear evolution
on super-horizon scales.
It is
this non-linear effect that is encoded by
the second derivative $N_{IJ}^{(t_k, t)}$~\cite{Lyth:2005fi}.
We distinguish these contributions by
schematically
writing $\fNL = \fNL^{(\alpha)} + \fNL^{(N_{IJ})}$.
If the intrinsic bispectrum is small,
then $\fNL^{(N_{IJ})}$ will be
comparable to
the total $\fNL$. 
For equilateral configurations evaluated on the peak scale, $\fNL^{(N_{IJ})}$ is
\begin{equation}
    \label{f_NL}
    \fNL^{(N_{IJ})}(k;t)
    \approx
    \frac{5}{6}\frac{N_{\phi\phi}^{(t_k,t)}}{\big[ N_\phi^{(t_k,t)} \big]^2}
    \equiv \fNL^{(N_{\phi\phi})}(k; t)\;,
\end{equation}
due to the fact that, over peak scales, the field fluctuation
dominates the velocity fluctuation.
Hence,
using Eq.~\eqref{N_phi} and its derivative,
$\fNL^{(N_{\phi\phi})}$ can be written~\cite{Cai:2018dkf}
\begin{equation}
    \label{fNL N_IJ}
    \fNL^{(N_{\phi\phi})} (k;t) = \frac{5}{2} \frac{\left[4\eta_{2}(\eta_{2}-3) +h^2 +4\eta_{2} h \right]}{\left( 2\eta_{2} +h -6 \right)^2} \;.
\end{equation} 
Both $N_\phi^{(t_k,t)}$ and $\fNL^{(N_{\phi\phi})}$ are constant over peak scales.%
    \footnote{Note that the total $\fNLeq(k;t)$ is constant over peak scales
    for models featuring a non-attractor phase \cite{Davies:2021loj},
    as it is also
    for models where the small-scale enhancement
    is produced (on super-horizon scales) by multi-field effects~\cite{Iacconi:2023slv}.}

After substitution
of Eqs.~\eqref{f_NL} and~\eqref{Pzeta peak scale}
in Eq.~\eqref{type 1 seven},
we find
\begin{equation}
    \label{12-type final}
    \Delta \dimlessP_\zeta(p;t)_\text{12} = \frac{2}{5} \left( \frac{\eta}{6}\right)^2 \epsilon(t_s)\,  \fNL^{(N_{\phi\phi})}(\bar k;t) \, \dimlessP_\zeta(\bar k)_\text{tree} .
\end{equation}
We have written power spectrum, and the
cubic nonlinearity parameter,
in terms of a representative scale $\bar{k}$
near the peak.
In Eq.~\eqref{12-type final} we
assume that $p/k$ is finite, so the behaviour of the
$p/k \rightarrow 0$ limit is no longer simple to extract.
    
In Eqs.~\eqref{fNL N_IJ}--\eqref{12-type final},
the evaluation time $t$
of the correlation function $\langle \zeta_{\vect{p}}\zeta_{-\vect{p}}\rangle_{12}$
is taken to be during the final slow-roll phase,
after the system has reached an adiabatic limit.
In particular,
the value of $\fNL^{(N_{\phi\phi})} (k;t)$ depends on the
character
of the USR$\to$SR transition~\cite{Cai:2018dkf}. 
During the ultra-slow-roll phase
we expect $\fNL^{(N_{\phi\phi})} (k;t)=\Or(1)$~\cite{Namjoo:2012aa}.
Later,
for a sufficiently smooth transition,
its amplitude
decreases and
becomes suppressed by a slow-roll parameter
in the subsequent slow-roll phase~\cite{Cai:2018dkf}.
If the transition is sudden then the evolution is more complicated
and $\fNL^{(N_{\phi\phi})}$ may not be completely suppressed.

The physical meaning of each factor appearing in Eq.~\eqref{12-type final}
can be understood as follows. 
First, the $(\eta/6)^2$ factor is due to the contribution of velocity fluctuations
associated with
the large-scale mode during ultra-slow-roll. 
Second, $\epsilon(t_s)$ acts as a book-keeper
of the background evolution:
for a perfectly de Sitter background ($\epsilon=0$)
the derivative in Eq.~\eqref{derivative of H wrt pi} vanishes,
so the small-scale modes do not react to the large-scale
mode $\zeta_{\vect{p}}$.
Therefore the 12-type loop-correction is zero.
Third, $\fNL^{(N_{\phi\phi})}(\bar k;t)$ accounts for non-linearity in the small-scale modes.
Finally, $\dimlessP_\zeta(\bar k)_\text{tree}$ tracks the growth of
perturbations on small scales, locking the amplitude of the 12-type loop
to the small-scale enhancement. 

Let us now comment on the magnitude of the 12-type loop correction,
subject to the caveats already given above.
The calculation outlined here is general,
meaning that Eq.~\eqref{12-type final} does not rely on assumptions
about the character of the USR$\to$SR transition.
After substitution of $\eta=-6$ in Eq.~\eqref{12-type final}
one sees that if $\dimlessP_\zeta(\bar k)_\text{tree} \lesssim 10^{-2}$,
the amplitude of $\Delta \dimlessP_\zeta(p;t)_\text{12}$
will be small, unless the equilateral non-Gaussianity on small
scales is sufficiently large to overcome suppression
by $\epsilon(t_s)$ and $\dimlessP_\zeta(\bar k)_\text{tree}$. 
For models featuring a smooth transition,
this expectation is not realistic \cite{Cai:2018dkf}.
On the other hand, for instantaneous transitions ($h=-6$)
one can have $\fNL^{(N_{\phi\phi})}(\bar k; t)=\Or(1)$,
yielding a 12-type loop-correction that is less suppressed.

We can compare Eq.~\eqref{12-type final} with the
result quoted by Firouzjahi \& Riotto for their 12-loop~\cite{Firouzjahi:2023ahg}.
Our answers both contain the amplitude factor $\dimlessP_\zeta(\bar k)_\text{tree}$
and the nonlinearity parameter $\fNL^{(N_{\phi\phi})}(\bar k;t)$,
which is $\eta_s$ in Ref.~\cite{Firouzjahi:2023ahg}.
The major differences are:
(i) the $(\eta/6)^2$ factor in Eq.~\eqref{12-type final};
(ii) an exponential suppression in Eq.~(III.42),
$\exp(-3\Delta N_\text{USR})$ with $\Delta N_\text{USR}$ being the duration of the USR phase,
absent in Eq.~\eqref{12-type final};
(iii) our $\epsilon(t_s)$, which is not present in Eq.~(III.42).
While our expressions differ in detail,
we agree with the conclusion of
Ref.~\cite{Firouzjahi:2023ahg}
that a smooth USR$\rightarrow$SR transition makes
the 12-type one-loop correction small. 
 
\subsection{22-loop}
\label{sec: 22 type loop}

The 22-type loop was discussed
from a general perspective in~\S\ref{sec:delta-N-back-reaction}.
It corresponds to the second diagram in Fig.~\ref{fig:one-loop-delta-N}.
The momenta entering the 4-point correlation function
in Eq.~\eqref{eq:prototype-22}
all correspond to peak scales,
where the velocity fluctuation is suppressed relative to the field
fluctuation.
Therefore Eq.~\eqref{eq:prototype-22} becomes
\begin{equation}
    \label{type 2 one}
    \langle \zeta_{\vect{p}} \zeta_{-\vect{p}} \rangle_\text{22}
    =
    \frac{[N_{\phi\phi}^{(t_{k},t)}]^2}{4} 
    \int \frac{\d^3q}{(2\pi)^{3}} \int \frac{\d^3r}{(2\pi)^{3}}
    \langle
        \delta \phi_{\vect{q}}{(t_{k})}
        \delta \phi_{\vect{p}-\vect{q}}{(t_{k})}
        \delta \phi_{\vect{r}}{(t_{k})}
        \delta \phi_{-\vect{p}-\vect{r}}{(t_{k})}
    \rangle \;.
\end{equation}
In the language of Fig.~\ref{fig:deltaN-diagrams},
this can be represented diagrammatically as 
\begin{equation}
    \langle \zeta_{\vect{p}} \zeta_{-\vect{p}} \rangle_{\text{22}}
    =
    \eqgraph{1ex}{0ex}{
        \begin{fmfgraph*}(100,50)
            \fmfleft{l}
            \fmfright{r}
            \fmf{plain}{l,v1}
            \fmf{dashes,left,tension=0.45}{v1,v2}
            \fmf{dashes,left,tension=0.45}{v2,v1}
            \fmf{plain}{v2,r}
            \fmfv{label=$N_{\phi\phi}^{(t_{{k}},,t)}$,label.angle=135,
                decor.shape=circle,decor.filled=full,decor.size=2thick}{v1}
            \fmfv{label=$N_{\phi\phi}^{(t_{{k}},,t)}$,label.angle=45,
                decor.shape=circle,decor.filled=full,decor.size=2thick}{v2}
        \end{fmfgraph*}
    }\quad.
\end{equation}
After performing the Wick contractions in Eq.~\eqref{type 2 one}
we find 
\begin{equation}
    \label{type 2 two}
    \dimlessP_\zeta(p;t)_\text{22} = \frac{p^3}{2\pi^2}\frac{[N_{\phi\phi}^{(t_{k},t)}]^2}{2}
    \int\limits_{k_s \lesssim q \lesssim k_e} \frac{\d^3q}{(2\pi)^3} \;  P^{\phi\phi}(q;t_k)^2 \;.
\end{equation}
Using Eq.~\eqref{massless correlator} for the small-scale power spectrum
and remembering that the Hubble rate is approximately constant
during ultra-slow-roll, Eq.~\eqref{type 2 two} yields 
\begin{equation}
    \label{type 2 three}
    \dimlessP_\zeta(p;t)_\text{22} =
    \frac{[N_{\phi\phi}^{(t_{k},t)}]^2}{2}
    \frac{H^4}{(2\pi)^4} \int_{k_s}^{k_e} {\d q} \;\frac{p^3}{q^4} \;.
\end{equation}
As explained in~\S\ref{sec: 12 type loop},
in some circumstances
the derivative ${N_{\phi\phi}^{(t_{k},t)}}$ can
be related to the peak-scale
equilateral non-Gaussianity parameter $\fNL^{(N_{\phi \phi})}(k;t)$.
Following the procedure described in that section, we find
\begin{equation}
\begin{split}
    \label{22-type final}
    \Delta \dimlessP_\zeta(p;t)_\text{22} &\equiv \frac{\dimlessP_\zeta(p;t)_\text{22}}{\dimlessP_\zeta(p;t)_\text{tree}}\\
    &=\frac{6}{25}
    \Big[ \fNL^{(N_{\phi\phi})}(\bar k, t) \Big]^2
    \frac{\big[ \dimlessP_\zeta(\bar k)_\text{tree} \big]^2}{\dimlessP_\zeta(p)_\text{tree}}
    \left(\frac{p}{{k_s}}\right)^3 \;.
\end{split}
\end{equation}
We conclude that,
as anticipated in~\S\ref{sec: loops in delta N},
the 22-type loop
has a volume suppression factor $N^{-1} = (p/k_s)^3$
that depends
on the hierarchy between the large and small modes.
This makes the 22-loop of relative order $(p/k_s)^3$ compared
to the 12-loop. One might worry that this is inconsistent,
because we should expect corrections of order $(p/k)^2$
in a separate universe framework.
However, this statement applies to the field operator
$\zeta$ and not each correlation function individually.
Instead, the separate universe formula will give the
leading contribution in gradients for each type of loop.
Gradient corrections to $\zeta$, via Eq.~\eqref{eq:zeta-separate-universe-fourier},
would generate corrections to \emph{both} Eq.~\eqref{12-type final}
and Eq.~\eqref{22-type final}
at relative order $(p/k_s)^2$.
Eq.~\eqref{22-type final}
is therefore a reliable estimate of the leading contribution to the
22-loop, but we should recognize that it is subleading to the
first gradient corrections to the 12-loop.

Recall that~\eqref{22-type final}
is valid for any value of $h$.
For a peak with amplitude $\dimlessP_\zeta(\bar k)_\text{tree}=\Or(10^{-2})$,
it yields
\begin{equation}
    \label{delta P 4pt corr 1}
    \Delta \dimlessP_\zeta(p;t)_\text{22} =\Or(10^{4})
    \Big[ \fNL^{(N_{\phi\phi})}(\bar k;t) \Big]^2
    \left(\frac{p}{{k_s}}\right)^3 \;.
\end{equation}
Even a hierarchy as small as $p/k_s \sim 10^{-1.5}$
is already sufficient to cancel out the large
$\Or(10^4)$ prefactor.
This type of loop correction is volume suppressed
for instantaneous and smooth transitions.

\subsection{13-loop}
\label{sec: 13 type loop}
The 13-type loop was discussed
from a general perspective in~\S\ref{sec:delta-N-back-reaction}.
It corresponds to the third diagram in Fig.~\ref{fig:one-loop-delta-N}.
In the Gaussian approximation where we retain only
disconnected contributions to
the phase-space 4-point function 
in Eq.~\eqref{eq:prototype-13},
the corresponding momentum configuration
is of the
``double-soft kite'' type described in
Ref.~\cite{Kenton:2016abp}.
Therefore,
as for the 12-loop, the 13-loop depends on a soft limit of
$\delta X$ correlations, which encode long--short mode couplings. 
The result is that the 13-loop is
\emph{not} automatically volume suppressed.
The connected contributions to the 4-point function that have been
dropped would generically involve only a single soft mode,
but they enter at two-loop level and cannot consistently be retained
at the accuracy to which we are working.

While peak-scale
modes contribute only
field fluctuations, the large-scale modes in Eq.~\eqref{eq:prototype-13} can be of
either field or velocity type.
Therefore we obtain
\begin{equation}
    \label{type 3 one}
    P_\zeta(p;t)_\text{13} = \frac{1}{3} N_I^{(t_{k}, t)} \left( {N_{J \phi\phi}^{(t_{k}, t)}}+{N_{\phi J \phi }^{(t_{k}, t)}}+{N_{\phi\phi J}^{(t_{k}, t)}} \right) P^{IJ}(p;t_k) \int \frac{\d^3q}{(2\pi)^{3}} P^{\phi\phi}(q;t_k) \;. 
\end{equation}
%The $J$ index here represents both field and velocity perturbations, as the corresponding phase-space fluctuation is associated with momentum ${\vect{p}}$.
Using Eqs.~\eqref{N_phi}-\eqref{N_pi} and~\eqref{Npi fct of Nphi},
one can derive a relation between the third derivatives
of $N$ appearing in Eq.~\eqref{type 3 one}, 
\begin{equation}
\label{relation between N third derivatives}
    N_{\pi\phi\phi}^{(t_{k}, t)} = N_{\phi \pi\phi}^{(t_{k}, t)} = N_{\phi \phi\pi}^{(t_{k}, t)} = \frac{N_{\phi \phi\phi}^{(t_{k}, t)}}{3} \;. 
\end{equation}
This relationship is a specific property of the 3-phase
ultra-slow-roll model
\emph{and} the precise
choice of initial time.
It does not apply in general.

Expanding the summation over $J$ in Eq.~\eqref{type 3 one},
and using Eq.~\eqref{relation between N third derivatives}, we obtain%
    \footnote{Note that the second term in Eq.~\eqref{type 3 two} was omitted
    in \texttt{v1} of the arXiv version of this paper.
    The missing contribution has been added from \texttt{v2} onwards.}
\begin{equation}
    \label{type 3 two}
    P_\zeta(p;t)_\text{13}= N_I^{(t_{k}, t)}{N_{\phi \phi\phi}^{(t_{k}, t)}} \left( P^{I \phi}(p;t_k) + \frac{1}{3}P^{I \pi}(p;t_k) \right) \int \frac{\d^3q}{(2\pi)^{3}} P^{\phi\phi}(q;t_k) \;. 
    %+ \frac{1}{3} N_I^{(t_{k}, t)} \left( {N_{\pi \phi\phi}^{(t_{k}, t)}}+{N_{\phi \pi \phi }^{(t_{k}, t)}}+{N_{\phi\phi \pi}^{(t_{k}, t)}} \right) P^{I\pi}(p;t_k) \int \frac{\d^3q}{(2\pi)^{3}} P^{\phi\phi}(q;t_k) \;. 
\end{equation}
In the language of Fig.~\ref{fig:deltaN-diagrams}, these four
contributions can be visualized diagrammatically as 
\begin{equation}
\begin{split}
    \langle \zeta_{\vect{p}} \zeta_{-\vect{p}} \rangle_{\text{13}}
    &=
    \eqgraph{1ex}{0ex}{
        \begin{fmfgraph*}(100,50)
            \fmfleft{l}
            \fmfright{r}
            \fmf{plain}{l,v1}
            \fmf{dashes}{v1,v2}
            \fmf{dashes,right,tension=0.75}{v2,v2}
            \fmf{plain}{v2,r}
            \fmfv{label=$N_{\phi}^{(t_{{k}},,t)}$,label.angle=90,
                decor.shape=circle,decor.filled=full,decor.size=2thick}{v1}
            \fmfv{label=$N_{\phi\phi\phi}^{(t_{{k}},,t)}$,label.angle=-45,
                decor.shape=circle,decor.filled=full,decor.size=2thick}{v2}
        \end{fmfgraph*}
    }
    \mspace{5mu}
    +
    \mspace{5mu}
    \eqgraph{1ex}{0ex}{
        \begin{fmfgraph*}(100,50)
            \fmfleft{l}
            \fmfright{r}
            \fmf{plain}{l,v1}
            \fmf{dbl_dots}{v1,v2}
            \fmf{dashes}{v2,v3}
            \fmf{dashes,right,tension=0.55}{v3,v3}
            \fmf{plain}{v3,r}
            \fmfv{label=$N_{\pi}^{(t_{{k}},,t)}$,label.angle=90,
                decor.shape=circle,decor.filled=full,decor.size=2thick}{v1}
            \fmfv{label=$N_{\phi\phi\phi}^{(t_{{k}},,t)}$,label.angle=-45,
                decor.shape=circle,decor.filled=full,decor.size=2thick}{v3}
        \end{fmfgraph*}
    } \\
    & \quad  + \frac{1}{3}\; \eqgraph{1ex}{0ex}{
        \begin{fmfgraph*}(100,50)
            \fmfleft{l}
            \fmfright{r}
            \fmf{plain}{l,v1}
            \fmf{dashes}{v1,v2}
            \fmf{dbl_dots}{v2,v3}
            \fmf{dashes,right,tension=0.55}{v3,v3}
            \fmf{plain}{v3,r}
            \fmfv{label=$N_{\phi}^{(t_{{k}},,t)}$,label.angle=90,
                decor.shape=circle,decor.filled=full,decor.size=2thick}{v1}
            \fmfv{label=$N_{\phi\phi\phi}^{(t_{{k}},,t)}$,label.angle=-45,
                decor.shape=circle,decor.filled=full,decor.size=2thick}{v3}
        \end{fmfgraph*}
    }
    \mspace{5mu}
    +
    \mspace{5mu}
    \frac{1}{3} \; \eqgraph{1ex}{0ex}{
        \begin{fmfgraph*}(100,50)
            \fmfleft{l}
            \fmfright{r}
            \fmf{plain}{l,v1}
            \fmf{dbl_dots}{v1,v2}
            \fmf{dbl_dots}{v2,v3}
            \fmf{dashes,right,tension=0.55}{v3,v3}
            \fmf{plain}{v3,r}
            \fmfv{label=$N_{\pi}^{(t_{{k}},,t)}$,label.angle=90,
                decor.shape=circle,decor.filled=full,decor.size=2thick}{v1}
            \fmfv{label=$N_{\phi\phi\phi}^{(t_{{k}},,t)}$,label.angle=-45,
                decor.shape=circle,decor.filled=full,decor.size=2thick}{v3}
        \end{fmfgraph*} 
    }
\end{split}
\end{equation}
Using Eq.~\eqref{our evolved sigma}, we find 
\begin{equation}
\label{type 3 three}
    P_\zeta(p;t)_\text{13} = N_I^{(t_{k}, t)}{N_{\phi \phi\phi}^{(t_{k}, t)}} \left( 1 + \frac{\eta}{6} \right) P^{I \phi}(p;t_k) \int \frac{\d^3q}{(2\pi)^{3}} P^{\phi\phi}(q;t_k)   \;.  
\end{equation}
Eq.~\eqref{type 3 three} shows that the 13-type loop
vanishes for ultra-slow-roll evolution, where $\eta=-6$.
However,
it should be noted that absence of the
13-type loop is \emph{not} a general property of models featuring a non-attractor phase.
In the present case it depends not only
on the condition $\eta=-6$, but also on Eq.~\eqref{relation between N third derivatives}
which relates the third derivatives of $N$.
As already explained,
this relation applies specifically
when the $\delta N$ calculation
is based at a time during (or at the end of) the USR phase.
If we were to base the $\delta N$ formula
at some later time during the final slow-roll phase,
the 13-type loop need not vanish.
This is expected because the total one-loop 
correction should be treated as a whole. We cannot usually
break apart the different topologies and assign
them a meaning
invariant
under changes in the calculational scheme.

Although the 13-loop will vanish in our scenario,
it is instructive to continue the computation.
This enables us to highlight
the structural ingredients necessary to produce
a result that is not volume suppressed. 
To this end, we isolate the vanishing prefactor $(1+\eta/6)$
and retain $\eta$ explicitly,
but substitute $\eta = -6$ elsewhere.
After further substitution
of Eqs.~\eqref{massless correlator} and~\eqref{Npi fct of Nphi},
the 13-type loop correction relative to the tree level scalar power spectrum
becomes
\begin{equation}
\begin{split}
\label{type 3 four}
    \Delta \dimlessP_\zeta(p;t)_\text{13} & \equiv \frac{\dimlessP_\zeta(p;t)_\text{13}}{\dimlessP_\zeta(p;t)_\text{tree}}\\
    &=2  \left( 1 + \frac{\eta}{6} \right) {N_{\phi\phi\phi}^{(t_{k}, t)} }
    \left[
        N_\phi^{(t_{k}, t)} \left( 1+\frac{\eta}{6}\right)
        -
        \frac{\eta}{6 \Mp [2\epsilon(t_{k})]^{1/2}} \right]
    \epsilon(t_{k})
    \Mp^2
    \left(\frac{H}{2\pi}\right)^2 \int_{k_s}^{k_e} \frac{\d q}{q} \\
    & = \left( 1 + \frac{\eta}{6} \right) {N_{\phi\phi\phi}^{(t_{k}, t)} }
    \frac{|\eta| \Mp}{6}
    \left[{{2\epsilon(t_{k})}} \right]^{1/2}
    \left(\frac{H}{2\pi}\right)^2 \Delta N_\text{USR}\;.
\end{split}
\end{equation}
To obtain the last line
have we set $\eta=-6$ in the first term within square brackets.
As in the case of the 12-loop,
this term will vanish.
Meanwhile,
the momentum integral evaluates to $\ln{\left({k_e}/{k_s}\right)}$,
which approximately coincides with the duration of the ultra-slow-roll phase
$\Delta N_\text{USR}=\Or(1)$.
Using Eqs.~\eqref{N_phi} and~\eqref{Pzeta peak scale},
identifying the amplitude of the tree-level $\dimlessP_\zeta$
on the small-scale plateau with its value at the peak scale, $\bar k$,
and working in the limit of a smooth transition $h \rightarrow 0$,
we find
\begin{subequations}
\begin{equation}
\label{type 3 physical 1}
    \Delta \dimlessP_\zeta(p;t)_{\text{13},\, h\to0}
    =2 \left( 1 + \frac{\eta}{6} \right) N_{\phi\phi\phi}^{(t_{k}, t)}
    \frac{|\eta| \Mp^3}{6} [ 2 \epsilon(t_{k}) ]^{1/2}
    \epsilon_2 \Delta N_\text{USR} \dimlessP_\zeta(\bar k)_\text{tree} .
\end{equation}
Meanwhile, for an instantaneous transition with $h = - 6$ we find
\begin{equation}
\label{type 3 physical 2}
    \Delta \dimlessP_\zeta(p;t)_{\text{13},\, h=-6}
    =
    8 \left( 1 + \frac{\eta}{6} \right) N_{\phi\phi\phi}^{(t_{k}, t)}
    \frac{|\eta| \Mp^3}{6} [ 2 \epsilon(t_{k}) ]^{1/2}
    \epsilon_2 \Delta N_\text{USR} \dimlessP_\zeta(\bar k)_\text{tree}
    \;.
\end{equation}
\end{subequations}
% Although we have left $\eta$ explicit in
% Eqs.~\eqref{type 3 physical 1}--\eqref{type 3 physical 2},
% we have derived these expressions assuming ultra-slow-roll
% dynamics and so it should be understood to have the value
% $\eta = -6$.
The combination
$|\eta|/6$ tracks the contribution of large-scale velocity fluctuations
at time $t_k$.
For consistency, we should regard this
factor as equal to unity because our results formally apply only
in the limit of ultra-slow-roll evolution where $\eta = -6$.

Excluding the vanishing factor $(1+\eta/6)$,
the remaining terms in Eqs.~\eqref{type 3 physical 1}--\eqref{type 3 physical 2}
clearly exhibit the essential
ingredients
needed to synthesize a non-vanishing 13-contribution. 
First, non-linear interactions among small-scale modes
are again required.
In this case, these interactions are measured
by the third derivative $N_{\phi\phi\phi}^{(t_{k}, t)}$.
Second, the slow-roll parameters $\epsilon(t_k)$ and $\epsilon_2$
act as
book-keepers
for the dependence
on the background dynamics,
as $\epsilon(t_s)$ does for the 12-loop in Eq.~\eqref{12-type final}. 
On a perfectly de Sitter background, the 13-loop vanishes. 
Third, the duration of the USR phase, $\Delta N_\text{USR}\sim \ln(k_e/k_s)$, measures
the
number of
small-scale modes that contribute in the loop integral. 
Finally, $\dimlessP_\zeta(\bar k)_\text{tree}$ locks the amplitude of the 13-loop to the enhancement of the small scales, as it does for the 12 loop~\eqref{12-type final}.

While Eqs.~\eqref{type 3 physical 1}--\eqref{type 3 physical 2} are very useful to
understand the physical interpretation of the 13-loop,
they depend on the character of the USR$\rightarrow$SR transition. 
For this reason, in the following we revert to the last line of
Eq.~\eqref{type 3 four}.
The third derivative $N_{\phi\phi\phi}^{(t_{k}, t)}$
can be expressed (approximately)
in terms of the dimensionless non-linearity parameter
$\gNL$,
defined in Eq.~\eqref{t NL and g NL def},
\begin{equation}
\label{g_NL}
    \gNL(k;t) \approx \frac{25}{54}\frac{N_{\phi\phi\phi}^{(t_{k}, t)}}{\big[ N_\phi^{(t_{k}, t)} \big]^3} \;,
\end{equation}
due to the fact that only field fluctuations contribute
on peak scales.
Employing Eq.~\eqref{N_phi} and its derivatives, we find 
\begin{equation}
\label{gNL smooth}
    \gNL(k, t) =
    \frac{25}{3}
    \frac{h^3 + 6 h^2 \eta_2 +12 (h-2){\eta_2}^2 +8 {\eta_2}^3 }{\left(2\eta_2 +h-6 \right)^3} . 
\end{equation}
It is suppressed for a smooth USR$\to$SR transition
for which
$\eta_2 \ll 1$
and $h \ll 1$,%
    \footnote{\label{footnote on h small expansion}
    Note that expanding Eqs.~\eqref{fNL N_IJ} and~\eqref{gNL smooth}
    for $h\ll1$ requires careful consideration.
    Specifically, both $h$ and $\eta_2$ are small
    and the final result is sensitive to the order in which the
    expansions for $h\ll1$ and $\eta_2\ll1$ are performed.
    This subtlety was not discussed in Ref.~\cite{Cai:2018dkf}.
    The result reported there, $\fNL\approx -5\eta_{2}/6$, was obtained by first expanding
    Eq.~\eqref{fNL N_IJ} for $h\ll 1 $ and then for $\eta_{2}\ll1$.
    While it is important to obtain a precise result,
    this does not affect the discussion in the main text.
    In both cases $\fNL$ and $\gNL$ are either slow-roll suppressed
    (since $\eta_2\ll 1$),
    or suppressed due to the smoothness of the transition ($h\ll 1$).}
and is scale invariant across the peak.
We therefore identify it with a representative value $\gNL(\bar{k}; t)$
on the fiducial scale $\bar{k}$.
Using~\eqref{Pzeta peak scale} and~\eqref{g_NL} in Eq.~\eqref{type 3 four}, we obtain 
\begin{equation}
    \label{13-type final 1}
    \Delta \dimlessP_\zeta(p;t)_\text{13} = 
    \left( 1+\frac{\eta}{6}\right) \frac{54}{25}  \gNL(\bar k;t) \, \frac{|\eta|}{6}
    \big[ N_\phi^{(t_k,t)} \Mp\big]
    [2\epsilon(t_k)]^{1/2} \, \Delta N_\text{USR} \,\dimlessP_\zeta(\bar k)_\text{tree} \;.
\end{equation}
Eq.~\eqref{13-type final 1} applies to any type of transition. 
Specialising to the case of a smooth transition using Eq.~\eqref{simplify N_phi} yields
\begin{equation}
    \label{13-type final}
    \Delta \dimlessP_\zeta(p;t)_{\text{13},\,h\to0} = 
    \left( 1+\frac{\eta}{6}\right) \frac{54}{25} \gNL(\bar k;t) \, \frac{|\eta|}{6}\,
    \left( \frac{\epsilon(t_k)}{\epsilon_2} \right)^{1/2}
    \Delta N_\text{USR} \,\dimlessP_\zeta(\bar k)_\text{tree} \;, 
\end{equation} 
where $ \gNL(\bar k;t)\ll 1$. 

While neither
the 12- and 13-type loop corrections are volume-suppressed,
the 12-diagram is the
only
non-vanishing correction with this property
in our scenario.
Note that the 13-type diagram corresponds topologically
to the one produced by quartic $\zeta$-interactions
within the in--in formalism. 
In Ref.~\cite{Firouzjahi:2023aum}, Firouzjahi calculated
the loop-level contribution from both
cubic and quartic interactions,
and
found that for $h=-6$ they have the same sign and are equal.
Understanding the reason for this mismatch
will require
a careful understanding of how to interpret the
separate universe calculation within the
in--in framework. We leave this for future work.

Before closing, let us compare our final expression~\eqref{13-type final}
for the 13-type diagram
with the corresponding result Eq.~(III.46) obtained
by Firouzjahi \& Riotto~\cite{Firouzjahi:2023ahg}. 
While we obtain a vanishing 13-type loop, their result is nonzero.
For our case, the 13-loop vanishes only if one includes all relevant contributions.
In particular,
see Eqs.~\eqref{type 3 two}-\eqref{type 3 three}, where the first and second term cancel.
Inspection of the non-vanishing factors in Eq.~\eqref{13-type final}
shows that we recover the
dependence on the duration of the ultra-slow-roll
epoch
found by Firouzjahi \& Riotto.
However, our expressions differ due to
the factors
$|\eta|/6$ and $[ \epsilon(t_k)/\epsilon_{2} ]^{1/2}$,
which are present in Eq.~\eqref{13-type final}
but do not appear in Eq.~(III.46). 
(Note that the ${\eta_s}^2$ factor in Eq.~(III.46) corresponds to $\gNL(\bar k;t)$.
However, this is different to our $\gNL(\bar k;t)$ because it is evaluated at a different time).
As for the 12-type contribution, we believe that
these differences are due to different implementations
of the $\delta N$ calculation (\S\ref{sec:delta-N-back-reaction}).
 
\section{The smoothing scale as a Wilsonian cutoff}
\label{sec:wilsonian-counterterms}
We now return to the question raised
below Eq.~\eqref{eq:zeta-separate-universe-fourier}
on p.~\pageref{eq:zeta-separate-universe-fourier}.
How does the analysis change if we
choose to smooth on the scale $\bigBox$
immediately after the $\smallBox$-size boxes
emerge from the horizon?

\subsection{The separate universe framework as an effective field theory}
At the outset, it is already clear that the
smoothing scale functions as a Wilsonian cutoff.
Indeed, the separate universe
framework can be interpreted
as an
almost-textbook application of the
logic of effective field theories.
Moving the smoothing lengthscale from
$\smallBox$
(relatively, an ultraviolet scale)
to
$\bigBox$
(relatively, an infrared scale)
corresponds to integrating out
degrees
of freedom.
There are no long-range effects
mediated by the modes between
$\bigBox$ and $\smallBox$,
so their influence should appear
local from the perspective of the $\bigBox$-size
regions. 
Therefore we expect that it can be
captured by a renormalization
of local operators.
It is this renormalization that allows
different versions of the computation,
performed with different smoothing scales,
to agree.
The smoothing/renormalization
process erases
most detailed information
about the modes between $\bigBox$ and $\smallBox$.
Only
a limited quantity of aggregated information
is retained,
carried by the renormalized operators.
As in any effective theory,
if we choose to work
up to a specified accuracy in the
ratio $\smallBox/\bigBox = p/k \ll 1$,
then only a finite number
of operators need be renormalized.

In this section we outline
the main steps in the renormalization procedure.
Our discussion applies quite
generally
and does not depend
on specific properties of the 3-phase ultra-slow-roll
model.
However, the precise choice of counterterms
is model-dependent.
In order to give an explicit discussion
it is helpful to work
in the context of the
momentum dependence
carried by the 12- and 22-type loops, see Eqs.~\eqref{12-type final} and~\eqref{22-type final}.
It was explained in~\S\ref{sec: 13 type loop}
that the 13-type loop
vanishes for the specific scenario
of~\S\S\ref{sec: SR to USR to SR}--\ref{sec:delta N loops calculation}.
However, our discussion will include suitable counterterms
to renormalize its contribution
in a scenario where it is present.
% induced by a plateau of enhanced modes vanishes, 
% we will also discuss counterterms 
% degenerate with the momentum dependence
% carried by (non-vanishing factors
% in) it, see Eq.~\eqref{13-type final}.}

\para{Counterterms}
Concretely, suppose we smooth
on the scale $\bigBox$ just after 
the time $t_k$ corresponding to horizon
exit of the mode $k \sim \smallBox^{-1}$.
We are then committed
to working with the smoothed fields
$\smoothBig{\delta X^I(\vect{x}, t)}$,
which correspond to the volume-averaged
fields calculated
in~\S\ref{sec:back-reaction-particle-creation}.

Smoothing on this scale averages over
injection of energy
due to particle creation at the horizon,
but then \emph{erases the newly-injected modes}.
The $\zeta$-field smoothed on the same
scale is $\smoothBig{\zeta(\vect{x}, t)}$.
It is given by Eq.~\eqref{eq:zeta-separate-universe},
and its Fourier modes
satisfy~\eqref{eq:zeta-separate-universe-fourier}
with the replacement $\smallBox \rightarrow \bigBox$.
In particular, the base time
$t_\ast=t_k$ in Eq.~\eqref{eq:zeta-separate-universe-fourier}
is not changed.
However, the modes between $p$ and $k$
are now missing from the convolution integral
in the quadratic part of~\eqref{eq:zeta-separate-universe-fourier}.
This happens because the
$\delta X$ fields in~\eqref{eq:basic-separate-universe}
and~\eqref{eq:zeta-separate-universe}
are smoothed before forming nonlinear
products, as emphasized in~\S\ref{sec:delta-N-back-reaction}.

There are several consequences.
First, consider the
$\delta N$ tree-level term in
Eq.~\eqref{eq:prototype-tree}.%
    \footnote{To be clear,
    by
    ``$\delta N$ tree-level'' we mean that this term is at
    tree-level in the $\delta N$ expansion.
    As explained in~\S\ref{sec: loops in delta N},
    the $\langle \delta X \delta X\rangle$ correlation
    function that appears
    in it should be computed to one-loop level.}
This includes a spatial average over
particle creation,
no matter where the cutoff is positioned.
If we work with the cutoff at $\smallBox$,
these effects are explicitly
included via the analysis of~\S\ref{sec:back-reaction-particle-creation}.
On the other hand, if we work with the
cutoff at $\bigBox$
the same effect is incorporated into the
definition of the smoothed $\delta X$ fields,
and there is no loop to include.
The situation is different
for
the 12-, 22-, and 13-type
loops~\eqref{eq:prototype-12}--\eqref{eq:prototype-13}.
These are dramatically modified.
With the cutoff
set to the lengthscale $\bigBox$,
each loop integral now has a cutoff of order $p$.
All information about
nonlinear interactions among the
peak-scale modes,
responsible for cascading power
into the deeper infrared modes,
is absent.

How can the final two-point function
for $\smoothBig{\zeta}$
be made to agree with that
already calculated for $\smoothSmall{\zeta}$?
Consider the 12-
and 13-type loops, Eqs.~\eqref{12-type final} and~\eqref{13-type final}.
Their $p$-dependence is the same as the tree-level
power spectrum.
We introduce
a quadratic counterterm,
which for simplicity we express in terms of $\zeta$.
The counterterm action has the form
\begin{equation}
    \label{eq:prototype-counterterm}
    \ActionCt
    =
    \int_{-\infty}^{\infty} \d t
    \int \frac{\d^3 q_1}{(2\pi)^3} \frac{\d^3 q_2}{(2\pi)^3}
    \, (2\pi)^3 \delta(\vect{q}_1 + \vect{q}_2)
    a(t)^3
    \Mp^2
    \deltact(t)
    \OpO_{\vect{q}_1}(t)
    \OpQ_{\vect{q}_2}(t) ,
\end{equation}
where $\OpO_{\vect{q}}(t)$
and $\OpQ_{\vect{q}}(t)$
symbolically stand for $\zeta_{\vect{q}}(t)$
or one of its derivatives,
each
possibly carrying tensor indices
under 3-dimensional rotations
that we have suppressed.
The normalization factor of $\Mp^2$
appears in explicit calculations
and guarantees that $\zeta$ is projected
out of the path integral in a limit
where gravity is decoupled via
$\Mp \rightarrow \infty$.
It is introduced here by hand
so that $\deltact$ has the expected
mass dimension
if~\eqref{eq:prototype-counterterm}
is expressed in terms of a canonically normalized
field $\delta\phi$.
The kernel $\deltact(t)$
should have mass dimension
$2 - \dim(\OpO) - \dim(\OpQ)$.
It may be an explicit function of time,
but does not depend on the momenta
$\vect{q}_1$, $\vect{q}_2$.
It is the analogue
of a Wilson coefficient.
Its amplitude and time evolution should depend only
on properties of the short modes.
From this perspective,
as for any Wilson coefficient,
Eq.~\eqref{eq:prototype-counterterm}
represents a factorization of physics
associated with the different scales $p$ and $k$.
We treat $\deltact$ as if it were a one-loop
term in the loop expansion; its tree-level graphs
should be included along with the one-loop
contributions described above.
The counterterm insertion in the two-point function
corresponds to the diagram
\begin{equation*}
    \label{eq:twopf-counterterm}
    \langle \zeta_{\vect{q}_1} \zeta_{\vect{q}_2} \rangle^{\text{ct}}
    \supset
    \mspace{10mu}
    \eqgraph{1ex}{0ex}{
        \begin{fmfgraph*}(60,30)
            \fmfleft{l}
            \fmfright{r}
            \fmf{plain,label=$\vect{q}_1$}{l,v}
            \fmf{plain,label=$\vect{q}_2$}{v,r}
            \fmfv{decoration.shape=cross,label=$\tau$,label.angle=90}{v}
        \end{fmfgraph*}
    }
\end{equation*}
where the cross denotes the counterterm vertex.
In the simplest case where $\OpO_{\vect{q}}$ and $\OpQ_{\vect{q}}$
are scalar operators,
this diagram gives a contribution to the two-point function
that is schematically of the form
\begin{equation}
    \langle \zeta_{\vect{p}} \zeta_{-\vect{p}} \rangle^{\prime,\OpO \OpQ}_\eta
    =
    \im \psi_p^\zeta(\eta)^2
    \int_{-\infty}^{\eta} \frac{\d \tau}{H^4(\tau) \tau^4}
    \; \Mp^2 \deltact(\tau)
    \psi^{\OpO\ast}_p(\tau)
    \psi^{\OpQ\ast}_p(\tau)
    +
    \text{c.c.},
\end{equation}
where $\eta$ is the conformal time of evaluation,
$\tau$ is the time at the counterterm vertex,
and `$+ \text{c.c.}$' indicates that we should add
the complex conjugate of the preceding term.
Recall that the prime $'$ attached to the correlation
function indicates that the conventional factor
$(2\pi)^3$ and the momentum conservation $\delta$-function
are not written.
The symbol `$\ast$' denotes complex conjugation.
We have
written the mode functions of the operators
$\zeta$, $\OpO$ and $\OpQ$
for momentum $\vect{p}$
as $\psi^\zeta_p$, $\psi^{\OpO}_p$, $\psi^{\OpQ}_p$, respectively.
The Wronskian normalization
for these mode functions
implies that
$\psi^\zeta_p$ can be written
as $(2p^3)^{-1/2}$ multiplied by a dimensionless
time-dependent function,
and
$\psi^{\OpO}_p$ and $\psi^{\OpQ}_p$ inherit their
normalization from $\psi^\zeta_p$.

In this section
we are working to lowest order
in the slow-roll expansion,
and we assume that each time integral is dominated
by contributions around horizon exit.
This means that a
(physical)
time or space derivative applied to
$\zeta_{\vect{p}}$
can be estimated
as roughly $\sim (p/a) \zeta_{\vect{p}}$,
because there is no other scale in the problem.
The analysis becomes technically
more involved when one includes
time evolution of the background, but is not
conceptually different.

For example,
if $\OpO = \OpQ = \dot{\zeta}$,
then
one can check that
$(\psi_p^\zeta)^2 \sim \dimlessP_\zeta(p) p^{-3}$
and
$\psi^{\OpO}_p \psi^{\OpQ}_p \sim \dimlessP_\zeta(p) p^{-1} / a^2$.
This counterterm preserves the shift symmetry of
the $\zeta$ action
and respects conservation of $\zeta$ outside
the horizon~\cite{Assassi:2012et,Senatore:2009cf,Pimentel:2012tw,Senatore:2012ya}.
The corresponding kernel $\deltact$ has dimension 0.
Hence, the counterterm contribution to the
two-point function can be estimated as
\begin{equation}
    \langle \zeta_{\vect{p}} \zeta_{-\vect{p}} \rangle^{\prime,\dot{\zeta}^2}_\eta
    \simeq
    \dimlessP_\zeta(p, \eta)
    \frac{\im}{p^4}
    \int_{-\infty}^\eta \frac{\d \tau}{H(\tau)^2 \tau^2}
    \Mp^2 \deltact(\tau)
    \dimlessP_\zeta(p, \tau)
    \times \cdots ,
\end{equation}
where `$\cdots$' denotes
time-dependent factors that we have not written explicitly.
If $\deltact$ contains no characteristic timescales
different to $1/p$
then the vertex integral can be estimated on
dimensional grounds as
\begin{equation}
    \label{eq:12-counterterm-time-vertex-estimate}
    \im \int_{-\infty}^\eta \frac{\d \tau}{\tau^2} \, \deltact(\tau)
    \times \cdots
    \sim p C ,
\end{equation}
where $C$ is some dimensionless constant determined
by $\deltact$.
We conclude
\begin{equation}
    \label{eq:kinetic-counterterm-estimate}
    \langle \zeta_{\vect{p}} \zeta_{-\vect{p}} \rangle^{\prime,\dot{\zeta}^2}_\eta
    \simeq
    \frac{C}{p^3} \dimlessP(p, \eta) .
\end{equation}
Evidently, the counterterm is
just a renormalization of the $\zeta$ kinetic term.%
    \footnote{For simplicity we have phrased the discussion in
    terms of $\dot{\zeta}^2$. However, at the order to which
    we are working, one could equally well express the analysis
    in terms of the counterterm $\zeta \partial^2 \zeta / a^2$.}
In a model with isocurvature modes one could also have a
mass counterterm $\zeta^2$ generated by $\OpO = \OpQ = \zeta$.
For this operator
the power counting for $p$ works in a similar way.
However, the time integral can be infrared divergent
which complicates the analysis.

\subsection{12- and 13-loop counterterms}
Clearly Eq.~\eqref{eq:kinetic-counterterm-estimate} has
$p$-dependence matching
the tree-level $\zeta$ power spectrum,
and therefore can be used to absorb
the 12- and 13-type loops.
Moreover, this can be done while respecting
the universality criterion that $\deltact$ depends only
on properties of the short modes.
The conclusion of this discussion is that, if we insist
on setting the smoothing scale equal to $\bigBox$,
we should add a counterterm
$\sim \dot{\zeta}^2$
of the form~\eqref{eq:prototype-counterterm}
with
$\deltact$ \emph{chosen}
to have the correct amplitude
and time dependence to reproduce the
explicit results~\eqref{12-type final} and~\eqref{13-type final}. 

In practice,
from the perspective of the
separate universe framework with an
$\bigBox$-scale cutoff,
this amplitude and time dependence is
not predictable because 
it depends on detailed information about
the behaviour of modes above the cutoff
that we have intentionally erased from
our model.
In the $\bigBox$-smoothed
version of the theory, we have no
choice other than
to accept the presence of the
counterterm and attempt to fit its
amplitude and time dependence
by comparison with observation.
This is exactly the same as
the renormalization procedure
required in the effective field theory
of large scale structure~\cite{Baumann:2010tm,Carrasco:2012cv,Pajer:2013jj,Abolhasani:2015mra,delaBella:2017qjy}.
The counterterm is needed only for modes that are softer than $k$,
and at times later than their horizon exit time $t_k$.

It follows that in the $\bigBox$-smoothed version of the theory,
the back-reaction
for peak-scale modes can be accommodated
but \emph{is not predicted}
because we have no adequate rationale to estimate
its time dependence.
This time dependence is critical to evaluate
the significance of any back-reaction.
By comparison, in the $\smallBox$-smoothed version,
detailed information about the interactions and behaviour
of the peak-scale modes is retained.
It is then true that
explicit computation of the loops,
as in Eqs.~\eqref{12-type final}, \eqref{22-type final}
and~\eqref{13-type final},
supplies a predictive estimate of the back-reaction
from these modes.
However, we cannot conclude from this that
the $\smallBox$-smoothed theory
is predictive as a whole,
because the back-reaction from peak scales
cannot be measured separately.
Specifically,
we should recognize that
the $\dot{\zeta}^2$ operator
will already receive renormalizations from all scales that
were integrated out in the smoothing procedure, i.e. modes $\gg k$.
Eqs.~\eqref{12-type final} and~\eqref{13-type final}
are therefore degenerate with
further unknown renormalizations
associated with deeper ultraviolet modes~\cite{Burgess:2014lza},
and
this conclusion applies to both the
$\smallBox$- and $\bigBox$-smoothed versions of the analysis.
It is these renormalizations that compensate for the explicit
ultraviolet cutoff dependence $q \lesssim k_e$
of the loop integrals reported in~\S\ref{sec:delta N loops calculation}.
Clearly, the final correlation functions must be independent of
any such arbitrary cutoff scale.%
    \footnote{One might wonder how the infrared cutoff scale is
    compensated (as it also must be).
    As explained in footnote~\ref{footnote:ir-convergence}
    on p.~\pageref{footnote:ir-convergence},
    this can be absorbed into a sufficiently
    precise definition of the
    expectation value that is being computed.}

The practical outcome is that
the one-loop shift represented
by Eqs.~\eqref{12-type final} and~\eqref{13-type final}
is not directly observable,
and should not be used to draw conclusions about any
putative
breakdown of the loop expansion. 
Only the \emph{running} of the one-loop
correction can be observed separately~\cite{Durakovic:2019kqq}.

There is a possible exception
for the specific scenario considered here.
Deeply subhorizon modes
do not feel the curvature
of spacetime and approximately restore the
Lorentz symmetry that is spontaneously broken by
the background.
One would therefore expect loop contributions from
these modes to renormalize the Lorentz-invariant
kinetic operator $\partial^a \phi \partial_a \phi$
rather than $\dot{\zeta}^2$ or $\zeta \partial^2 \zeta /a^2$
separately. This could provide a possible rationale to
regard running generated, e.g., by
Eq.~\eqref{12-type final}
%and~\eqref{13-type final}
(or more refined estimates of the same loops)
as the largest contribution.
This would be interesting to pursue, but we leave
a detailed analysis for future work.

The soft three-point function
appearing in the 12-loop~\eqref{type 1 two}
or the (double) soft four-point function in the 13-loop~\eqref{type 3 one}
will have subleading contributions
proportional to powers of $(p/k)^2$.
As has been explained,
if the $\Or(p/k)^0$ contribution should be
subtracted to isolate the physical correlations
then the order $(p/k)^2$ term will be the
leading contribution.
This momentum dependence can be absorbed by
higher-derivative counterterms for $\zeta$.
If we take $\OpO = \OpQ = \ddot{\zeta}$
in~\eqref{eq:prototype-counterterm}
then we expect $\psi_p^{\OpO} \psi_p^{\OpQ} \sim \dimlessP_\zeta(p) p/a^4$.
Also, $\deltact$ should now have mass dimension $-2$,
which we expect can be associated with
a suppression $\sim 1/k^2$
by the `heavy' scale that has been integrated out.
In the counterterm~\eqref{eq:prototype-counterterm}
this should appear as the corresponding \emph{physical} scale
$k/a$.
Therefore we write
$\deltact = \tildeDeltact / (k/a)^2$,
with $\tildeDeltact$ taken to be dimensionless.
It follows that 
we can estimate the contribution from this operator
to the $\zeta$ two-point function,
\begin{equation}
    \langle \zeta_{\vect{p}} \zeta_{-\vect{p}} \rangle^{\prime,\ddot{\zeta}^2}_\eta
    \simeq
    \dimlessP_\zeta(p, \eta)
    \frac{\im}{p^3}
    \int_{-\infty}^{\eta}
    \frac{\d \tau}{H(\tau)^2 \tau^2} \, \frac{\Mp^2}{k^2} \tildeDeltact(\tau)
    \dimlessP_\zeta(p, \tau)
    p \times \cdots .
\end{equation}
The estimate for the time integral is the same as above,
viz.,
$\im\int_{-\infty}^\eta \d \tau \, \tau^{-2} \, \tildeDeltact(\tau) \times \cdots
\sim pC'$.
[In this expression, $C'$ is a different constant to the quantity $C$ appearing
in~\eqref{eq:12-counterterm-time-vertex-estimate}.]
Hence,
\begin{equation}
    \langle \zeta_{\vect{p}} \zeta_{-\vect{p}} \rangle^{\prime,\ddot{\zeta}^2}_\eta
    \simeq
    \frac{C'}{p^3} \frac{p^2}{k^2}
    \dimlessP_\zeta(p, \eta) .
\end{equation}
As expected, this has the correct momentum dependence
to match the order $(p/k)^2$ term from the 12- and 13-type loops.
Higher powers of $p/k$ can be handled in exactly the same way,
by introducing increasingly irrelevant operators (in the technical sense)
built from higher derivatives of $\zeta$, and suppressed by
higher powers of $k/a$.

\subsection{22-loop counterterm}
A similar analysis can be given for the 22-loop,
although the kinematic structure
of this loop
produces differences.
In this section we provide only a sketch of the renormalization
procedure, leaving a complete analysis to future work.

It was explained in~\S\ref{sec:back-reaction-quasiparticles}
that,
from the perspective of the long mode $\vect{p}$,
the 22-loop can be regarded as injection of uncorrelated
noise. We therefore expect the counterterm needed to absorb the
22-loop to have a stochastic character.
To model it,
we introduce an auxiliary field $\xi$
coupled to $\zeta$.
The precise interaction must be chosen to reproduce
the momentum dependence of~\eqref{22-type final}.
(See, e.g., Refs.~\cite{Gell-Mann:1992wkv,Calzetta:1996sy}.)
Its contribution to the $\zeta$ two-point function
involves the insertion of a $\xi$ line,
\begin{equation*}
    \langle \zeta_{\vect{q}_1} \zeta_{\vect{q}_2} \rangle^{\prime, \xi \xi}_\eta
    \supset
    \eqgraph{1ex}{0ex}{
        \begin{fmfgraph*}(90,30)
            \fmfleft{l}
            \fmfright{r}
            \fmf{plain,label=$\vect{q}_1$}{l,v1}
            \fmf{plain,label=$\vect{q}_2$}{v2,r}
            \fmf{zigzag}{v1,v2}
            \fmfv{label=$\tau$,label.angle=90}{v1}
            \fmfv{label=$\nu$,label.angle=90}{v2}
        \end{fmfgraph*}
    }
\end{equation*}
The auxiliary field has an imaginary action
that can be regarded as a probability distribution
on $\xi_{\vect{q}}$.
Working in a Gaussian approximation, the variance
of this measure
would determine the noise correlator
$\langle \xi_{\vect{q}_2} \xi_{\vect{q}_1} \rangle$.
Therefore, when used in an effective
equation of motion, $\xi_{\vect{q}}$ can be regarded
as a stochastic variable.
Alternatively, we can attempt to
integrate out $\xi$ to produce an effective counterterm
involving only the long-wavelength
field $\zeta_{\vect{p}}$.
This is a Hubbard--Stratonovitch
transformation~\cite{Stratonovich1957SPhD,Hubbard1959PhRvL}.
However,
it does not yet seem clear what type of counterterms
would be produced.

\section{Discussion}
\label{sec:discussion}
In this paper we have reconsidered
the evaluation of loop corrections to the large-scale
primordial power spectrum.
We are motivated by scenarios in which there is a peak
in the power spectrum on short scales, but the general
framework we have assembled
is not specific to this case.
The same framework could be used to evaluate
back-reaction in many scenarios---%
including those where the short-scale
power is broadly distributed
over a range of scales rather than being concentrated
in the vicinity of a peak.
However, in this case the connection to the
toy model of~\S\ref{sec:toy-model}
becomes substantially more complicated.
Instead of a single mosaic of $\smallBox$-sized
boxes, we would need to deal with a
distribution of box sizes.
Moreover, if the power spectrum no longer has a significant
blue tilt these boxes would
develop complex spatial correlations.
One would therefore have to track all relevant
correlation lengths. The necessary information is embedded within
the higher-order correlation functions,
for which a suitable prescription would have to be given.

\subsection{Advantages of the separate universe framework}
This toy model is a useful source of intuition
regarding the types of back-reaction that can occur,
and their dependence of the scale hierarchy $p/k \ll 1$.
In particular, it emphasizes that the back-reaction
effects being computed are \emph{classical}.
For this reason the separate universe framework
provides a very convenient method with which to
perform loop computations.
The complexities of the in--in formalism
are not needed,
and may even prove to be an unhelpful
distraction.
Concretely, one can regard a formula 
such as Eq.~\eqref{eq:zeta-separate-universe-fourier}
as a more complete version of the
toy-model spatial average~\eqref{eq:corrected-average-zetaL-field}.
We argue that the separate universe framework
offers
a number of important simplifications when compared
with alternative frameworks.

First, the time dependence of each correlation function
can be evaluated accurately, even accounting for transitions
between different eras---such as the transition
between ultra-slow-roll and
ordinary slow-roll, discussed in~\S\ref{sec: SR to USR to SR}.
This is critical
in order to obtain reliable estimates for the magnitude
of each loop.
The same thing
can be done using in--in, but one needs the time dependence
of the propagator to be valid to arbitrarily late times.
This is possible, but is not always easy to achieve, for example if one
uses the slow-roll approximation to solve for the mode functions~\cite{Dias:2012qy}.

Second, a separate universe expression
such as~\eqref{eq:zeta-separate-universe-fourier}
makes explicit how the different back-reaction effects
add coherently (or fail to do so) over the large $\bigBox$-size
box.
Correlations depending on a soft limit
of a $\delta X$ correlation function,
such as the 12-loop~\eqref{eq:prototype-12}
and the 13-loop~\eqref{eq:prototype-13},
can add coherently.
This is one of our primary results:
loop contributions
\emph{can} be present without volume suppression.
The same conclusion
(with some differences of detail)
was reached by Riotto~\cite{Riotto:2023hoz}.
The unsuppressed parts
are the 12- and 13-type diagrams in Fig.~\ref{fig:one-loop-delta-N}.
For the specific 3-phase ultra-slow-roll model
of~\S\ref{sec: SR to USR to SR} we focus on the effect
of modes populating a small-scale plateau
in the tree-level
$\dimlessP_\zeta$.
In this approximation we
find that the 12-loop is always suppressed,
regardless of the character of the transition
from ultra-slow roll to the final slow-roll phase,
and that the 13-type loop vanishes.
Meanwhile,
the 22-type loop is volume suppressed.
Therefore the total large-scale, one-loop correction
is small relative to the tree-level power spectrum.
This conclusion applies for models with both smooth
and instantaneous transitions.
Of course,
it must be equally
possible
to arrive at the same understanding
using the in--in approach.
The $\delta N$ version
should
be regarded merely as a reorganization of the
calculation that makes these conclusions more manifest.
For example, in the $\delta N$ calculation it is
clear that the 22-loop~\eqref{eq:prototype-22}
should be suppressed by the central-limit volume factor
$N^{-1} = (p/k)^3$.
This is substantially harder to see in a direct in--in calculation,
where the outcome depends on cancellations between diagrams with
different vertices and is afflicted by technicalities involving
boundary terms.

Third, in the $\delta N$ calculation
it is straightforward to see how powerful factorization
principles, such as the Maldacena
limit~\cite{Riotto:2023hoz,Tada:2023rgp}
or the soft factorization formulae
developed in Ref.~\cite{Kenton:2015lxa},
can be used to evaluate the soft limits
needed for the loop integrands.
These factorizations are not at all easily visible
within pure in--in perturbation theory.
Further, the appearance of soft limits
in these integrands
emphasizes that one should be concerned about
subtractions that may be necessary
to isolate physical correlations,
as opposed to correlations that are merely gauge artefacts.
Many authors have suggested that,
in a single-field, adiabatic model,
the physical correlations in the squeezed limit
decay like
$(p/k)^2$~\cite{Tanaka:2011aj,Creminelli:2011rh,Pajer:2013ana,dePutter:2015vga}.%
    \footnote{For a dissenting view, see
    Matarrese {\etal}~\cite{Matarrese:2020why}.}
If this is the case, then the order $(p/k)^0 \sim 1$
contribution
to an integral over a soft limit
such as~\eqref{eq:prototype-12}
would apparently be subtracted.
This would make such loop corrections exponentially
small, because
the surviving contribution would presumably
be of order $(p/k)^2 \ll 1$.
This is at the same order in $p/k$
as the loop contributions
found by Tasinato using a large $|\eta|$
expansion~\cite{Tasinato:2023ukp},
and provides one possible interpretation of those terms.
However, if that is the case,
it is not clear why the calculation of
Ref.~\cite{Tasinato:2023ukp}
would not capture the $\Or(1)$ contribution
from the Maldacena factorization in the squeezed limit.

If soft limit subtractions make the
loop of order $(p/k)^2$, then
unfortunately we would be obliged to conclude
that the separate universe formula is
\emph{not} adequate to evaluate it.
This is because a formula such as~\eqref{eq:prototype-12}
should be corrected by gradient effects.
In a model with two scales (here, $p$ and $k$)
it is possible for these
corrections to scale like $(p/k)^2$~\cite{Jackson:2023obv}.
We would therefore be unable to capture these effects
accurately, unless one included the first gradient
corrections to the separate universe framework.
A formalism for doing so was elaborated by Tanaka \& Sasaki~\cite{Tanaka:2006zp,Tanaka:2007gh}.
Whether or not one chooses to pursue this possibility,
because the surviving one-loop term that
does \emph{not} depend on a soft limit
is the incoherent 22-loop
which is suppressed by a volume factor,
we would still be able to conclude that
the one-loop corrections are negligible in a single-field,
adiabatic model.
This is already sufficient
for models with a short-scale peak that are relevant
for primordial black hole formation,
because two-loop terms
would presumably be negligible even if they are not
volume suppressed.
Nevertheless,
as a point of principle,
it would be very valuable
to extend the conclusion to all orders
in the loop expansion.

A similar discussion
applies to the 13-loop~\eqref{eq:prototype-13}.
In the Gaussian approximation to the 4-point function used
in this paper, the integral in~\eqref{eq:prototype-13}
reduces to a double-soft limit
involving the insertion of two soft $\delta X$
modes into the hard $\delta X$ two-point function.
If all limits of this kind are subtracted then one
must conclude that a loop contribution that is \emph{not} volume
suppressed would require the presence of at least one
isocurvature mode.
In models containing such modes,
the leading loop effects would
return to lowest order in $p/k$,
and therefore
the formulae presented in~\S\ref{sec:separate-universe-deltaN}
could be used to evaluate their magnitude. 

It is not yet entirely clear how these considerations
should be applied to the loop
contributions of~\S\ref{sec: loops in delta N}
and in particular how
any subtraction should be implemented in a way
that is independent of any infrared regulator scale.
It is also unclear how (or whether) the subtraction
operation commutes with computation of the loop
integrals. We believe that both of these issues
require an adequate resolution before
we are in a position to evaluate squeezed
contributions to each loop.
However, we leave these issues for future work.

In the main text of the paper, we have not performed explicit subtractions.
This enables comparison of our results
with previous estimates that have been reported in the literature.

\subsection{The role of nonlinear couplings}
The appearance of soft limits
clarifies how the loop
depends on nonlinear mode couplings.
Previous discussions have generally emphasized
the dependence of the one-loop correction on powers of $|\eta| \sim 6$---%
see the $(\Delta \eta)^2=|\eta|^2$ factor in the final result of Ref.~\cite{Kristiano:2022maq}---%
because this determines whether the effect is significant.
We have argued above that this emphasis can obscure
the fact a soft limit is involved,
with possible nontrivial scaling as $p/k \rightarrow 0$.
In our analysis, some of
these powers of $\eta$
are proxies for a nonlinear coupling---%
either a soft limit
of a higher-order correlation
function such as $\langle \delta X^I \delta X^J \delta X^K \rangle$,
or a higher-order time-dependent coefficient such
as $N_{IJ}$.

Each loop involves at least one of these
two types of nonlinearity.
They reflect different aspects of the physics.
Coefficients such as $N_{IJ}$ or $N_{IJK}$
reflect the nonlinear dependence of
$\zeta$ on the early-time $\delta X$
fluctuations.
This type of nonlinearity
is responsible for the interactions
that allow short-wavelength modes
at the base time $t_\ast$
to cascade power into long-wavelength
infrared modes
at a later time,
but says nothing about whether this power can
add up coherently over a large volume.
When these coefficients appear
in a contribution to $\zeta$,
they organize themselves into
combinations such as $N_{IJ} N^I N^J / (N^K N_K)^2$
or $N_{IJK} N^I N^J N^K / (N^L N_L)^3$
that can be recognized as
the dominant contribution to
the reduced bispectrum $\fNL(k_1, k_2, k_3)$,
or a higher nonlinearity parameter such as $\gNL$,
when these parameters are
large~\cite{Lyth:2005fi,Seery:2006js,Byrnes:2006vq}.
These nonlinearity parameters are
evaluated for an equilateral momentum configuration
$k_i \sim k$
with characteristic scale $k$ corresponding to
the short-scale modes,
and at a time equal to the time of evaluation of the
loop.
One can regard each such parameter
as a measure of the nonlinearity of the short-scale modes
among themselves.
(We emphasize that nothing is being said about the
\emph{shape} of the correlations, only that the nonlinearity
parameter is evaluated on an equilateral \emph{configuration}.)
If these parameters are significantly suppressed, the
loop is likewise significantly suppressed.%
    \footnote{When the reduced bispectrum $\fNL(k_1, k_2, k_3)$
    becomes sufficiently small,
    one must distinguish between the combination
    $N_{IJ} N^I N^J / (N^K N_K)^2$
    and the equilateral configuration of the
    reduced bispectrum. There is a difference from
    the intrinsic non-Gaussianity of the $\delta X$
    fluctuations.
    It is the combination $N_{IJ} N^I N^J / (N^K N_K)^2$ that controls
    the amplitude of the loop.}
This conclusion is likewise difficult to see
from a direct in--in calculation, where it is obscured by
the time integrations that appear at each vertex
of the loop.

On the other hand, soft limits of
correlation functions such as
$\langle \delta X^I \delta X^J \delta X^K \rangle$
measure the coupling between long- and short-wavelength modes.
This is an independent effect,
responsible for allowing power cascaded
from short-wavelength modes to add up coherently
over large regions.
We can interpret the corresponding physics, as follows.
Transfer of power from small to large scales
can depend only on local environmental
conditions in each $\smallBox$-scale box.
To generate a coherent effect, the local environment
must first \emph{react}
to the presence of the long mode.
This biases nearby boxes to behave in a correlated
way, effectively tracing the skeleton
laid down by the long-wavelength modes.
It is this correlated reaction to the presence of the
long mode that allows \emph{back}-reaction onto
the long mode itself.

The long--short coupling
corresponds to a nonlinearity parameter
evaluated in a squeezed momentum configuration.
In our framework the nonlinearity
parameters that appear are usually not related to $\fNL$, $\gNL$, etc.,
in a simple way,
because the soft limit involves $\delta X$ correlations functions rather
than those of $\zeta$.
(The relation may be more direct in a single-field model.)
In principle it is possible to have one
type of nonlinearity without the
other, especially where the underlying correlations have
significant scale dependence.
However, as we have emphasized in~\S\S\ref{sec:separate-universe-deltaN}--\ref{sec:delta N loops calculation},
\emph{both} types of nonlinearity
are needed to generate large loop contributions
that are not suppressed by central-limit-like effects.
The 22-type loop involves only short-scale
nonlinearities
and therefore depends on the square of the equilateral-mode
$\fNL$ evaluated on the corresponding scales.
However, the 12- and 13-type loops
involve both types of nonlinearity.

This complex structure, involving nonlinear
parameters evaluated on
equilateral and squeezed configurations
of higher-order correlation functions,
is again not easily visible in the pure in--in framework.
There, the different contributions to each nonlinear
parameter are broken up between many interaction
vertices and must be reassembled to obtain
a full picture.
In the separate universe framework we can deal
with these correlations as a single entity from the outset.

\subsection{The Wilsonian EFT description}
\label{sec:discussion-wilsonian-eft}
In~\S\ref{sec:wilsonian-counterterms}
we have shown how to relate loop calculations
in the separate universe framework
when the smoothing scale is varied.
To calculate back-reaction from the peak,
our
preferred prescription (see~\S\ref{sec:delta-N-back-reaction})
involves smoothing
on the scale $\smallBox\sim 1/k$
associated with substructure generated by the
peak in the power spectrum.
This version of the calculation involves explicit loop
contributions,
described in~\S\ref{sec:back-reaction-quasiparticles}.
The loops have a natural cutoff at the
smoothing scale $\smallBox$.
If, instead, we choose to smooth on a larger scale,
some or all of the peak-scale
modes are removed from
our effective description.
This depopulates the degrees of
freedom available to circulate in the loops.
Eventually, they become merely vestigial
when the smoothing scale is sufficiently large
that no peak modes remain.
The effects they previously
described must instead be absorbed
into renormalization of local operators
in the $\zeta$ effective description.
This is the usual outcome from varying the cutoff
(in a Wilsonian sense)
in any effective theory.

In~\S\ref{sec:wilsonian-counterterms}
we argued that
effects from the 12-type loop
(which are not volume suppressed)
can be absorbed
into a renormalization of the $\zeta$ kinetic
operator $\dot{\zeta}^2$.
The same discussion applies to the 13-type loop,
which shares the same momentum dependence.
Corrections at higher orders in $(p/k)^2$
can be absorbed into higher-derivative
operators that are
suppressed by the `heavy' scale $k$
associated with the peak.
The resulting effective description, including loops,
organizes itself into an expansion in powers
of the small parameter $p/k \ll 1$.
It furnishes a very convenient tool with which
to understand
the significance of back-reaction in this model.
However, its utility depends on a separation
of scales between the short peak modes $\sim k$
and the long CMB modes $\sim p$.
The peak does not need to be monochromatic,
but it must be sufficiently well separated
from the modes whose effective
description we are trying to find.

The 3-phase ultra-slow-roll model was used to
calculate explicit estimates for the loop
contributions in~\S\ref{sec:delta N loops calculation}.
The EFT analysis raises the interesting possibility
that one could perhaps even
dispense altogether
with a dynamical model
for the short-scale modes,
instead accounting for their influence
through parameters of the effective description.
The primary difficulty one would encounter
in this approach is to relate the EFT parameters
to the height and width of the short-scale
spike, which would be needed to relate the back-reaction
to phenomenological considerations such as the
abundance of primordial black holes.

\subsection{Degeneracy with UV counterterms}
Taken together with
the discussion in~\S\ref{sec:discussion-wilsonian-eft},
the conclusion of~\S\ref{sec:wilsonian-counterterms}
is that local counterterms are needed
to compensate for the dependence of each loop integral
on the smoothing scale.
The smoothing scale functions as a natural cutoff
on each integral,
although (of course) we can impose a lower cutoff by
hand if we wish.
In this section we denote the cutoff by $\Lambda$.
In the language of~\S\ref{sec:delta N loops calculation}
the cutoff scale
is the upper limit of integration, $\Lambda = k_e$.

Provided $\Lambda$
is much larger than any wavenumber of interest,
we expect that
changes in its value can be absorbed by adjusting the
coefficients of local operators.
These are the counterterms.
In flat space one can see this as follows.
(In the following, $q$ and $p$ temporarily label
4-vectors.)
Consider a wavenumber $q$ contributing to the loop
near the cutoff.
We focus on the two-point function, so there is
only one relevant wavenumber, $p$,
and assume a Wick rotation to Euclidean signature.
By assumption $q \gg p$,
and therefore
we can Taylor expand the loop integrand
as a series in $p/q$.
The series expansion is regular if
there are no infrared singularities
for $p \rightarrow 0$.
It follows that the entire integral
can also be expressed as a series in $p^2$.
This is exactly the behaviour that would be produced
by insertion of one or more local operators built from the fields
and their derivatives,
as many as are needed to reproduce
the terms in the series.%
    \footnote{For simplicity, and because it is the case we need,
    the discussion in this paragraph has been framed as if there is
    only one loop momentum $q$.
    This implicitly restricts to one-loop contributions.
    However, the argument can be generalized to higher loops.}
Contributions that are \emph{not}
an analytic function of $p^2$,
and which therefore are not degenerate
with insertion of local operators,
must come from the region of the integral where
$q \sim p$
and the Taylor expansion in $p^2$ does not apply.
Any such nonanalytic pieces cannot be modified by
unknown ultraviolet effects and constitute reliable
low-energy predictions of the model.
For more details, see (e.g.) the classic papers
by Donoghue~\cite{Donoghue:1993eb,Donoghue:1994dn}.

A similar argument can be applied to
the loop integrands encountered in cosmology.
Moreover, this argument explains why it is meaningful
to compute loop corrections in an
effective field theory at all.
By construction, the ultraviolet behaviour of the effective
description is intentionally different to the
original parent theory. This is certainly true for the
separate universe framework, which predicts entirely
fictitious subhorizon behaviour
because gradients are neglected.
But
in any field theory, the loops run over all momenta.
Therefore one might worry that there are relevant effects,
generated by the contribution of subhorizon quanta,
that could be captured
using the in--in formalism
but could never be reproduced
by loops based on the $\delta N$ formula.%
    \footnote{Of course, the usual cosmological framework
    based on a local quantum field theory description of the
    matter fields, and a semiclassical gravitational field,
    can itself only be an effective description.
    Therefore we should not give
    unwarranted deference its ultraviolet predictions.
    The same argument applies to loops computed using in--in,
    just with the cutoff at a higher (subhorizon) scale.}
The resolution is that, if such effects are present, they
can only renormalize the coefficients of
local operators.

Clearly, the necessity to include these counterterms
complicates the interpretation
of loop integrals such
as those computed in~\S\ref{sec:delta N loops calculation}.
By choosing a suitable cutoff we were able to retain
sufficient dynamical information about the short
scale modes
to obtain reliable estimates of the back-reaction effect.
This procedure led to our final expressions for the 12-, 22-
and 13-type loops, Eqs.~\eqref{12-type final},
\eqref{22-type final}
and~\eqref{13-type final}.
Also,
in~\S\ref{sec:wilsonian-counterterms}
we demonstrated
explicitly
that at least the 12- and 13-type loops
had momentum dependence, aside from the
normalization $1/p^3$, that is analytic in $p$
and could be absorbed into a renormalization
of $\dot{\zeta}^2$.
(We have now reverted to our usual notation in
which $\vect{p}$ is a CMB-scale wavenumber and $p$
is its magnitude.)
Clearly this is a consequence of the large hierarchy $p/q$
between the mode $\vect{p}$ of interest and the
modes $\vect{k}$ that form the peak.
We should expect almost the entire loop correction
to $\vect{p}$, from modes of order $\vect{k}$,
to be degenerate with counterterms.
Only if $\vect{p}$ is changed to be of order the peak scale
should we expect the loop to produce significant nonanalytic
terms.

Unfortunately, this means that
our results~\eqref{12-type final} and~\eqref{13-type final}
must be combined with unknown ultraviolet
contributions before they
are used to predict observables.
By themselves, their meaning is ambiguous.
There are limited circumstances in which
it may be possible
to extract meaningful information.
First, if logarithmic running is present
then this can be regarded
as unambiguous,
because the logarithm is nonanalytic and
cannot be compensated by ultraviolet
contributions.
This strategy could be used to obtain
a prediction from the
renormalization of the $\dot{\zeta}^2$ kinetic term.
The running predicted in this way
would have to compete with running from
other unknown sources, but a variant
of the usual argument
about cancellation of ultraviolet effects
would apply.
Either the running from unknown ultraviolet effects
is small, in which case the contribution from the loop is
dominant,
or else it is large, in which case we expect running at least
as large as the low-energy prediction. Either way,
absent an unexpected tuning,
the
low-energy value gives a \emph{lower bound} on the expected
behaviour.

Second, if the leading contribution
enters at higher order in $p/k$
(perhaps because we are working in a model
for which the contributions at $(p/k)^0$
are not physical and must be subtracted)
then the loop renormalizes
a higher-derivative operator
such as $\Mp^2 \deltact \ddot{\zeta}^2$,
where the Wilson coefficient
$\deltact$
would be suppressed by the heavy scale $(k/a)^2$
as explained in~\S\ref{sec:wilsonian-counterterms}.
In this case there is a rationale to regard the
loop contribution obtained by integrating
over the peak
as the largest part of $\deltact$.
Contributions from quanta lying deeper in the ultraviolet
would presumably be suppressed by an even larger scale.

These considerations have not yet been applied to loop
computations based on the separate universe framework.
In this paper we present our results in a form that
can easily be compared with
previous expressions reported in the literature.
However, before it is possible to
evaluate the significance of one-loop corrections in
this model (and others),
there is clearly a critical need
to develop measures of the loop effect that are insensitive
to ultraviolet effects.
We hope to return to this in future work.

\section*{Acknowledgments}
The authors are grateful to Chris Byrnes for very helpful comments and questions.
LI would like to thank Matteo Braglia, Andrew Gow, Joseph Jackson, Alfredo Urbano and David Wands for many interesting discussions. LI was supported by a Royal Society
funded postdoctoral position for much of this work and acknowledges current
financial support from the STFC under grant ST/X000931/1.
DJM is supported by a Royal Society University Research Fellowship.
DS was supported by STFC grants ST/X001040/1 and ST/X000796/1.

\appendix
\section{Numerical results for a 3-phase ultra-slow-roll model}
\label{sec: numerical example}
In this Appendix we give
explicit numerical results for a 3-phase ultra-slow-roll model of the
type discussed in~\S\ref{sec: SR to USR to SR}.
This is one example of a model with a smooth USR to SR transition, 
which allows us to compare the analytical results of~\S\ref{sec: SR to USR to SR}
with numerical counterparts calculated from an explicit inflationary potential.
The results discussed in this Appendix are obtained by using the public
code
\texttt{PyTransport}~\cite{Dias:2016rjq,Ronayne:2017qzn, Mulryne:2016mzv}.

\begin{figure}
    \centering
    \captionsetup[subfigure]{justification=centering}
    \begin{subfigure}[b]{0.49\textwidth}
    \includegraphics[width=\textwidth]{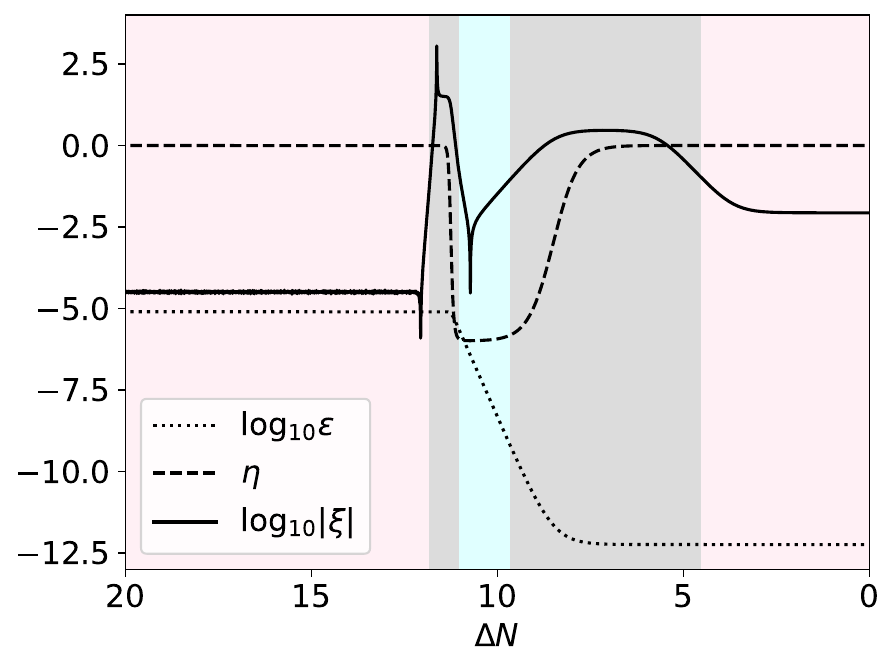}
    \end{subfigure}
    \begin{subfigure}[b]{0.48\textwidth}
    \includegraphics[width=\textwidth]{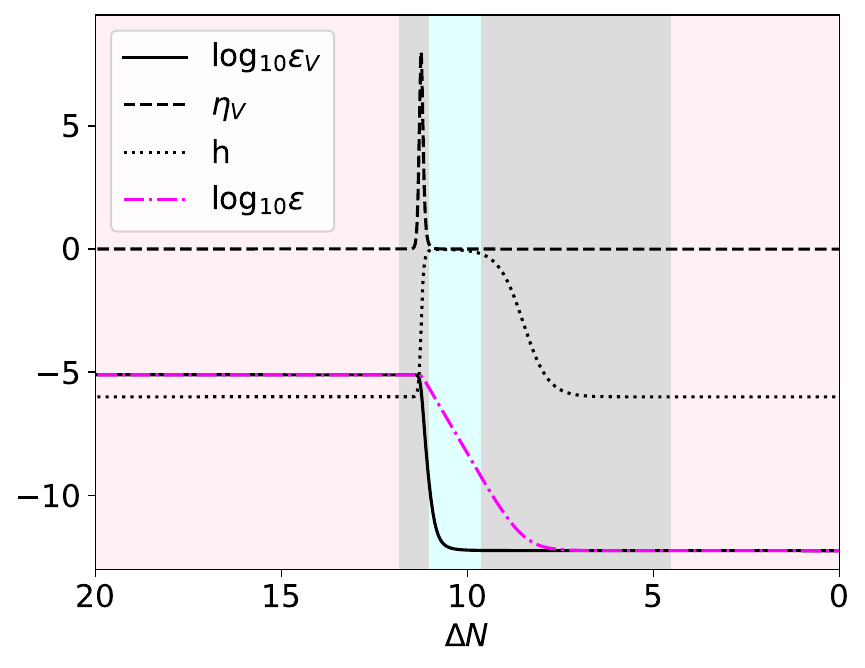}
    \end{subfigure}
    \caption{Slow-roll parameters evolution. \textbf{Left panel:} Evolution for the first three Hubble slow-roll parameters against $\Delta N\equiv N_\text{end}-N$.  
    \textbf{Right panel:} Evolution of the first two potential slow-roll parameters, $\epsilon_V$ and $\eta_V$, against $\Delta N$. We also include the time evolution of the $h$ parameter, see Eq.~\eqref{h def}, and $\log_{10}\epsilon$. In both panels, times corresponding to SR, USR evolution and transitions between the two are highlighted in pink, blue and gray respectively. We define the transitions times as the times when $|\xi|$ crosses $0.1$ (see the left panel).}
    \label{fig:background}
\end{figure}

We employ a single-field inflationary model, with potential 
\begin{equation}
\label{potential}
    \frac{V(\phi)}{p_0} =
    p_1 + p_2 
    \Big[ 
    \ln \big(
            \cosh p_3\phi
        \big)
        + \big(
            p_3 + p_4
        \big) \phi
    \Big]\;.
\end{equation}
The parameters are
\begin{subequations}
\begin{align}
    p_0 & = 4\times 10^{-12} , \\
    p_1 & = 1 , \\
    p_2 & = 5\times 10^{-7} , \\
    p_3 & = 4\times 10^3 , \\
    p_4 & = 2 .
\end{align}
\end{subequations}
\begin{figure}
    \centering
    \includegraphics[width=0.6\textwidth]{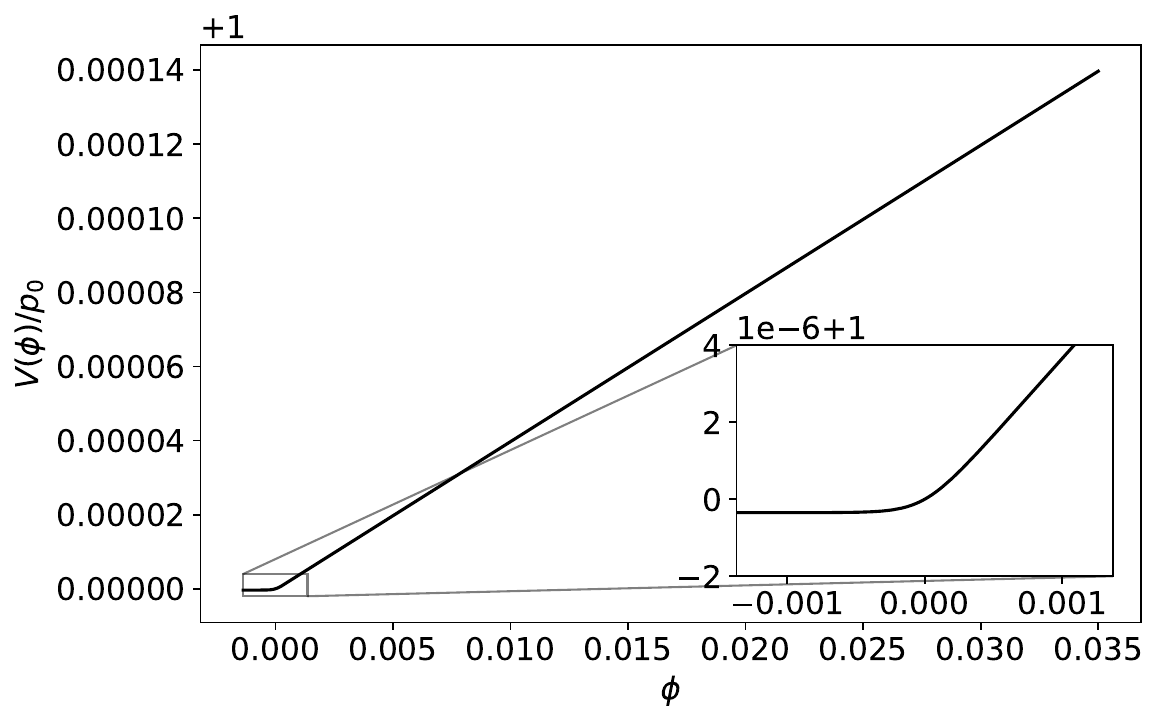}
    \caption{Shape of the (normalized) toy-model potential in Eq.~\eqref{potential}. The x-axis range corresponds to the values that the inflaton, $\phi$, acquires in the 20 e-folds following the horizon crossing of the CMB scale. The inset panel shows the potential shape at the transition between the two constant slope regions.}
    \label{fig:potential}
\end{figure}
The potential in Eq.~\eqref{potential} features two regions with constant slope, separated by a smooth transition region, see Fig.~\ref{fig:potential}. In this sense, it can be regarded as a smoothed version of the Starobinsky potential \cite{Starobinsky:1992ts}. 
We note that this should not be considered as a realistic model of inflation, see e.g. the evolution of $\epsilon$ in the left panel of Fig.~\ref{fig:background}, rather as a toy-model providing a simple realisation for the 3-phase ultra-slow-roll dynamics. 

Our fiducial background evolution is obtained by setting the initial condition $\phi_\text{in}=0.075$ and SR initial velocity. We stop the background evolution at e-folding time $N_\text{end}=30$, and identify the large-scale mode $p$ as the one crossing the horizon 20 e-folds before the end of inflation, $\Delta N_p\equiv N_\text{end}-N_p=20$. In realistic models $p$ cannot be the CMB mode, $0.05\,\text{Mpc}^{-1}$, as we would expect $\Delta N_p\in [50,60]$ in this case \cite{Liddle:2003as, Planck:2018jri}, as well as a larger separation of scales between the CMB and peak scales. Nevertheless, for the purpose of the current numerical calculations, this choice of large-scale mode is acceptable, as (i) $p/\bar k \simeq 10^{-5} \ll 1$, coherent with our working assumption of a large separation of scales between the large-scale mode $p$ and the peak scales; (ii) $\zeta_{\vect{p}}$ is approximately constant on super-horizon scales. With this in mind, we will refer to $p$ as the CMB scale.  

We represent in the left panel of Fig.~\ref{fig:background} the evolution of the first three Hubble slow-roll parameters, see Eq.~\eqref{slow roll paramters}, against $\Delta N\equiv N_\text{end}-N$. The time-evolution of $\epsilon$, $\eta$ and $\xi$ shows the sequence of SR$\to$USR$\to$SR phases. We choose to define the times at which the SR$\rightarrow$USR and USR$\rightarrow$SR transitions occur as the times when $|\xi|$ becomes larger than 0.1\footnote{This choice is arbitrary (within reason), e.g. one could instead define the transition times as the times when $|\xi|$ becomes larger than $0.05$. The choice of transition times will in turn influence the value of $h$, see Eq.~\eqref{h def}. For a smooth transition, as it is for the potential \eqref{potential} (see the evolution of $\eta$ in the left panel of Fig.~\ref{fig:background}), $h\to0$. For the time-evolution of $h(N)$, see the right panel of Fig.~\ref{fig:background}. In order to employ the analytic results derived in Ref.~\cite{Cai:2018dkf} to describe the dynamics of our toy model, we need therefore to select a transition time such that $h\to0$. This is the case for the criterion $|\xi|\geq 0.1$, which yields $h\simeq -0.2$. The choice $|\xi|\geq 1$ instead leads to $h\simeq -2$, which is clearly not appropriate if we want to apply the analytic results of Ref.~\cite{Cai:2018dkf}.}. 
This choice in turn identifies the start and end of the USR phase, therefore fixing the values of $N_s$ and $N_e$ respectively.  
In the first SR phase, $\epsilon$, $\eta$ and $\xi$ are small and the inflaton slowly rolls downs its potential. During this first SR phase, the CMB scale crossed the horizon, $\Delta N_p=20$. During the subsequent USR phase, $\epsilon$ decreases as $a^{-6}$ and $\eta\simeq -6$. During the USR$\to$SR transition, $\eta$ decreases in magnitude and $\epsilon$ becomes constant again. The time-evolution of $\epsilon$ shows that the potential \eqref{potential} doesn't yield to the end of inflation (happening when $\epsilon=1$), which explains why we artificially stop the background numerical evolution, after the system has evolved back to SR. 

In the right panel of Fig.~\ref{fig:background} we display the potential slow-roll parameters $\epsilon_V$ and $\eta_V$, see Eq.~\eqref{potential slow roll paramters}, against $\Delta N$. The parameter $\eta_V$ displays a spike during the SR$\to$USR transition, a feature common to potentials supporting a USR phase. We note that $\epsilon$ and $\epsilon_V$ are very close to one another during slow roll \cite{Liddle:1994dx}, while $\epsilon\gg \epsilon_V$ during USR. In the same panel, we also represent the time-evolution of $h(N) \equiv 6 \sqrt{2\epsilon_V}/\pi$, whose value at the end of the USR phase, $h$, see Eq.~\eqref{h def}, defines the character of the USR$\to$SR transition. For our choice of $N_e$, we have $h\simeq -0.2$, consistent with a smooth transition \cite{Cai:2018dkf}.

\begin{figure}
    \centering
    \includegraphics[width = 0.6\linewidth]{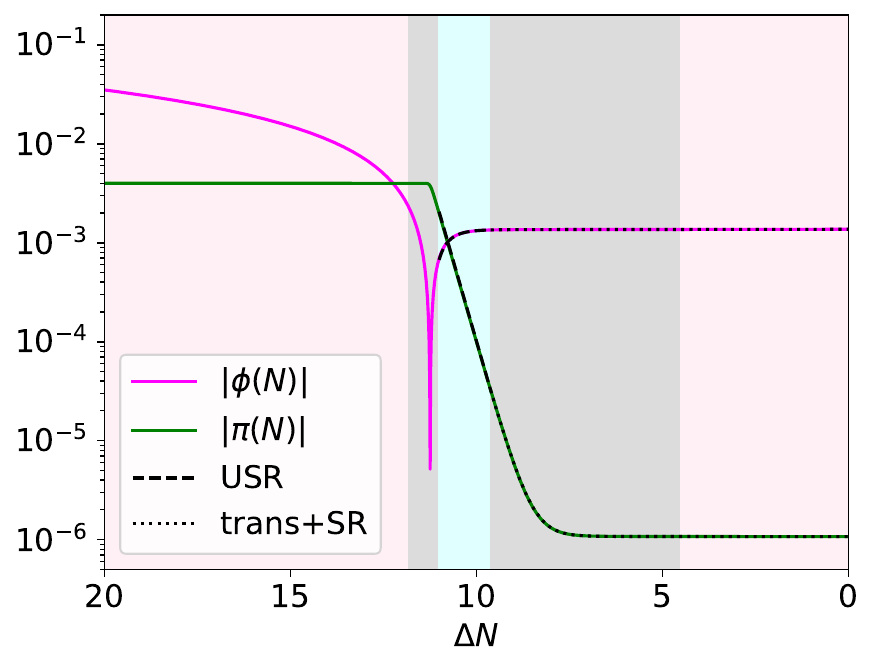}
    \caption{Numerical evolution of $|\phi(N)|$ and $|\pi(N)|$ against $\Delta N$. The black dashed and dotted lines represent the analytic solutions derived in~\S\ref{sec: delta N peak mode from hc} for the USR and USR$\rightarrow$SR transition and following SR phase respectively. Times corresponding to SR, USR evolution and transitions between the two are highlighted in pink, blue and gray respectively.}
    \label{fig:analytic approximations}
\end{figure}
In Fig.~\ref{fig:analytic approximations}, we compare  the numerical solutions for the inflaton evolution and its velocity, $|\phi(N)|$ and $|\pi(N)|$, with the analytical expressions derived in~\S\ref{sec: delta N peak mode from hc}, see Eqs.~\eqref{pi and phi USR} (black, dashed lines) and Eqs.~\eqref{phi N tr}-\eqref{pi N tr} (black, dotted lines). Fig.~\ref{fig:analytic approximations} shows an excellent agreement between numerical results and analytical expressions.  

\begin{figure}
    \centering
    \includegraphics[width=0.6\textwidth]{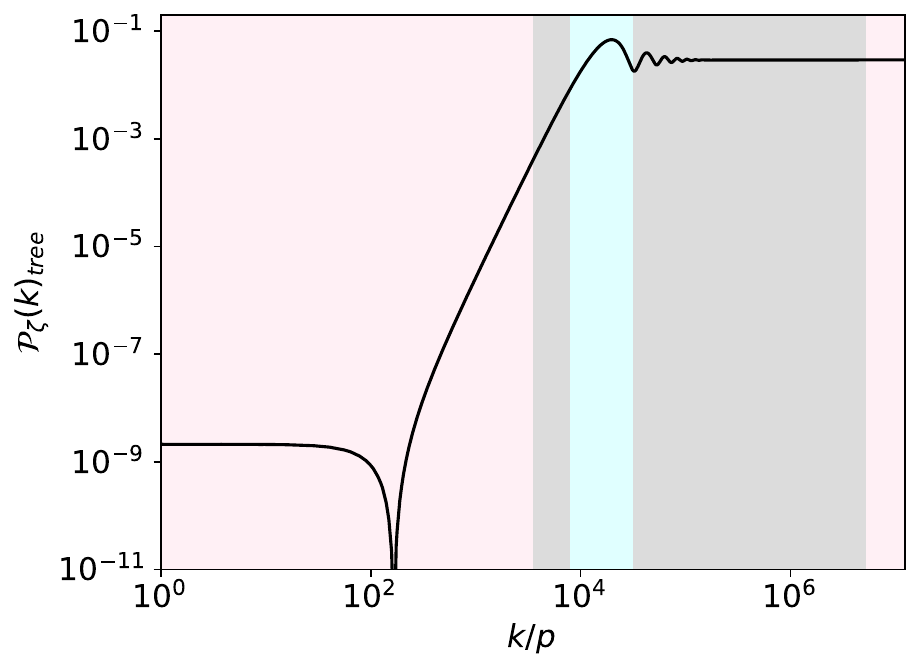}
    \caption{Tree-level, dimensionless scalar power spectrum, $\dimlessP_\zeta(k)_\text{tree}$, calculated for the potential \eqref{potential}. The CMB mode, with comoving wavenumber $p$, is defined by $\Delta N_p = 20$. We shade regions of $\dimlessP_{\zeta}(k)_{\text{tree}}$ for comoving wavenumbers that crossed the horizon during the SR, USR phases, and transition between these, in pink, blue and gray respectively. The transition times are defined as in Fig.~\ref{fig:background}. The amplitude at the peak is $\dimlessP_\zeta(\bar k)_\text{tree}\simeq0.07$.}
    \label{fig:Pz}
\end{figure}
The numerical, tree-level scalar power spectrum, $\dimlessP_\zeta(k)_\text{tree}$, is shown in Fig.~\ref{fig:Pz}---%
note that here we use $k$ to indicate a generic comoving wavenumber, while in the main text it usually refers to the peak scales. The behavior of $\dimlessP_\zeta(k)_\text{tree}$ can be understood in terms of the super-horizon ($k\ll aH$) evolution of the curvature perturbation, 
\begin{equation}
    \label{super horizon zeta}
    \zeta(N)_{k\ll aH}
    =
    c_1
    + c_2 \int \d N \;
    \exp\bigg(
        {-\int} \d N' \; \Big[ 3-\epsilon(N') + \eta(N') \Big]
    \bigg) ,
\end{equation}
where the first solution is often referred to as the \textit{growing} mode and the second one as the \textit{decaying} mode. During SR inflation ($\epsilon, \eta \ll 1$), the second solution quickly decays and therefore the curvature perturbation is constant on super-horizon scales. On the other hand, during USR, in which case $\epsilon\ll1$ and $\eta=-6<-3$, the decaying mode grows exponentially, and $\zeta$ is no longer constant on super-horizon scales. For a large-scale mode that crossed the horizon long before the onset of the USR phase, the decaying mode is so small at that time that its exponential growth during USR is not appreciable. This is the case, e.g., for the CMB mode $p$. While modes corresponding to the portion of $\dimlessP_\zeta(k)_\text{tree}$ growing as $k^4$ also crossed the horizon during SR, they did so close to the onset of the USR phase, such that the exponential growth of the decaying mode is appreciable at these scales. 

\begin{figure}
    \centering
    \captionsetup[subfigure]{justification=centering}
    \begin{subfigure}[b]{0.49\textwidth}
    \includegraphics[width=\textwidth]{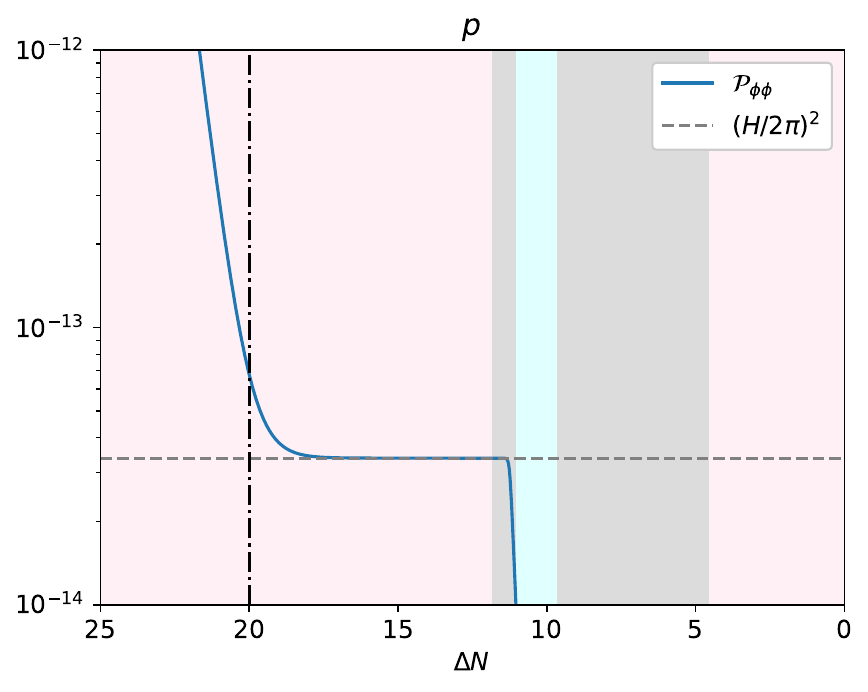}
    \end{subfigure}
    \begin{subfigure}[b]{0.49\textwidth}
    \includegraphics[width=\textwidth]{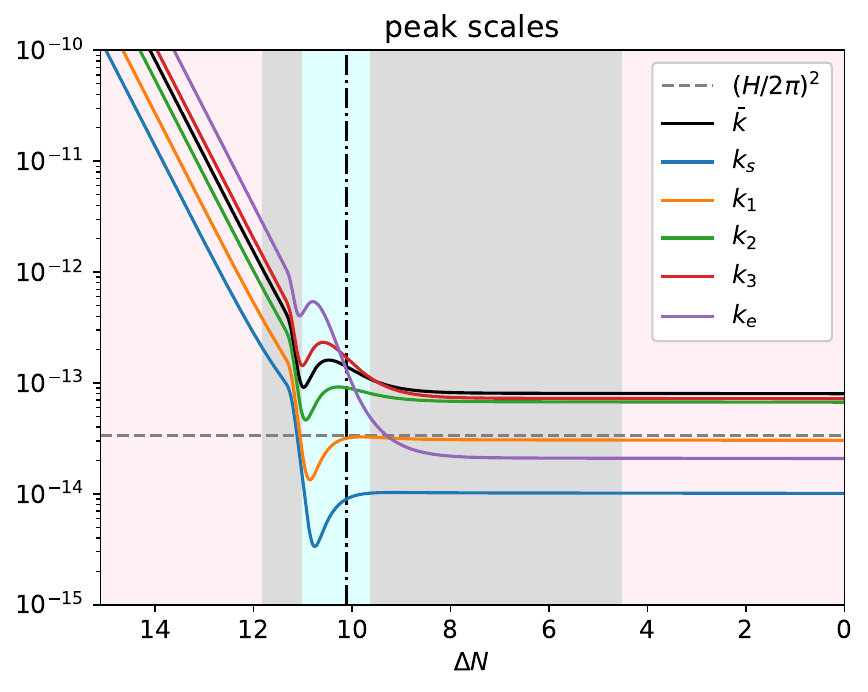}
    \end{subfigure}
    \caption{Field fluctuations around horizon-crossing. \textbf{Left panel:} Evolution of $\dimlessP^{\phi\phi}$ for the CMB mode $p$ against $\Delta N\equiv N_\text{end}-N$. The horizon-crossing time of the CMB mode is highlighted with a vertical, dot-dashed, black line. The value of $H^2/(2\pi)^2$ at horizon crossing is represented with a dashed, gray line. \textbf{Right panel:} Evolution of $\dimlessP^{\phi\phi}$ for the peak scale, $\bar k$, and 5 more peak scales, $k_s\leq k \leq k_e$, against $\Delta N\equiv N_\text{end}-N$. Each scale considered crossed the horizon at a different time between $N_s$ and $N_e$. In particular, we highlight the horizon-crossing time of $\bar k$ with a dot-dashed, black line. The dashed, gray line represents the value of $H^2/(2\pi)^2$ when $\bar k$ crossed the horizon. We expect this value to be the same for $k_s\leq k \leq k_e$ since $H\simeq \text{const}$ during USR. In both panels we highlight times corresponding to SR, USR evolution and transitions between the two in pink, blue and gray respectively. }
    \label{fig:horizon crossing values}
\end{figure}

Modes crossing during the USR phase correspond to the peak in the scalar power spectrum, which is characterized by oscillations. These are due to the localized spike in $V_{\phi\phi}$ just before the onset of USR, see the plot of $\eta_V$ in the right panel of Fig.~\ref{fig:background}. For an analytic treatment of these oscillations in a related model see, e.g., \cite{Martin:2011sn}. 
The $\delta N$ calculation in~\S\ref{sec: delta N peak mode from hc}, see Eq.~\eqref{Pzeta peak scale}, yields $\dimlessP_\zeta(\bar k)_\text{tree}\approx 0.027$, which is of the same order of magnitude as the numerical value $\dimlessP_\zeta(\bar k)_\text{tree}\approx 0.07$. The discrepancy between the $\delta N$ and numerical results is due to the use of Eq.~\eqref{massless correlator} in the $\delta N$ calculation, i.e. the oscillations in the power spectrum are not taken into account. For more details, see Fig.~\ref{fig:horizon crossing values} and the discussion below. 

We note that on very small scales the peak in $\dimlessP_\zeta(\bar k)_\text{tree}$ does not decrease in amplitude, as expected within realistic models, but rather settles into a plateau. 
 
In Fig.~\ref{fig:horizon crossing values} we represent the e-folding evolution around horizon-crossing of $\dimlessP^{\phi \phi}$ for the CMB mode $p$ (left panel) and for six peak-scale modes (right panel). We compare these with the horizon-crossing value predicted for massless, non-interacting fields on de Sitter, see Eq.~\eqref{massless correlator}, which we use in the analytical calculations of~\S\ref{sec: SR to USR to SR}  and~\S\ref{sec:delta N loops calculation}. The results in the left panel show that working with Eq.~\eqref{massless correlator} is justified when considering the CMB mode $p$, as $\dimlessP^{\phi\phi}(p)$ evolves towards $H(t_p)^2/(2\pi)^2$ soon after horizon crossing. On the other hand, $\dimlessP^{\phi\phi}$ for peak scales oscillates around $H(t_{\bar k})^2/(2\pi)^2$, resulting into the $\dimlessP_\zeta(\bar k)_\text{tree}$ oscillations seen in Fig.~\ref{fig:Pz}. Since the numerical values of  $\dimlessP^{\phi\phi}$ soon after horizon crossing are approximately of the same order of magnitude as $H(t_{\bar k})^2/(2\pi)^2$, we employ Eq.~\eqref{massless correlator} in~\S\ref{sec:delta N loops calculation}. 
Nevertheless, the results in the right panel of Fig.~\ref{fig:horizon crossing values} flag up the limitations of analytical approximations (for a related discussion, see Ref.~\cite{Jackson:2023obv}).
In particular, the use of Eq.~\eqref{massless correlator} implies that the effect of the oscillations at peak scales is not included in the analytical calculation of~\S\ref{sec:delta N loops calculation}.  

\begin{figure}
\centering
\captionsetup[subfigure]{justification=centering}
\begin{subfigure}[b]{0.49\textwidth}
\includegraphics[width=\textwidth]{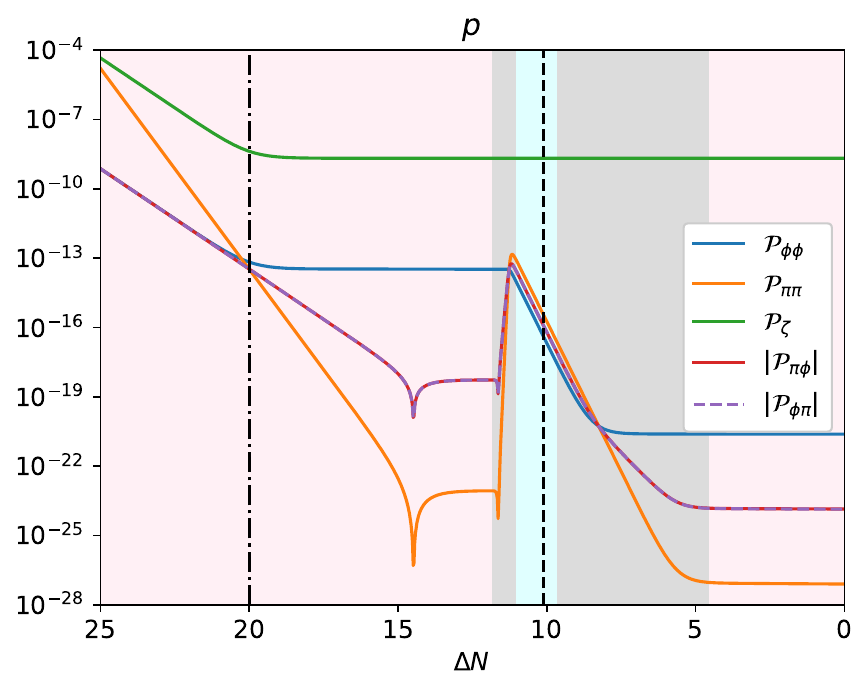}
\end{subfigure}
\begin{subfigure}[b]{0.49\textwidth}
\includegraphics[width=\textwidth]{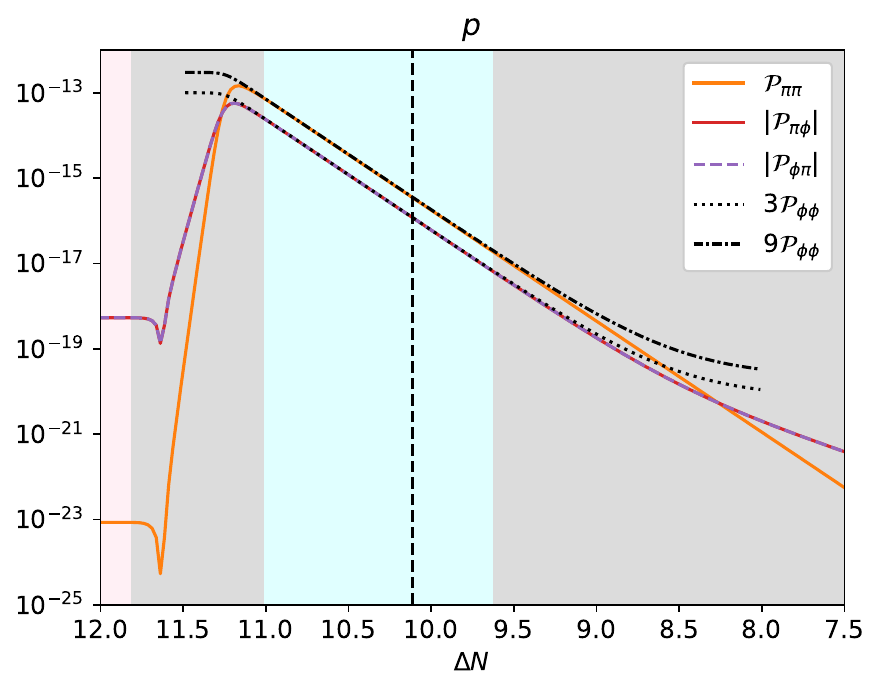}
\end{subfigure}
\begin{subfigure}[b]{0.49\textwidth}
\includegraphics[width=\textwidth]{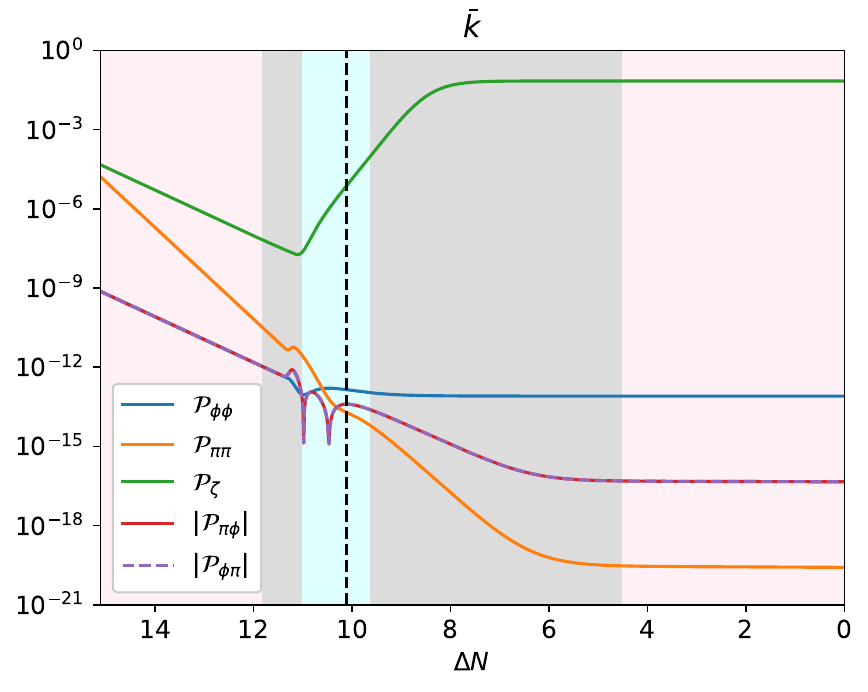}
\end{subfigure}
\caption{E-folding evolution of the field, velocity and curvature perturbation dimensionless correlators and field-velocity cross-correlators for the CMB mode $p$ (top line) and peak scale $\bar k$ (bottom line). The top right panel allows us to zoom into the USR and transition regions. Here, we represent the CMB dimensionless correlators $\dimlessP^{\pi\pi}(p)$, $\dimlessP^{\phi\pi}(p)$ and $\dimlessP^{\pi\phi}(p)$, and compare their numerical solutions with the analytical approximations in Eq.~\eqref{our evolved sigma}. In each panel we highlight times corresponding to SR, USR evolution and transitions between the two in pink, blue and gray respectively. We signal the horizon crossing time of the CMB (peak) scale with a vertical, dot-dashed (dashed), black line.}
\label{fig:correlators CMB and peak}
\end{figure}
In Fig.~\ref{fig:correlators CMB and peak} we display the evolution of the field, velocity correlators and field-velocity cross-correlators for the CMB mode $p$ and the peak scale mode $\bar k$. 

First, let us comment on the evolution of the correlators around the time of horizon crossing. Both for the CMB and peak scale modes, the $\dimlessP^{\pi\pi}$, $\dimlessP^{\phi\pi}$ and $\dimlessP^{\pi\phi}$ correlators decay soon after horizon crossing, leaving $\dimlessP^{\phi\phi}$ as the only sizeable correlator at this time. This shows with a numerical example what discussed in~\S\ref{sec: SR to USR to SR}.
We also note that the $\dimlessP^{\pi\pi}$, $\dimlessP^{\phi\pi}$ and $\dimlessP^{\pi\phi}$ correlators for the large-scale mode $p$ (see the top, left panel of Fig.~\ref{fig:correlators CMB and peak}), reach a constant value at about $\Delta N=14$. 
This is due to the fact that, at that time, the decaying mode ($\dimlessP^{\pi\pi}\propto (k/aH)^4$ and $\dimlessP^{\phi\pi},\, \dimlessP^{\pi\phi}\propto (k/aH)^2$) becomes smaller than the growing mode. The latter can obtained from Eq.~\eqref{eq:pi-SR-attractor}, yielding $\delta \pi_p(t) = -\eta\,\delta \phi_p(t)/2$, valid on super-horizon scales during SR.

On the other hand, when the $\delta N$ calculation of $\dimlessP_\zeta(p)_\text{tree}$ is initialized during the USR phase, see~\S\ref{sec: delta N CMB mode from USR time}, all correlators must be taken into account. This is due to the fact that during USR the correlators $\dimlessP^{\phi\phi}$, $\dimlessP^{\pi\pi}$, $\dimlessP^{\phi\pi}$ and $\dimlessP^{\pi\phi}$ are all of comparable magnitude, i.e. one \textit{cannot} neglect the velocity perturbations. In particular, the numerical results displayed in the top-right panel show an excellent agreement with the analytical expressions \eqref{our evolved sigma}. 

As discussed below Eq.~\eqref{super horizon zeta}, the large separation between the horizon-crossing of the $p$ mode and the onset of USR makes the super-horizon evolution of $\zeta_p(N)$ not appreciable, as shown by the numerical solution to $\dimlessP_\zeta(p;N)_\text{tree}$ in the top-left panel. On the other hand, $\dimlessP_\zeta(\bar k;N)_\text{tree}$ grows from the onset of the USR phase, settling to a constant value when the system has evolved back to SR, see the bottom panel.  

\begin{figure}
    \centering
    \includegraphics[width=0.6\textwidth]{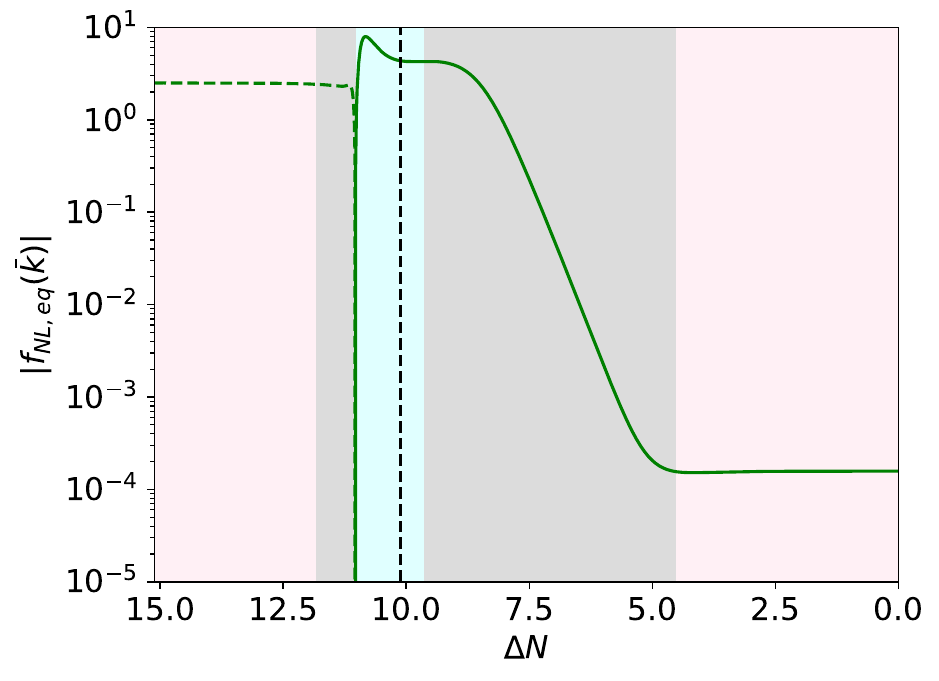}
    \caption{Evolution of $|\fNL|$ equilateral for the peak scale ($k_1=k_2=k_3=\bar k$) against $\Delta N\equiv N_\text{end}-N$. 
    Positive (negative) $|\fNL|$ is represented with a continuous (dashed) line. 
    The black, dashed vertical line marks the horizon crossing time of $\bar k$. We highlight times corresponding to SR, USR evolution and transitions between the two in pink, blue and gray respectively. }
    \label{fig:fNL}
\end{figure}
Before closing, let us comment on non-Gaussianity. The numerical results for the e-folding evolution of the non-linearity parameter $\fNL$, see Eq.~\eqref{f NL def}, are displayed in Fig.~\ref{fig:fNL}. In particular, we consider $\fNL$ in the equilateral configuration $k_1=k_2=k_3=\bar k$. As expected for a model featuring a smooth transition from USR to SR, while $\fNL$ is sizeable soon after horizon crossing, it then decreases during the transition, acquiring a slow-roll suppressed value in the subsequent SR phase \cite{Cai:2018dkf}.   

\end{fmffile}

\bibliography{refs} 
\bibliographystyle{JHEP}

\end{document}